\documentclass[conference]{IEEEtran}
\ifCLASSINFOpdf
\else
\fi
\usepackage{epsfig}
\usepackage{endnotes}
\usepackage{times}
\usepackage{balance}
\usepackage{flushend}
\usepackage{amsfonts}
\usepackage{epsfig}
\usepackage{amssymb}
\usepackage{amscd}
\usepackage{amsmath}
\usepackage{microtype}
\usepackage{array}
\usepackage{graphics}
\usepackage{caption}
\usepackage{multirow}
\usepackage[table,xcdraw]{xcolor}

\usepackage{caption}
\usepackage[hyphens]{url}
\usepackage[
  breaklinks = true,
  unicode = true,
  urlcolor = blue,
  colorlinks = true,
  citecolor = blue,
  linkcolor = red]{hyperref}
\usepackage{soul}
\usepackage{breakurl}
\usepackage{tabularx}
\usepackage{epstopdf}
\usepackage{threeparttable}
\usepackage{xcolor}
\usepackage{subfig}

\newcommand{\ignore}[1]{}
\newcommand{\revised}[1]{}
\newcommand\comment[1]{}

\usepackage{lipsum}
\newcommand{\minus}{\vspace{-0.05in}}
\newcommand{\minusminus}{\vspace{-0.1in}}

\makeatletter
\renewcommand\footnoterule{%
  \kern-3\p@
  \hrule\@width.4\columnwidth
  \kern2.6\p@}
  \makeatother

\begin{document}
%
\title{Guardian of the HAN: Thwarting Mobile Attacks on Smart-Home Devices Using OS-level Situation Awareness}




%
\author{\IEEEauthorblockN{Soteris Demetriou\IEEEauthorrefmark{1},
Nan Zhang\IEEEauthorrefmark{2},
Yeonjoon Lee\IEEEauthorrefmark{2}, 
Xiaofeng Wang\IEEEauthorrefmark{2},
Carl Gunter\IEEEauthorrefmark{1},
Xiaoyong Zhou\IEEEauthorrefmark{3}
and Michael Grace\IEEEauthorrefmark{3}}
\IEEEauthorblockA{\IEEEauthorrefmark{1}Department of Computer Science, University of Illinois at Urbana-Champaign}
\IEEEauthorblockA{\IEEEauthorrefmark{2}School of Informatics and Computing, Indiana University, Bloomington}
\IEEEauthorblockA{\IEEEauthorrefmark{3}Samsung Research America}
\{sdemetr2, cgunter\}@illinois.edu, \{nz3, yl52@, xw7\}@indiana.edu, zhou.xiaoyong@gmail.com, m1.grace@samsung.com
}


\maketitle

\begin{abstract}

A new development of smart-home systems is to use mobile apps to control IoT devices across a Home Area Network (HAN).  Those systems tend to rely on the Wi-Fi router to authenticate other devices; as verified in our study, IoT vendors tend to trust all devices connected to the HAN.  This treatment exposes them to the attack from malicious apps, particularly those running on authorized phones, which the router does not have information to control, as confirmed in our measurement study.  Mitigating this threat cannot solely rely on IoT manufacturers, which may need to change the hardware on the devices to support encryption, increasing the cost of the device, or software developers who we need to trust to implement security correctly. 

In this work, we present a new technique to control the communication between the IoT devices and their apps in a unified, backward-compatible way. Our approach, called \textit{Hanguard}, does not require any changes to the IoT devices themselves, the IoT apps or the OS of the participating phones. Hanguard achieves a fine-grained, per-app protection through bridging the OS-level situation awareness and the router-level per-flow control: each phone runs a non-system userspace \textit{Monitor} app to identify the party that attempts to access the protected IoT device and inform the router through a \textit{control plane} of its access decision; the router enforces the decision on the \textit{data plane} after verifying whether the phone should be allowed to talk to the device. Hanguard uses a role-based access control (RBAC) schema which leverages  \textit{type enforcement} (TE) and \textit{multi-category security} (MCS) primitives to define highly flexible access control rules. We implemented our design over both Android and iOS ($>95\%$ of mobile OS market share) and a popular router. Our study shows that Hanguard is both efficient and effective in practice.

\end{abstract}


%
\IEEEpeerreviewmaketitle

\section{Introduction}
\label{sec:introduction}


The pervasiveness of Internet of Things (IoT) devices has brought in a new wave of technological advances in home automation. According to Gartner~\cite{gartnerIoT}, 6.4 billion IoT devices will be online in 2016, among which a significant portion are \textit{smart-home} systems like smart thermostats~\cite{nestThermostat,honeywell}, fitness trackers, refrigerators, etc., and the number is expected to go above 20 billion by 2020. Examples of such devices include: the Belkin NetCam~\cite{belkinNetcam}, a camera for streaming surveillance video to a mobile phone; the iBaby monitor~\cite{iBaby}, a device for remote babysitting; the Family Hub refrigerator~\cite{familyHub}, which enables online checking of the fridge's contents. Increasingly, these devices are designed to communicate not only with their servers in the cloud but also with other IoT devices and the user's phone over the Home Area Network (HAN), which is typically built around a Wi-Fi router.  For example, Nest Protect Fire sensors~\cite{nestProtect} are capable of propagating an alarm across multiple sensors installed in different rooms of a house. For the convenience of management, such interconnected IoT equipment often relies on the secure connections of HAN (Wi-Fi authentication) for protection and trusts all the computing systems on the same network. This treatment, however, completely exposes the device to the attacks from compromised local systems, a threat becoming increasingly realistic.

\vspace{5pt}\noindent\textbf{Menace of local threats}.  Indeed, it has been reported that high-profile WiFi-enabled smart home devices, including the WeMo Switch and motion sensor~\cite{wemoSwitchMotion, kellyWemo, nobleWemo, smith1, smith2}, Belkin NetCam~\cite{securityFocus}, baby monitoring devices~\cite{storm1, storm2, rapidReport} and smart light bulbs~\cite{goodinPhilipsHue}, are all vulnerable to a local attack: an adversary within the same HAN is shown to be able to control those devices or steal sensitive user information, e.g., live video streams~\cite{securityFocus}, from them. Several studies further reveal that this is possible since such devices have poor---or no--authentication mechanisms~\cite{2014notra, hpStudy, cctvVuln, shekyan2013, kimCCTV, greene2014, hill2015, proofpoint, pierluigi, proofpoint} and therefore easily fall prey to a local attacker. 


Defending against such attacks becomes particularly challenging when the IoT devices are controlled by phones: once the same phone also carries malware (even when the app has nothing but the network privilege), protecting the device it controls becomes impossible at the network level, as the phone is completely legitimate to access the device though the malicious app running on it is not. Given the high smartphone penetration rates~\cite{poushter2016smartphone}, the millions of available mobile applications on both official and third-party markets~\cite{appsGP}, and the ease of distribution of such applications~\footnote{Android applications can be self-signed.}, devices that can be reached through mobile apps can also become an easy target to adversaries. Unfortunately, such adversaries are not only realistic; they are on the rise~\cite{malrise1, malrise2, malrise3}. Because of that they become the main subject of study of many other academic works~\cite{arp2014drebin, Zhou:2012:DAM:2310656.2310710, Grace:2012:RSA:2307636.2307663, Zhou:2012:DRS:2133601.2133640} while concerns are also raised on public communication channels~\cite{malpubcon1, malpubcon2, malpubcon3}. \ignore{In our research, we analyzed the official apps of popular \textit{phone-controlled} smart-home devices and confirmed that a worrisome percentage of them do not have any authentication protection at all (Section~\ref{subsec:analysis}).} In our research we verified that IoT vendors tend to trust the local network (Section~\ref{subsec:security}). This makes them vulnerable to a mobile adversary as we illustrate with attacks on real-world IoT devices, including the \textit{WeMo Switch}, \textit{WeMo Motion}, \textit{WeMo in.sight.AC1} and \textit{My N3rd}. The demos of these attacks can be found on a private website~\cite{hanguardgsites}.

\ignore{We further implemented attacks on 4 of such devices, including the \textit{WeMo Switch}, \textit{WeMo Motion}, \textit{WeMo in.sight.AC1} and \textit{My N3rd}, and demonstrated both the\ignore{devastating} feasibility and consequences of the exploits.\ignore{: XXX.} The demos of these attacks can be found on a private website~\cite{hanguardgsites}.}

Addressing the issue here cannot solely rely on device manufacturers: unfortunately business factors such as time to market and keeping the cost of the device low but also operational factors such as low power consumption, lead to the production of devices without encryption capabilities~\cite{square}.  In such cases, response to threats can only be reactive and it would entail manufacturing a new version of the device which would still leave users with the old version susceptible to attacks.  To make things worse, device manufacturers can be slow in responding~\cite{slow1,slow2} to security and privacy threats. Router vendors have already identified this threat. New hubs and routers pushed onto the market are increasingly armed with various IoT protections (e.g. Microsoft Azure IoT hub~\cite{azureIotHub}, Google's OnHub router~\cite{onHub}. Integrating protection and management capabilities in the router has significant benefits as the infrastructure is already in place in most households and it enables unified policy management. However, as mentioned above, security control at the router level cannot succeed without knowledge of the OS-level situation within an authorized mobile phone, particularly whether a request to a target device comes from its official app or an unauthorized party.  Fundamentally, a practical solution to the problem needs to bridge the gap between the OS-level observation (apps making network connections on a phone) and the network-layer view (requests from the phone for accessing an IoT device), with minimum modifications on the HAN infrastructure and all the systems involved.

\vspace{5pt}\noindent\textbf{Situation-aware device access protection}.\ignore{ With smart-home systems quickly becoming the mainstream, reaching millions of families, the security risks they pose need serious effort to address. A straightforward way to address this problem is to employ intrusion detection techniques at the router, which will attempt to analyze the network traffic and infer the application responsible for a data flow. However, this would be an expensive and non practical process since for every flow, (1) the router will need to first determine the traffic pattern before making a decision and (2) the application will not be able to connect to the IoT device during the inference.}
A simple solution to the problem is just inferring the identity of the app communicating with an IoT device according to its traffic fingerprint.  This approach, however, is unreliable and can be easily defeated by, for example, a repackaged app that closely mimics the authorized program's communication patterns. Also, individual apps' fingerprints need to be reliably generated, deployed and continuously updated, and further to be checked on the router against each communication flow it observes, which adds cost to both the router developer and the user. In this paper, we present a different approach, a new technique that achieves fine-grained, situation-aware access control of IoT devices over a home area network.  Our approach, called \textit{Hanguard}, distributes its protection logic across mobile phones and the Wi-Fi router for jointly constructing the full picture of an IoT access attempt during runtime, which is then utilized to control the access on the network layer. More specifically, on the phone side, the information about the app making network connections is collected and passed to the router; on the router side, security policies are enforced to ensure that only an authorized app can touch a set of functionalities the device provides. In this way, malware on network-authenticated phones can no longer endanger the operations of the IoT devices, even when the IoT devices are not equipped with proper authentication and encryption protection.

Hanguard is designed to directly work on the existing HAN infrastructure, without modifying mobile operating systems or IoT devices. To deploy the system, one only needs to install a \textit{Monitor} app with non-system privileges on mobile phones and update the firmware of the Wi-Fi router with a security patch. A key technical challenge here is how to gather situation information (processes making network connections) on mobile phones, which is not given to a third-party userspace app on both Android and iOS. Although all these systems provide VPN support, the app using the service still cannot observe the process generating traffic and will significantly slow down the network communication of the whole system (Section~\ref{subsec:phone}). To address the issue, we leverage side channel information for lightweight discovery of runtime situation on Android and utilize the VPN to only mark out authorized apps' traffic on iOS (Section~\ref{subsec:phone}). Such information is then delivered to the router through a separate control channel, which is synchronized with the traffic generated by the app (over a data channel) and used by the router to determine whether the communication should be allowed to proceed. 

We implemented our design over both Android and iOS which cover more than 95\% of the mobile OS marketshare~\cite{idc:mobileos}, and a TP-Link WDR4300v1 Wi-Fi router.  Our evaluation shows that Hanguard easily identified and blocked all unauthorized attempts to access IoT devices with negligible overhead in the common case. (Section~\ref{sec:evaluation}).

\vspace{3pt}\noindent\textbf{Our contributions}.  The contributions of the paper are summarized as follows:

\vspace{3pt}\noindent$\bullet$\textit{ New understanding}. We found that \textit{IoT vendors treat the HAN as a trusted environment}. This treatment leaves the devices vulnerable to a new type of confused-deputy problem, when a malicious app utilizes an authorized phone to gain unauthorized access to IoT devices through the HAN. We further demonstrate the grave consequences of such attacks on four real IoT devices. Our findings highlight the need for proactive protection built within the HAN.
\ignore{
including the \textit{WeMo Switch}, \textit{WeMo in.sight.AC1},  \textit{WeMo Motion} and the \textit{My N3rd} device
We analyzed popular phone-controlled home IoT devices and found that a worrisome percentage of them are open to attacks from mobile malware.  Our study reveals the threat posed by this new type of confused-deputy problem, when a malicious app utilizes an authorized phone to gain unauthorized access to IoT devices through the HAN, and its serious consequences, which are demonstrated on 4 real IoT devices including the \textit{WeMo Switch}, \textit{WeMo in.sight.AC1},  \textit{WeMo Motion} and the \textit{My N3rd} device. The findings highlight the urgent need for effective protection against the threat.
}

\vspace{3pt}\noindent$\bullet$\textit{ New access control mechanisms}. We have utilized \textit{type enforcement} and \textit{multi-category security} principles to design a new fine-grained access control mechanism for WiFi devices.

\vspace{3pt}\noindent$\bullet$\textit{ New system techniques}. \ignore{We developed Hanguard, the first practical, backward compatible protection against attacks on smart-home devices.} Hanguard employs a new software-defined networking (SDN) approach applied on existing infrastructure of home area networks: it features a new controller system architecture distributed across HAN phones and the router. To the best of our knowledge we are the first to use phones as Monitors for local area SDN. Our design can have applications in enterprise settings, peer-to-peer networks and others.
\ignore{Our system is designed to efficiently gather situation information across phones and the network, and utilize it for enforcing an SELinux-like policy model constructed transparently to the users for fine-grained access control. The system is characterized by its negligible performance impact and its convenience to deploy over today's HAN.}

\vspace{3pt}\noindent$\bullet$\textit{ Implementation and evaluation}. We implemented Hanguard on both Android and iOS phones, and a commercial router, and evaluated it against attacks on real-world IoT devices and on various performance metrics.  Our study demonstrates the practicality and efficacy of the new system.

\vspace{5pt}\noindent\textbf{Roadmap}. The rest of the paper is organized as follows: Section~\ref{sec:background} motivate the work and Section~\ref{sec:threat} presents our study on popular smart-home devices; Section~\ref{sec:hanguard} presents Hanguard and Section~\ref{sec:evaluation} our evaluation; Section~\ref{sec:analysis} conducts a security analysis of our system and Section~\ref{sec:discuss} discusses Hanguard.\ignore{'s  flexibility in switching between monitoring traffic techniques}; Section~\ref{sec:related} reviews related prior research and Section~\ref{sec:conclude} concludes the paper.

\section{Motivation}
\label{sec:background}
\noindent\textbf{IoT on HAN}. Home automation systems today are increasingly connected to the Internet and to each other. Examples of such devices include smart cameras~\cite{belkinNetcam, canaryCam, arloCam, dropCam, homeboy}, various sensors~\cite{nestProtect, wemoSwitchMotion, kornersafe, bleepbleeps, netatmo, quirky}, smart door bells~\cite{ring, skybell} (with HD video, motion sensing and bidirectional audio capabilities), smart cooking appliances like Mr Coffee~\cite{mrcoffee} (for remote control of coffee brewing), smart gardening products such as OpenSprinkler~\cite{opensprinkler} (for remote management of irrigation) and more. These devices are typically connected to a HAN through its Wi-Fi router. To make this happen, one uses a smartphone to communicate with a temporary access point created by the IoT device to configure the device, entering login credentials for the device to establish a Wi-Fi connection with the router.  Through the connection, the device can talk to its cloud service and receive commands from the service and the user's phone. A conventional way to do that is to let the phone control the device through the cloud service, even when both the phone and the IoT devices are within the same HAN.

Although this treatment simplifies the IoT-control mechanism, restricting an authorized app to always communicate with the IoT device in the same way regardless of whether they are on the same network, it comes with availability and performance penalties. In fact this has already caused a lot of trouble: e.g., the Ring doorbell~\cite{ciprianiRing} is reported to take 30 seconds to deliver the notification to the user in some cases; Canary~\cite{brownCanary} and Scout alarm~\cite{scoutCommunity} was found to be unable to send out an alarm once the Internet is down.  Also users of the Lutron bridge~\cite{lutronForum} and the Chamberlain garage opener~\cite{gibbsChamberlain} complained that they need the Internet to turn on the light and open the garage door. In response to such concerns, support for direct phone-device communication through the HAN becomes a requirement. For example Lightify claims that from version 1.0.3b11, its users will be able to control lights offline~\cite{lightifyFaq}. Another prominent example is Samsung's Smarthings---a leading IoT hub: Smartthings reported their plans to support local computation, which enables its systems to work together even without Internet~\cite{smarthingsLocal}; in fact apps and devices which previously existed in the cloud, are now moved to the local hub~\cite{localConnectionsSupportSmarthings}.

\vspace {3pt}\noindent\textbf{Smart-home security}. The popularity of the smart-home devices also comes with new security risks. As mentioned earlier, several studies show that most IoT devices today are not adequately protected, leaving the door widely open to different kinds of attacks. A prominent example is Shodan~\cite{shodan, cctvVuln, shekyan2013}, a search engine for IoT, which has discovered a lot of vulnerable webcams online (ports made open to the public and with only weak authentication protection), exposing private information such as video streams of a sleeping baby~\cite{shodanBaby}. Moreove, a recent Mirai-based botnet, took advantage of smart-home devices weak authentication to launch a massive DDoS attack in the US~\cite{vergeDDOS}. In addition to such remote attacks, smart-home devices are also found to be subjected to local threats. It has been reported that through a laptop or a desktop running in the same network, one can gain unauthorized access to the Wemo Switch and Motion sensor~\cite{kellyWemo, nobleWemo, smith1, smith2}, the Belkin NetCam~\cite{securityFocus}, the Philips Hue smart light bulbs~\cite{goodinPhilipsHue}, etc.  Our research further shows that such attacks can be launched from a malicious app running on a smartphone and the problem becomes particularly serious when the phone is authorized to access the IoT device while the app is not (Section~\ref{subsec:security}).

Such mobile malware have been established as one of the most prevalent cyber-threats. On the one hand some reports show low mobile device infection rates by malicious applications in the US~\cite{lever2013core}, but on the other hand these reports are somewhat variable~\cite{faruki2015android}. Furthermore, there are increasingly more potentially harmful applications in other parts of the world e.g. in China. More importantly, the mere presence of mobile malware given the ease of their distribution (self-signed apps, third-party markets) and the rise of mobile malware~\cite{malrise1, malrise2, malrise3}, render it a security threat. Mobile malware became the main subject of study of many previous academic works which have already illustrated the prevalence and severity of this problem~\cite{arp2014drebin, Zhou:2012:DAM:2310656.2310710, Grace:2012:RSA:2307636.2307663, Zhou:2012:DRS:2133601.2133640}.

\vspace {3pt}\noindent\textbf{Need for proactive device-independent protection}.
One way to mitigate the mobile adversary on home IoT devices is by fixing all the security vulnerabilities on the IoT devices. However, it is generally accepted that fixing the security issues within embedded devices is difficult~\cite{slow1, slow2, square}. Unlike traditional computers, whose security-critical vulnerabilities are often addressed through patching, no clear path is available for patching and upgrading IoT devices once they leave the manufacturer's warehouse\ignore{~\cite{}}. The problem is caused by a lack of computing power on some devices or the patching infrastructure, or in some other cases, by the use of third-party hardware, software and other resources, which are hard to patch by the device manufacturers themselves. Consider for example, the case of the square dongle which is used for secure payments. Driven by time to market goals and low cost requirements, the vendor shipped the earlier version of the dongle with no hardware support for encryption~\cite{square}. To fix the problem, Square had to produce a new dongle device. Moreover, previous work~\cite{naveed2014inside} demonstrated that various Bluetooth medical devices also suffered from similar issues, lacking authentication and encryption. As a result of such practices, many faulty systems are still in use even after their problems have been reported, leaving them even more vulnerable since their weaknesses already become public knowledge. Lastly, many reports indicate that these problems extend to IoT devices~\cite{hpStudy, cctvVuln, shekyan2013, kimCCTV, greene2014, hill2015, proofpoint, pierluigi, proofpoint}. We believe that the answer to this challenge for smart home WiFi devices, is a network-level protection that utilizes distributed app-level awareness from existing user-managed devices. The system should be deployable on the existing HAN infrastructure; it should also enable access control management of smart-home devices and safeguard them and their users even when the devices themselves are vulnerable.

\section{Operations and Existing Trust Model}
\label{sec:threat}
\begin{figure*}[!t]
    \centering
    \null\hfill
    \subfloat[IoT Products Functionality Categorization.]{\includegraphics[width=0.3\textwidth]{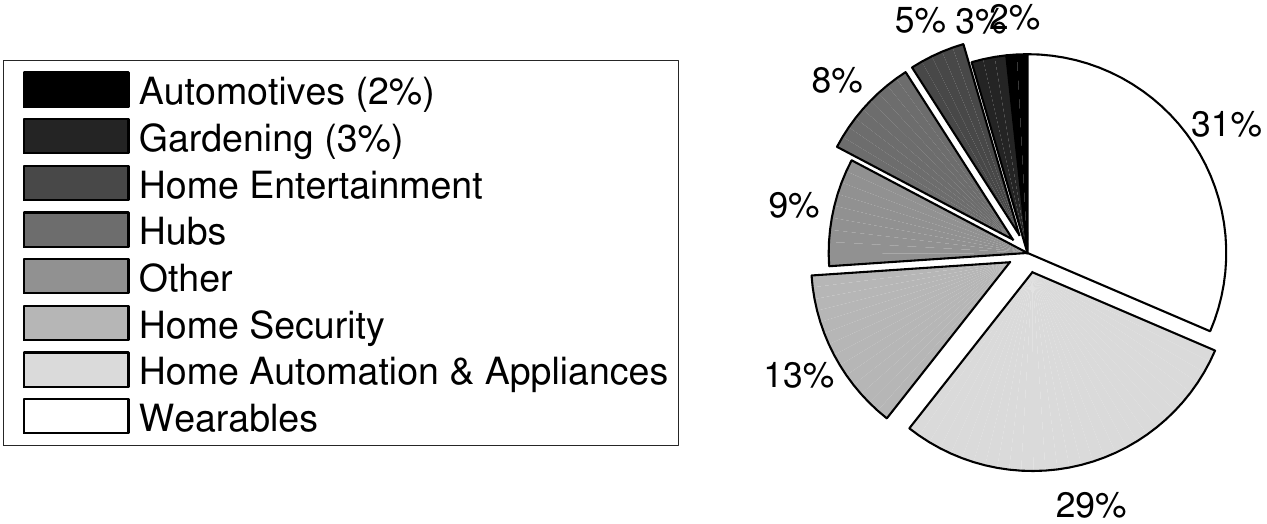}\label{fig:func_all_mirror}}
    \hfill
    \subfloat[Setup Phase of WiFi Devices.]{\includegraphics[width=0.3\textwidth]{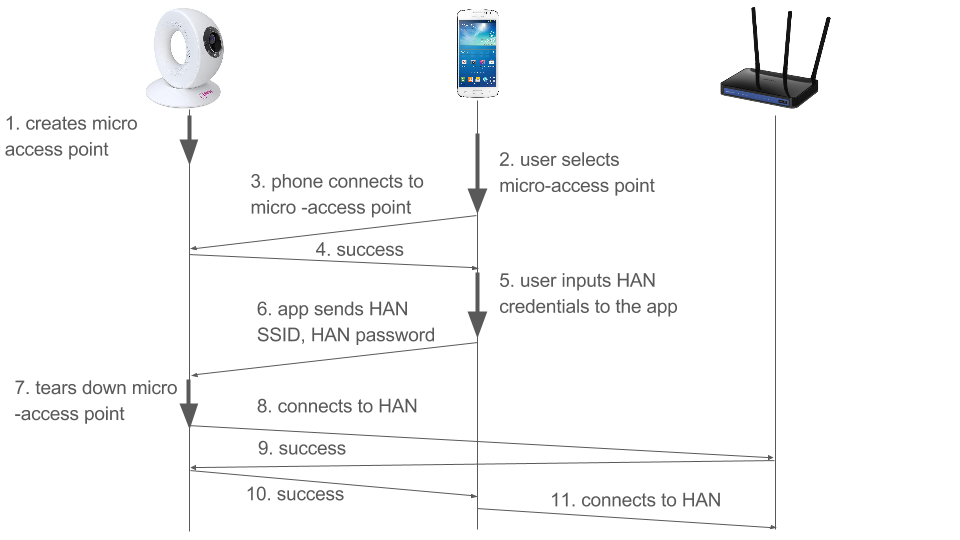}\label{fig:setupPhase}}
    \hfill
    \subfloat[Communication Phase of WiFi Devices.]{\includegraphics[width=0.3\textwidth]{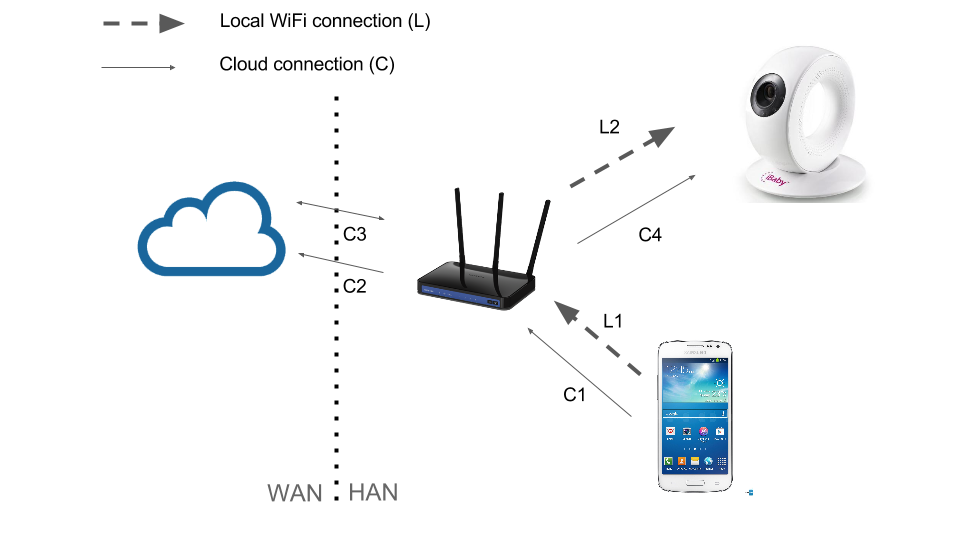}\label{fig:communicationPhase}}
    \hfill\null
    \caption{\label{fig:shape_snapshots}Operations of Smart Home Devices.}
    \minusminus
\end{figure*}

We performed an analysis of devices' operations, their HAN trust model and the implications stemming from a mobile adversary. Our findings informed Hanguard's design decisions.


\vspace {5pt}\noindent\textbf{Methodology.} One approach for our study would be to investigate the IoT devices' firmware. That would entail---after identifying such devices--finding images of their firmware or, for each device, buying the device and extracting its firmware. Subsequently, each firmware needs to be analyzed, which is a non trivial task~\cite{costin2014large}. However, most of these devices are now controlled by mobile apps. Thus their control mechanisms can be examined by analyzing the apps instead of the firmware. Note that, our approach has multiple benefits over analyzing firmware: (1) we can easily acquire Android apps, (2) there is no monetary cost, (3) it is generally easier and faster to analyze mobile apps than an embedded device's firmware. 

The most straightforward way to find the Android apps of the IoT devices, is to search for them at Google Play using keywords such as ``home automation'' and ``internet of things'', which, however, turned out to be less effective: through manual inspections of search outcomes, we found that many apps identified this way were not related to any IoT systems and in the meantime, popular IoT apps fell through cracks.  Our solution is to crawl \url{iotlist.co}, a popular site for discovering IoT products. From the list, the crawler we ran collected the meta-data of 353 products~\footnote{\small all listed products at the time of the study}, including ``Title'', ``Description'', ``Product Url'', ``Purchase Url'' and others. Such data was further manually checked to identify a list of package names for the official apps of these devices.  Searching Google Play using the list, our crawler downloaded the apps and their meta-data from the Play store. Out of the 353 products we found that 63\% (223) of them have apps on Google Play, 2\% (7) are iOS only and the rest are mostly unfinished products (listed on \url{kickstarter.com} and \url{indiegogo.com}) or\ignore{those that} are no longer available. This indicates that indeed most IoT devices today are controlled by smartphones. The APK files of these apps were decompiled (using apktool~\cite{apktool}) and their \texttt{.smali}~\footnote{smali is a human readable representation of Android bytecode (.dex).} and manifest files~\footnote{A manifest file indicates what security permissions an app requires and various application building blocks.} were extracted. Further, each app's machine code was converted to Java code using dex2jar~\cite{dex2jar} for manual analysis.

\subsection{Operations of Smart-Home Devices}
\label{subsec:analysis} 
To better understand the operations of smart home devices, we manually went through (1) the meta-data of the collected products, (2) their online documentations and websites, and (3) through their apps' source code when available. Figure~\ref{fig:func_all_mirror}\ignore{Table~\ref{tab:functionalities}} illustrates our manual categorization of the IoT products based on their functionality. Note that the \textit{Wearables} category (31\%) embodies mostly fitness and location trackers, smartwatches and personal medical devices. We call such devices \textit{personal} devices; these commonly use Bluetooth to connect to a smartphone app. Previous work has already studied the security of personal devices, they found problems with encryption and authentication and proposed solutions~\cite{naveed2014inside, demetriouNDSS2015}. From the figure, we can also see that most of the listed IoT devices (55\%) are smart home automation/entertainment/security/hub systems, which are the focus of our study (see Appendix~\ref{sec:appendix} for a complete categorization). We call these \textit{shared} devices. Such devices could directly benefit from an access control scheme built within the HAN. Previous work on shared devices, was focused on a single IoT integration platform (hub)~\cite{fernandes2016security, fernandes2016flowfence}.

In our research we wanted to identify the workflow of the WiFi-enabled smart-home devices irrespective of the existence or non existence of a proprietary integration platform (hub). We identified 2 high level phases of operation: the \textit{setup} phase and the \textit{communication} phase. During the setup phase, the system provides the user with the means of communicating the WiFi credentials to the smart device. Then the smart device uses the credentials to authenticate itself to the local network. Figure~\ref{fig:setupPhase} illustrates the most common setup scenario among the WiFi devices. In \textbf{Step 1}, the smart device creates a micro access point and advertises its SSID. This is is usually initiated by the user pushing a specific sequence of buttons on the device. In \textbf{Steps 2,3 and 4}, the user utilizes a smart phone app to directly connect to the device. The app searches for the advertised SSID and guides the user to select the device network. Once the smartphone connects to the micro-access point, the app asks the user to input the HAN WiFi credentials to the app (\textbf{Step 5}). Then the app transfers the credentials to the device (\textbf{Step 6}). In \textbf{Step 7,8, and 9}, the device tears down the micro-access point and attempts to connect to the HAN WiFi network. The user is notified through the app regarding the result of the attempt (\textbf{Step 10}). Once the device is connected to the HAN the app asks the user to connect her phone to the HAN as well to interact with the device (\textbf{Step 11}). Other IoT vendors choose to complete the setup phase through a \textit{tethered mode}: the user is expected to connect her smartphone with the device through a cable (e.g. audio jack, usb). The credentials are passed by the app using the tethered channel. Evidently the setup phase can be cumbersome to the users but is expected to happen mostly once per device. Some vendors do try to streamline this process. For example some devices use a \textit{no-network mode}. In this case the user could input the WiFi credentials to the app and then show the smartphone screen to an optical sensor on the device. The device recognizes and parses the credentials.

During the communication phase, the user can send commands to the smart device through the app. This can happen through 2 main avenues as illustrated in Figure~\ref{fig:communicationPhase}. The app can (a) attempt to connect to the cloud service of the device. In this case the cloud service is responsible for authenticating the app and relaying the command to the appropriate IoT device. Alternatively (b) the app connects directly to the device through WiFi. Obviously the former requires constant Internet connection and comes with latency penalties. Thus vendors (e.g. Samsung Smartthings) are increasingly turning into using the cloud only for remote control and turning to WiFi for local control.

\subsection{HAN Trust Model of Smart-Home Devices}
\label{subsec:security}
In our research, we further investigated the trust model of WiFi smart-home devices. Prior research already demonstrated that the interaction between smartphone apps and the cloud is alarmingly unguarded~\cite{chen2014oauth, wei2014amandroid}. On the other hand, the local communication between the apps and the devices is not as well understood. In fact \textit{it is unclear whether app developers and IoT device manufacturers treat the local network and everybody connected to it as trusted entities} and whether such treatments leave the devices susceptible to attacks from both local adversaries and remote adversaries that gain access to the HAN. Moreover, even though it has been reported that IoT devices come with serious problems ~\cite{hpStudy, cctvVuln, shekyan2013, kimCCTV, greene2014, hill2015, proofpoint, pierluigi, proofpoint}, little has been done to understand the security risks stemming from malicious mobile apps. This is particularly important since, IoT devices are controlled by apps which send commands either through the cloud or the local network. Here, we aim to bridge these gaps in knowledge. Our findings build on to the existing evidence which collectively support the need for a unified security and management system built within the HAN to safeguard today's smart-home devices.

To facilitate our manual analysis we further built a parsing tool which simply looks for the presence of password requests in the layout files of the apps. Android enables developers to create the layout of a screen statically using xml layout files. In such files, one can specify a hint to a text field using \texttt{android:hint="my\_hint"}. Developers can also explicitly associate a field with a password using \texttt{android:password="true"} for API level 3 and earlier, or \texttt{android:inputType="textPassword"} for later versions. Our tool looks in all xml files under the layout directories of an app under analysis for the presence of these definitions; where \texttt{my\_hint} we used \texttt{password} and \texttt{passphrase}.

Our study aims to achieve the following goals: (a) Find out whether vendors and developers of WiFi smart home devices/apps erroneously treat the home area network as a trusted environment; (b) Find out how easy or hard it is for a mobile adversary to take advantage of unguarded local smart home devices in practice.

\hspace{-7pt}\textbf{HAN Trust Model.}  We performed a statistical significance test focused on the following \textit{null} hypothesis ($N_0$): \textit{HAN apps with only remote connections are equally likely to perform authentication compared to HAN apps with only local connections}. To answer this question we separated our collected IoT apps into two groups. Apps with only remote connections and apps with only local WiFi connections. We used 55 unique Android applications with WiFi/Internet only connections to HAN IoT devices. A full list of the apps selected for HAN trust model analysis is provided in the Appendix~\ref{sec:appendix}. 


To separate the apps into the two groups, we manually went through (1) their online documentations and websites, (2) public forums, and (3) their Java Android code. We found that 22 (40\%) do perform some internet socket connection with local discovered devices or fixed local IPs~\footnote{Note that if the connection is construed to be performed for the \textit{setup} phase, we do not regard it as a local connection.}. 25 (45\%) were found without local WiFi connections, 5 (9\%) we could not determine, for 2(4\%) decompilation failed,  and 1 (2\%) was by that time removed from Google Play. For each of the 2 sets (local; no local) we analyzed them further to discover whether they perform any authentication. For the ones that perform only remote connections we used our parsing tool that searches for password requests in the layout files of the apps. We found that out of 25 apps, 16 do request a password and 9 do not. Since, our tool could miss password requests that are not defined statically, we manually went through the 9 apps flagged as performing no authentication and found that actually a password was used in some respect in all 9 of them. In particular, 7 of them were web apps using libraries such as the \texttt{cordova} library that allows developing apps with web technologies (e.g. Javascript). The remaining two were constructing the user interface element responsible for the password field in code.

For the 22 apps with local WiFi connections we could not simply use the above tool since it would reveal little to no information on whether a password is used for a connection with the IoT device or the cloud. Thus we manually went through their code looking for network API calls responsible for local connections (e.g. creation of sockets connecting to local IPs, or UPnP discovery). We examined the calls to such APIs and found that 9 of the apps do not authenticate to the IoT device. \ignore{The results are summarized on Table~\ref{tab:g2}.}

\ignore{
\begin{table}[]
\scriptsize
\centering
\caption{Contingency table of HAN IoT apps.}
\label{tab:g2}
\begin{tabular}{l | c | c}
                           & \multicolumn{1}{l}{\textbf{w/ authentication}} & \multicolumn{1}{l}{\textbf{w/o authentication}} \\ \hline
\textbf{w/ local WiFi connections}  	& 13                                    & 9                                      \\
\textbf{w/o local WiFi connections} 	& 25                                    & 0                                     
\end{tabular}
\vspace{-1em}
\end{table}
}

To determine whether apps with local connections are less likely to perform authentication one could perform a $\chi^2$-test of independence. A challenge we had was the small absolute number of relevant available apps derived from \url{iotlist.co}. Because of this, the $\chi^2$-test of independence might not derive  statistically significant results. To overcome this we used the Fisher's exact test~\cite{fisher1922interpretation}. This is a common approach to derive statistically significant results when the sample size is small. We leveraged a tool by Carlson et. al.~\cite{carlson2009estimating} to perform such a test on the our \textit{null} hypothesis ($N_0$). A 2-sided P value less than 0.05 was considered significant. 


The test yielded a 2-sided P-value of $0.00036 < 0.05$ and thus we can reject $N_0$. Therefore, we can now confidently say that \textit{HAN apps with local connections are less likely to get authenticated by smart-home devices}. This validates an important intuition that IoT vendors consider the HAN to be a trusted environment. However, given the fact that phones are an integral part of such a network and that phones can carry self-signed apps from third-party markets, this treatment becomes detrimental to the security of HAN IoT devices. This result further highlights the need for an access control system that can be integrated in home area networks with minimal changes to the existing infrastructure, that is backward compatible, independent of vendor and developer practices and that allows the users the flexibility to manage and control who should communicate to which device.

\hspace{-7pt}\textbf{HAN mobile adversary.} The previous finding is particularly alarming. Next we attempt to illustrate how a weak mobile adversary can take advantage of this problematic trust model and compromise smart home devices. Towards this end, we cherry-picked four devices with local connections and authentication issues and attempted to perform real-world, practical attacks. The devices we picked are listed on Table~\ref{tab:effectiveness}. Our targets include the \textit{WeMo Switch} and \textit{WeMo Motion}~\cite{wemoSwitchMotion}, the \textit{WeMo in.sight.AC1}~\cite{wemoInsight}, and \textit{My N3rd}~\cite{myn3rd}. The \textit{WeMo} devices are examples of popular plug-and-play devices.  Just on Android, the official app of the WeMo devices was download 100,000--500,000 times~\footnote{\small This is a conservative number as people can download the app from alternative Android app markets or from iTunes for iOS devices.}.  Note that all the WeMo devices are manufactured by a single vendor. By focusing on three WeMo devices we want to showcase how an erroneous trust model by a vendor can spread across various of its devices. This suggests that trusting the local network was a design decision and not an implementation issue manifesting in an isolated device. My N3rd, while not yet popular, it is chosen to showcase a new category of do-it-yourself (DIY) devices. It allows one to connect it to any other device enabling turning on/off that device from the My N3rd mobile app. Increasingly more such projects appear on the market with Arduino-based projects taking the lead. While exciting for users, such devices tend to inherit the problematic trust model and allow an adversary to take full control of ones devices.

In our experiments we consider a mobile adversary that tries to get unauthorized access to the IoT devices. The mobile adversary can perform an attack from an unauthorized phone, or from an unauthorized app on an authorized phone~\footnote{Note that the case of an unauthorized app on an unauthorized phone trivially reduces to the first case we consider.}. To test the above cases, we use 2 Nexus phones. The first one is assumed to be untrusted and the second one is assumed to belong to one of the HAN users. We then tried to access the target IoT devices using both phones. Unfortunately we found that the adversary can trivially connect and control all devices. The video demos of our attacks can be found online~\cite{hanguardgsites}.


\begin{table}[]
\scriptsize
\centering
\caption{Example devices picked for real-world attack demonstrations.}
\label{tab:effectiveness}
\begin{tabular}{lcc}
\textbf{Target Device} & \textbf{Description} & \textbf{\# App Installations}  \\ \hline
WeMo Switch            &  Actuator & 100K - 500K  \\
WeMo Motion            &  Sensor   & 100K - 500K  \\
WeMo Insight Switch    &  Actuator & 100K - 500K \\
My N3rd			       &  Actuator & 100 - 500 
\end{tabular}
\vspace{-1em}
\end{table}

\section{Hanguard Design \& Implementation}
\label{sec:hanguard}
In this section, we elaborate on our design of Hanguard and its implementation over HAN and mobile platforms.

\subsection{Design Overview}
\label{subsec:overview}
\noindent\textbf{Adversary model}. As mentioned in Section~\ref{sec:threat}, IoT devices are controlled through smartphone apps. These devices are designed to act blindly on the commands from authorized phones  (based upon their authentication with the HAN router). This treatment becomes increasingly problematic: while the smartphone may indeed belong to a rightful user, the applications that it runs can come from less known places (e.g., third-party app stores) and less trustworthy developers (e.g., malware authors). Given smartphone penetration~\cite{poushter2016smartphone}, prevalence and ease of distribution of mobile applications~\cite{appsGP}, adversaries can now find their way to the HAN through a legitimate phone with minimal effort. Moreover---as demonstrated on Section~\ref{subsec:security}---given the erroneous threat model of today's IoT devices, which trusts all the requests issued from a trusted source (a router or phone), such malicious applications can easily gain unauthorized control of IoT devices (e.g. turning on/off an actuator, or reading the collected data of a sensor)\ignore{~\ref{sec:threat}}.

Thwarting such attacks is inherently hard. A straightforward solution is to implement a unified security logic in the router, since traffic from applications to IoT devices goes through it. However, the router alone does not have enough information to make any \textit{application level access control} decisions. One could resort to traffic fingerprinting techniques to infer the application generating the traffic. The approach can (1) be easily evaded by a malware repackaged from an authorized app, (2) bring in false alarms and (3) impacts the performance of the router.

Hanguard is designed to address the issue through bridging network and application level semantics, binding an app's identity to its traffic to enable a fine-grained access control on IoT devices. In the meantime, it does not modify both software and hardware of these devices, the operating systems of smartphones, and does not make assumptions about the router hardware. For this purpose, our adversary model is focused on the situation where a malicious app is installed on a smartphone device authenticated to the HAN. The adversary is considered to already know the communication protocol used by the victim IoT. We further assume that the smartphone hosting the app has not been compromised at the OS or hardware level, which limits the adversary to the user land, at the app level.  Note that though outside our adversary model, Hanguard can also provide coarser-grained protection against guest phones and compromised phones, remote adversaries and more traditional WiFi attacks. To avoid confusion we discuss how this can be done separately (Section~\ref{subsec:beyond}).

\begin{figure}[t]
\centering
\includegraphics[width=8cm]{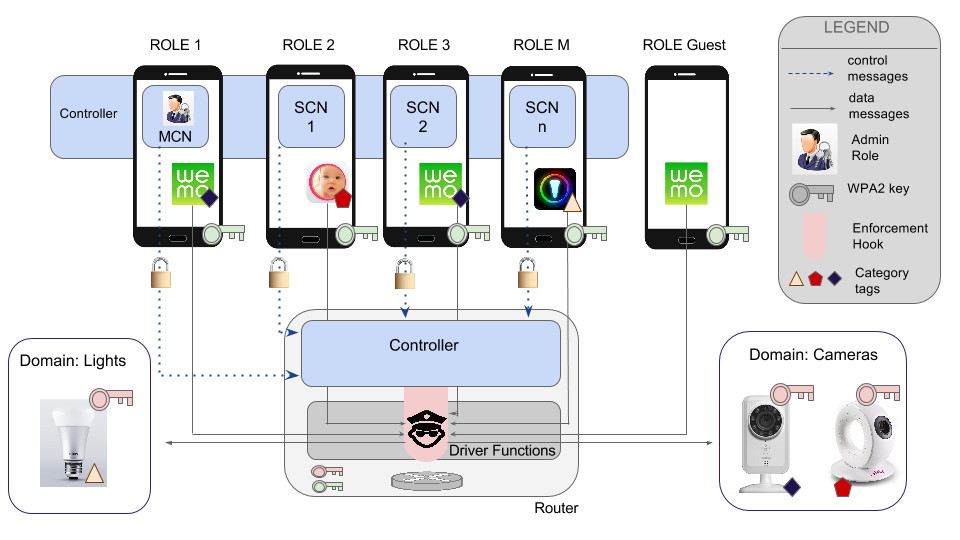}
\caption{Hanguard high level architecture.}
\label{fig:hanguard_arch}
\minusminus
\minusminus
\end{figure}

\vspace {5pt}\noindent\textbf{Idea and architecture}. Figure~\ref{fig:hanguard_arch} illustrates the architecture of Hanguard.  Our design is partially inspired by software defined networking (SDN) (see~\cite{sdnSurvey} for a survey), which separates the network traffic (\textit{data}) from its management (\textit{control}).  In the meantime, Hanguard is meant to be easily deployed to today's HAN.  Serving this purpose is a distributed security control architecture that includes a \textit{Controller} on a HAN router for policy enforcement and a \textit{Monitor} on the user's phone for collecting its runtime situation and making access decisions (which are enforced by the router). To avoid changing the mobile OS, the Monitor is in the form of a user-space app. It detects the app making network communication and its compliance with security policies, and then pushes the access permit to the router's Controller through a secure control channel (Section~\ref{subsec:phone}). The router utilizes that information to enforce the policy (Section~\ref{subsec:routerenforce}): only the traffic with a permit from the Monitor is allowed to reach IoT devices.

In essence, this design preserves the data channel within which unmodified information from smartphone apps is propagated to the router, and creates an independent control channel for security decisions. Such a separation, it comes with obvious performance benefits: no extra headers to be processed by the router on a per packet basis in the data channel. It can also guarantee that control information is always transmitted through a secure channel, and allows the router to further enforce policies and ensure, even in periods of heavy congestion, that security decisions are delivered in a reliable manner. In addition, our design allows for a clear separation of tasks: the security policies can be easily managed by the user through a mobile app interface; \ignore{the app-level control logic is placed on Monitors northbound from the router;} the router reduces to simply enforcing the flow decisions. This keeps the router as simple as possible and allows for readily updating the security logic with a mere application upgrade.


\vspace {5pt}\noindent\textbf{Policy Model}. Using the SELinux~\cite{peter2001integrating} type-enforcement (TE) scheme, one tags subjects (e.g. processes) with \texttt{domains} and objects (e.g. files) with \texttt{types}. Then a policy rule can be written (usually by security experts) that specifies which domain can access which type. By default all interactions are forbidden unless a policy rule is in place that allows the interaction. The same concept was introduced for the Android OS as well (SEAndroid~\cite{smalley2013security}) where apps are assigned domains and resources are assigned types~\footnote{Since version 4.4 SEAndroid is incorporated in enforcing mode on Android devices}. In both SELinux and SEAndroid, one could use the concept of multi-category security (MCS): MCS allows for tagging a subject and an object with one or more category tags. When an SELinux policy is enforced, the system first checks the TE rules to decide whether the interaction is explicitly allowed. Then, the category check is applied to determine whether the subject and object also belong to the same category. Since the MCS check is applied after the TE check, it can only further restrict security. For example it can be used to segregate departments in an enterprise setting. On Android it is used to enforce the multi-user functionality. 

Hanguard implements an RBAC (\textit{role-based access control}) policy model which leverages \textit{type-enforcement} and \textit{multi-category security} primitives. It uses them in a unique way to create SELinux-like policy rules, to protect smart-home devices. However, Hanguard does not need security experts to create the policies; policies are generated at runtime and transparently to the user. In particular, the user is only expected to perform simple mappings between a finite set of IoT apps, IoT devices and HAN users. Default policies are automatically created at \textit{setup} phase to further reduce users' burden. Hanguard's access control model parses such mappings and assigns a \textit{category} tag to each app and its respective IoT device. Further, each IoT device is labeled with a \textit{type}. \textit{Types} can be organized in overlapping groups called \textit{domains}. Each mobile phone is assigned a \textit{role} and each role can be configured to access a number of domains. For example, the iBaby camera can be labeled with the \textit{type} ``babyMonitor\_t''. A \textit{domain} ``cameras\_d'' can be created to encompass the ``babyMonitor\_t'' \textit{type} device among others. Lastly, the \textit{role} of a HAN user's phone (e.g. ``Adult'') that is supposed to be able to access the cameras, can be configured as eligible to access the ``camera\_d'' \textit{domain} and in extend the  ``babyMonitor\_t'' \textit{type} device. This is analogous to the \textit{type-enforcement} scheme in SELinux which binds processes to resources. Here, the relation between the \textit{role} and the \textit{domain} ensures that an untrusted phone (e.g., a visitor's phone) cannot touch protected devices and even an authorized phone, once compromised, cannot communicate with the IoT devices it is not supposed to talk to (see Section~\ref{sec:analysis}). At the same time and orthogonally to the type-enforcement scheme, the iBaby camera and its official app, can be assigned the \textit{category} ``iBaby''.  The \textit{category} here binds a specific app on a phone to the device the phone is authorized to access. For example, the role ``Adult'' can be configured to access the domain ``cameras\_d''; while that stipulates that the adult's phone can control the baby cameras, access is not granted unless the app on her phone and the actual baby camera that it tries to reach are tagged with the same category. Note that more than one category tags can be associated with a domain. This enables the generation of a policy rule which allows an app (subject) to access multiple devices (resources) of the same type. 

By default, a phone registered with the HAN is assigned the role ``HAN user'', which is allowed to access the ``Home'' domain. The latter encompasses every newly installed IoT device (which is assigned a unique \textit{type}). However, the access can only succeed when the app on the phone is given the same category tag as the device it attempts to reach. Such an app-device binding is established when the app is used to configure the device, which is established through a special device, a phone or a PC, that takes the role of an \textit{Admin}. This role can configure the router, register other user phones, access all domains and update security policies. During a policy update, new domains, roles and access relations between them can be generated. The policy model also handles unregistered phones (e.g., those belonging to visitors), which connect to the Han as a ``Guest'', a \textit{role} not allowed to interact with the devices in the ``Home'' \textit{domain}.

A security policy is stored on both the phone side and the router side. Although its enforcement happens on the router, its compliance check is performed jointly by the router and the phone. The former ensures that only the authorized phone, as indicated by its \textit{role}, can access the \textit{domain} involving the device.  The latter runs the \textit{Monitor} to inspect the app and the target device's \textit{category} tags and asks the router to let their communication flows go through only when the category tags are the same. This policy model, is extremely flexible and can instantiate a diverse set of relations between users, applications and IoT devices. \ignore{Further, its similarity to a well understood and documented model renders it easier to deploy by router vendors while at the same time Hanguard defines and enforces it transparently to the users.} We describe how individual components of the system work in the follow-up sections.

\begin{figure}[t]
\centering
\includegraphics[width=8cm]{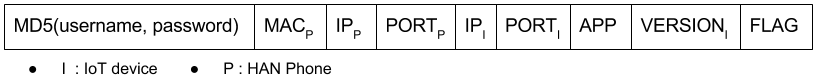}
\caption{Hanguard Control Message delivered over TLS: A control message contains the following information: a hash of the user credentials (username, password); the phone's MAC address; an identifier for the detected flow; the identifier of the requesting app; the policy version used; and a flag indicating flow validation/invalidation. \ignore{The credential hash is being utilized by the router to authenticate the user of the phone. The SSL certificates used by the control channel authenticate the Monitor. The MAC address identifies a phone. All three together describe a \textit{role}. The flow identifier is a unique representation of the source IP address, source port, destination IP address and destination device Port. The requesting App is used to inform the router about the \textit{category} used. The policy version is utilized by the router to reject control messages derived from an outdated policy, and the validation Flag indicates a valid/invalid flow.}}
\label{fig:controlMessage}
\minusminus
\end{figure}

\subsection{Phone-side Situation Monitoring}
\label{subsec:phone}

In our distributed access-control system, the \textit{Monitors} are deployed as user-space apps with limited privileges. They are aiming at identifying the subject of a communication attempt, whether the party trying to access an IoT device across the HAN is an authorized one. Such information is delivered through a control message to the \textit{Controller module} running on the router, informing it the context of the access attempt (since the router cannot see the app initiating the communication), which helps the router enforce appropriate security policies. Note that we designed the system in a way that the workload on the router is minimized, which is important in maintaining the performance level needed for serving the whole local network. More specifically, the \textit{Monitor} launches at boot time to establish an ongoing secure connection with the \textit{Controller module} on the router. Through the channel, the situation on the phone is \textit{pushed} to the router, enabling it to perform a \textit{per-flow} (instead of \textit{per-packet}) access control. Further, the security policies (Section~\ref{subsec:overview}) are broken into two parts: the \textit{Monitor} checks whether an app is authorized to access a device and asks the router to enforce its decision, while the router implements a \textit{phone-level} policy check as a second line of defense, which protects the smart-home devices even when a phone is fully compromised (see Section~\ref{sec:analysis}). 

\ignore{ Once an unauthorized access attempt is detected, the \textit{Monitor} informs the user about the event. Moreover, the router sends an out-of-band message to the admin user to guarantee that even in the case of compromised user phones, the admin receives the notification~\footnote{In our system the router sends an email to the admin user.}.
}

The communication between the \textit{Monitor} and the router goes through a TLS control channel. The control message delivered through the channel is in the format illustrated in Figure~\ref{fig:controlMessage}. For example, it includes a hash of the user credentials (username, password), the sender phone's MAC address, an identifier for the detected flow (IP/port), an identifier for the app making the request, the policy's version number and a flag indicating whether this flow should be allowed or not. The negative flag is used to invalidate flows (Section~\ref{subsec:routerenforce}): this happens when a TCP FIN packet is observed for a TCP flow or when the
target app stopped sending UDP packets with the same flow signature.

\ignore{A control message contains the following information: a hash of the user credentials (username, password); the phone's MAC address; an identifier for the detected flow; a flag indicating flow validation/invalidation; and the policy version used. The credential hash is being utilized by the router to authenticate the user of the phone. The SSL certificates used by the control channel in tandem with the MAC address authenticate the phone to the router. The flow identifier is a unique representation of the source IP address, source port, destination IP address and destination device Port. The policy version is utilized by the router to reject control messages derived from an outdated policy.}

\begin{figure}[t]
\centering
\includegraphics[width=5cm]{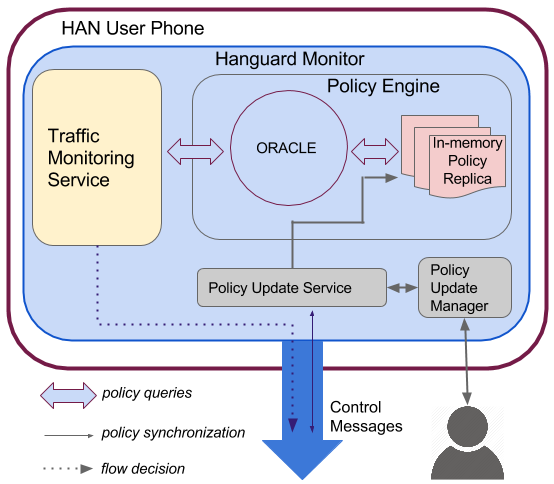}
\caption{Hanguard \textit{Monitor} architecture.}
\label{fig:monitor}
\minusminus
\minusminus
\end{figure}

The anatomy of a mobile phone \textit{Monitor} is depicted on Figure~\ref{fig:monitor}. Every registered phone on the HAN, can be assigned \textit{roles} instantiating an RBAC (Role-Based Access Control) scheme on the router. Furthermore, the phone used to configure the router is by default designated as the \textit{Master Controller Node} (MCN) and every other phone is designated as the \textit{Slave Controller Node} (SCN). A HAN user can update the policy through the \textit{Policy Update Manager} running in her phone's \textit{Monitor}. A \textit{Monitor} accepts policy updates only when it is running on a master node and after verifying its user's credentials. A distributed \textit{Policy Update Service} (see~\ref{subsec:routerenforce}) intermediates policy synchronization and replication in the system. Every connected (reachable) node gets the latest policy replica as soon as it connects to the network or when there is an update. Unregistered devices are automatically assigned the ``Guest'' role as soon as they connect to the network. Each \textit{Monitor} has a local in-memory replica of the policy base, that allows it to make decisions for its own traffic efficiently, alleviating the router from further processing. Having the policy also at the phone side is an important decision in SDN-like systems since it allows for efficient decision making by the Monitors, reduces the bandwidth on the control channel and keeps the routers simple and fast~\cite{sdnSurvey}. For example, Monitors need to only send their per-flow decision to the router instead of continuously sending all the mobile OS-situation measurements. In the last case, the number of control messages in the HAN would exponentially increase while the router would need to process all the measurements before making a decision, with severe performance degradation.

\ignore{
An alternative would be to send control messages for every flow (both legitimate and illegitimate) and let the router make the access decision. That would entail more messages in the HAN and thus more bandwidth consumed while at the same time it would force the router to store and enforce decisions for a significantly larger set of flows.
}

\vspace {5pt}\noindent\textbf{Situation awareness on  iOS}. As mentioned earlier, the \textit{Monitor} is designed to find out which app is talking to an IoT device under protection. Such information, however, is not directly given to a non-system app on both iOS and Android. To tackle this we utilize a new iOS capability that allows developers to proxy network traffic. Once this functionality is enabled by an app and approved by the user, all network packets from all apps will traverse the network stack and instead of being sent through the physical interface to the remote destination, they end up in a virtual interface (tunnel). The tunnel will redirect those packets to the proxy app running the VPN functionality.

iOS offers developers the capability to proxy network traffic with the \texttt{NEVPNManager} APIs). However, blindly tunneling apps' traffic through the VPN is very expensive, often slowing down the mobile system's network performance by an order of magnitude. This workflow is illustrated in Figure~\ref{fig:iosMonitor}: when an app makes a network call this would entail, for every packet, a userspace-kernel context switch, traversing the network stack, trapping the traffic through the tunnel interface and context-switching to userspace again to deliver the network packets to the proxying app. Then the proxying app needs to process the network headers (essentially performing layer 3-4 translations) and then resending the packet.

\ignore{
\begin{figure}[t]
\centering
\includegraphics[width=6cm]{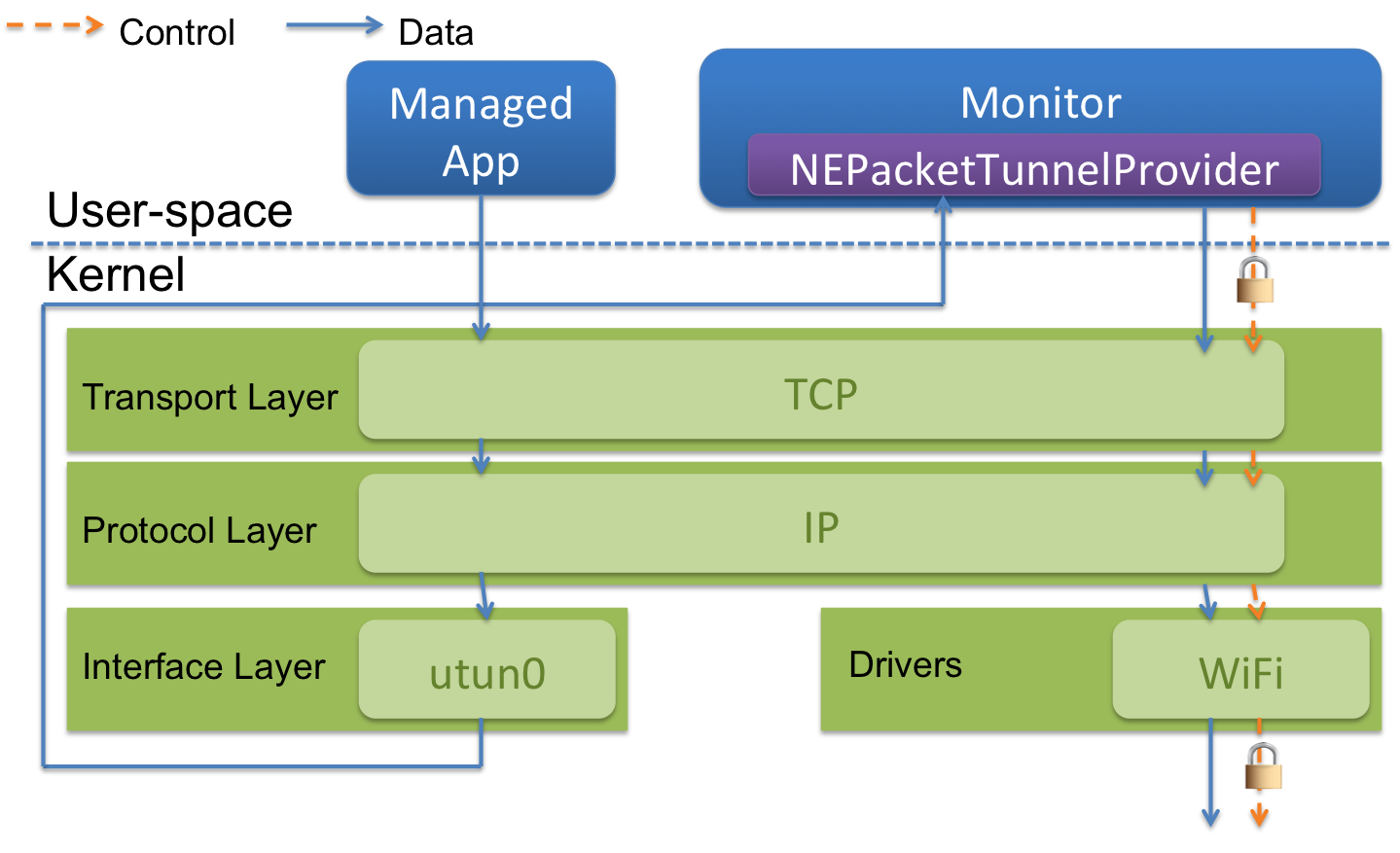}
\caption{Traffic monitoring by a Hanguard iOS \textit{Monitor}.}
\label{fig:iosMonitor}
\vspace{-1em}
\end{figure}
}

Our solution is to utilize the VPN in a unique way: instead of running the iOS Monitor to proxy the traffic of all apps (through the \texttt{NEVPNManager} APIs), which is expensive, requires a remote VPN server and gives little information about the identity of the app generating traffic, our iOS \textit{Monitor} uses the \texttt{NEPacket}\texttt{Tunnel} \texttt{Provider} APIs with a per-app VPN configuration, to tunnel the traffic only from \textit{authorized apps} (the official apps of the IoT devices), while leaving all other traffic outside the tunnel to avoid unnecessary delays. Furthermore, over the tunnel, our iOS \textit{Monitor} does not change the data: it merely acquires packet header information and forwards the packet to its original destination. The whole purpose is that through authenticating itself to the \textit{Controller module} on the router through TLS and its credentials, the \textit{Monitor} informs the router that the flow in the tunnel is authorized. Other flows towards the IoT devices from the phone are by default considered illegitimate and will all be dropped at the router. In this way, we can strike a balance between the protection of legitimate IoT management traffic and the performance impact of the security control.

\vspace {5pt}\noindent\textbf{Situation awareness on Android}.  A straightforward way to capture traffic from other apps on Android is to follow a similar process with iOS and utilize the closely equivalent \texttt{VPNService}~\cite{vpnservice} API, introduced in Android 4.0. However, the implementation of VPN on Android is similar to the one in iOS and would entail similar overheads. To collect the situation information in a more lightweight manner, Hanguard leverages side channels on Android an approach which results in astounding performance benefits.
\ignore{
However, when an app uses the \texttt{VPNService} it does not know exactly where the traffic comes from. Further, blindly tunneling apps' traffic through the VPN is very expensive, often slowing down the mobile system's network performance by an order of magnitude. To collect the situation information in a more lightweight manner, on Android, we leveraged its side channels.
}

\begin{figure}[t!]
    \centering
    \null\hfill
    \subfloat[iOS.]{\includegraphics[height=2.5cm]{figures/iosMonitor.png}\label{fig:iosMonitor}}
    \hfill
    \subfloat[Android.]{\includegraphics[height=2.5cm]{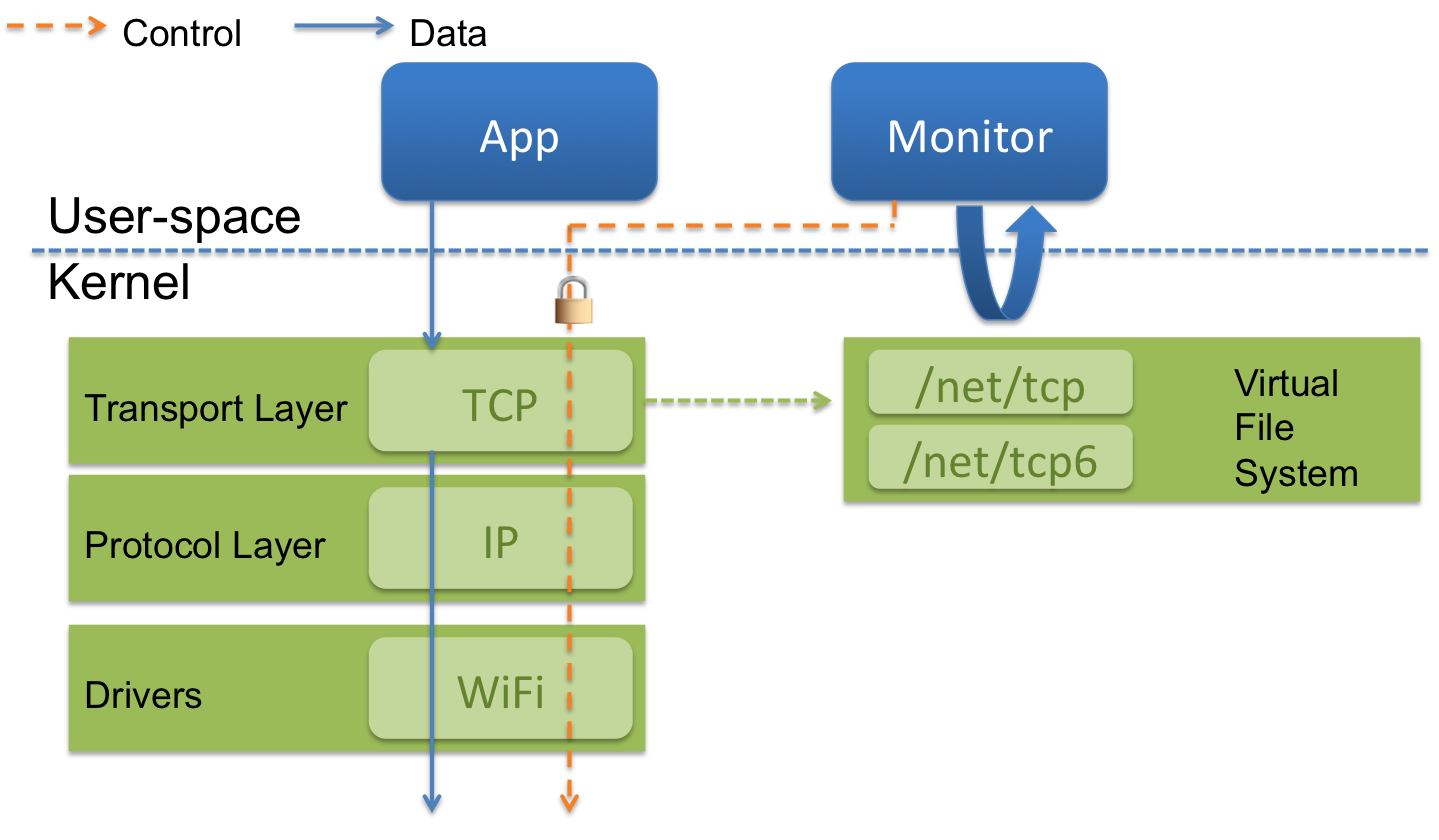}\label{fig:androidMonitor}}
    \hfill\null
    \label{fig:monitorTraffic}
    \caption{Traffic Monitoring Workflow.}
    \minusminus
    \minusminus
\end{figure}

The Android \textit{Monitor} we implemented continuously looks at the \texttt{procfs} file system (see Figure~\ref{fig:androidMonitor}). \texttt{procfs} is a virtual file system which exposes the current status of an Android phone's kernel internal data structures. Particularly the files \texttt{proc/net/tcp}, \texttt{proc/net/tcp6}, \texttt{proc/net/udp} and \texttt{proc/net/udp6} disclose the ongoing TCP and UDP connections between the phone and a remote destination, including the source/destination IP addresses of the ongoing connection and its port numbers, the status of the connection etc~\footnote{Note that iOS does not reveal to an app the information about other processes through its \texttt{procfs} file system. Before iOS 9, one could use the system call \texttt{sysctl} to access such information. This channel has been closed since then.}. The addresses here can be either IPv4 and IPv6 (with the suffix ``6''). These connections are also associated with a specific UID that the \textit{Monitor} can map to an installed app. To minimize operation overheads, the \textit{Monitor} does not open and parse a file for each access. Instead it just checks the file's metadata (i.e. the last modified time or \texttt{mtime} in UNIX terms) to determine whether the file has been changed since the last visit. A complication here is that Android often fits an IPv4 address into the IPv6 format before reporting it to the user. Such an address is automatically captured by the \textit{Monitor} and converted back to the IPv4 form. As an example, consider an app on a phone with an IPv4 address \textit{192.168.1.189} that connects to an IoT device with the address \textit{192.168.1.32}. During the app's runtime, the connection may not show up in \texttt{proc/net/tcp} but appears inside \texttt{proc/net/tcp6} instead with \texttt{0000000000000000FFFF} \texttt{0000BD01A8C0} for the source IP  and \texttt{0000000000000000} \texttt{FFFF0000200} \texttt{1A8C0} for the destination. It is clear that the IPv4 address is enclosed in the 32 least significant bits~\footnote{in little-endian order, presented using four-byte hexadecimals} and the 96 remaining bits are fixed. The \textit{Monitor} detects the address from its fixed part and converts the rest to an IPv4 format before communicating the app's identity to the router through a control message. Note that Android suffers from the repackaged apps problem~\cite{Zhou:2012:DRS:2133601.2133640}. To address this the Android Monitor uses a package's signature to verify apps claiming the identity of policy-controlled apps.

\subsection{Router-side Policy Enforcement}
\label{subsec:routerenforce}

\begin{figure}[t]
\centering
\includegraphics[width=6cm]{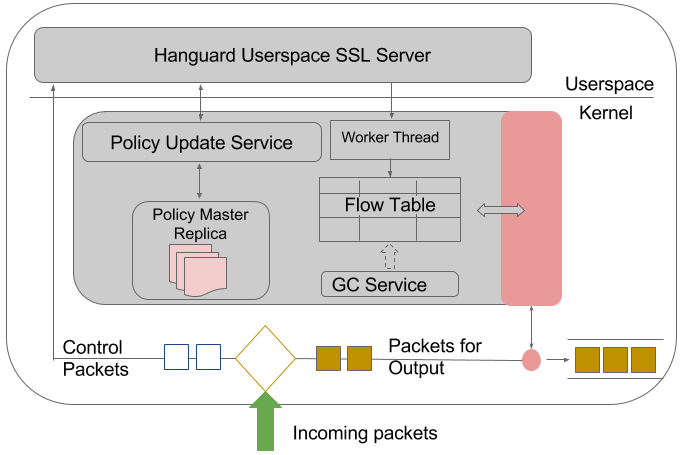}
\caption{Hanguard Router Controller Module}
\label{fig:router}
\minusminus
\minusminus
\end{figure}

The design of the controller module mainly focuses on synchronizing security policies across all the systems within the HAN and enforcing these policies on the router, as illustrated in Figure~\ref{fig:router}. More specifically, the module maintains a \textit{Master Policy Replica}, and runs a \textit{Policy Update Service} responsible for updating the policies and distributing them across registered \textit{Monitors}. Further, the \textit{Controller module} introduces a \textit{Per-Flow Decision Cache} (PFDC) for keeping the access decisions (on the app level) pushed by the \textit{Monitors}, and a \textit{Garbage Collection Service} (GCS) for maintaining the cache. It also hooks on the router's packet flow for the policy enforcement. 

\vspace {3pt}\noindent\textbf{Policy synchronization}. It is critical to ensure that all \textit{Monitors} have a consistent view of the global policy, otherwise flows that need to be blocked might be allowed by the router, or legitimate flows could be blocked.\ignore{ Moreover, note that, based on the CAP theorem, a distributed system cannot ensure both strong consistency and availability in the face of network partitions. In other words, if we want to tolerate the fact that mobile phones can be disconnected from the Internet at any point of time, we have to choose between ensuring that all of them have a consistent view of the policies or allowing the MCN user to update the policies at any given point in time. Since this is a security system, we opted in strongly consistent policies.} In our system, the \textit{MCN} node is allowed to change policies only when the HAN router can be reached. In particular, the update is pushed to the router's \textit{Controller module} through a write-through model to update the policies both in volatile and persistent storage. After a successful update, the \textit{Controller module} sends an acknowledgement along with a new policy version number to the \textit{MCN} node and pushes the new policies to all other reachable \textit{Monitors}. The \textit{Monitors} then upgrade their local policy base along to the current version. However, in case a \textit{Monitor} is not reachable, it could miss the update. As a result, a \textit{Monitor} could allow an app on its phone to access an IoT device no longer allowed to access, as the app-level decisions are made by the \textit{Monitor}. To tackle this, the Monitors are designed to asynchronously sense network changes that happen to the phones through the \texttt{ACCESS\_NETWORK\_STATE} permission (for Android): whenever the phone is disconnected from the network or switched to a different network, the app receives a notification and as soon as the connection to the router is restored, it checks the current policy version and performs an update when necessary. On iOS, this can be done by using the \texttt{Reachability} callback.

\vspace {3pt}\noindent\textbf{Receiving decisions}. As mentioned earlier, app-level access control on the router relies on the decision made by the \textit{Monitor} and delivered to the router through the control channel. To effectively enforce such a decision on a traffic flow, the \textit{Controller module} is designed to efficiently authenticate and process the control messages from the \textit{Monitor} to avoid holding up the legitimate interactions with the target IoT device. Specifically, the \textit{Controller module} maintains TLS connections with the \textit{Monitors} through a userspace program. When a decision from a \textit{Monitor} arrives, the router checks the policy version and the sender user's credentials\ignore{It ensures that the decision is received from a valid user and that the source device belongs to that user.}, and once validated, passes the decision's \textit{flow ID} (source IP and port, destination IP and port) to the kernel that updates the \textit{PFDC} using the flow ID as the key to record the validation/invalidation decision on the flow, which is then enforced by the router. Note that the flow id is used for efficient app-level enforcement. However, data flows are first checked against a phone-level policy which ensures that the flow comes from a valid HAN phone (see Section~\ref{sec:analysis}).

Supporting this decision-making process requires an efficient userspace to kernel communication mechanism (for the router). Although this can be achieved through \texttt{system calls}, \texttt{ioctl} calls or \texttt{procfs} files, these approaches are either complicated to implement or unable to handle asynchronous interactions. Our solution employs the \texttt{netlink} socket IPC mechanism for the user-kernel communication, which can be easily built (without changing the kernel) and are asynchronous in nature: it queues incoming messages and notifies the receiver through a handler callback. In our implementation, the callback spawns a \textit{worker thread} that processes the message and updates the \textit{PFDC}, either by inserting a valid flow or removing an invalid flow.

The \textit{PFDC} is loaded at the router's boot time from its persistent storage. It holds the following information per-flow: the \textit{flow ID}, the \textit{flow validation/invalidation flag}, the \textit{requesting app} and the \textit{data last seen time}. This cache is used for enforcing app-level policies (whether a specific app is allowed to access a device), for the purpose of enhancing the existing flow-control capability of the router, which cannot differentiate two flows from the same IP and port but produced by different apps. By searching the cache, the router can apply the app-level access decision upon the whole flow, instead for every individual packet, an advantage over deep packet inspection and traffic fingerprinting techniques.
To limit the amount of the resource the cache uses (given that the router is a resource-limited device), a \textit{Garbage Collector Service} (GCS) is run to remove the obsolete records with the oldest data last seen time. Also, a per-phone limit is applied to prevent a \textit{Monitor} from using too much resource of the cache. Note that this is a measure against DoS. Flow invalidation decisions are made by the Monitors.


\vspace {3pt}\noindent\textbf{Enforcement}. The router enforces \textit{phone-level} and \textit{app-level} policies. For the former, it checks every packet to determine whether it originates from a phone that is allowed to access a particular IoT device. Phones and IoT devices are identified based on their MAC addresses. For the latter it checks with the \textit{PFDC} cache to determine whether the flow is generated from a valid app.

A technical challenge in implementing the protection is where to place the security control within the existing router infrastructure. On a Linux-enabled system used by the router, once a packet is received, it is put by the link layer into a backlog queue from which the IP layer pulls packets for checksum checking and routing decisions. If the packet is destined for the current machine, it is then passed to the transport layer. If not the packet is forwarded. Apparently, the security control should happen on the IP layer (e.g. in the \texttt{ip\_forward()} function). However, a packet might follow a different path within the kernel depending on whether the current system is configured to run as a bridge or a router. For example, in a bridge mode, no layer 3 operation is involved and as a result the aforementioned function will never operate on the packet. Our solution is to place the Controller hook in \texttt{dev\_queue\_xmit()}, a generic driver function, which ensures that no packet bypasses the check.

To minimize the impact on the communication unrelated to the smart-home devices, the Hanguard-enhanced router quickly inspects each packet it receives to determine whether further attention is needed. Specifically, a TCP flow is considered interesting if its destination MAC address is associated with a protected IoT device. Packets not fitting this description are forwarded on without a delay, and others are first handled according to the phone-level policy (whether the phone can access the IoT device) stored at the router, and then the app-level policy (whether the app can do that) which is based upon the validation flag set by the Monitor. For the packet allowed to go through, its flow's last seen time is updated to the packet's arrival time. Hanguard helps its users detect and react to spurious access attempts with its \textit{notification mechanism}: Hanguard (1) keeps a log, and (2) sends out-of-band notifications to the admin user when a violation or tampering of the policy is attempted.

\section{Empirical Evaluation}
\label{sec:evaluation}
\ignore{We evaluated the prototype of Hanguard, in an attempt to answer the following questions:}

We implemented a prototype of Hanguard on top of a TP-Link WDR4300v1 router with a gigabit NIC and a wireless network at the 2.4 GHz band (300Mbps) running OpenWRT Chaos Chalmer with a Linux 4.1.16 kernel, and also Nexus phones running Android 5 (Lollipop) and an iPhone 4S running iOS 9. Our work answers the following research questions: 

\ignore{(RQ1) \textit{Is Hanguard effective in thwarting attacks from malicious applications?} (RQ2) \textit{What is the performance impact and resource consumption of the Monitors on the phone side?} (RQ3) \textit{What is the overall overhead of our system?}}

\vspace {5pt}\noindent$\bullet$ \textit{RQ1:} Is Hanguard effective in thwarting attacks from malicious applications?

\vspace {5pt}\noindent$\bullet$ \textit{RQ2:} What is the performance impact and resource consumption of the \textit{Monitors} on the phone side?

\vspace {5pt}\noindent$\bullet$ \textit{RQ3:} What is the overall overhead of our system?

\ignore{The Vanilla version of the router runs OpenWRT Chaos Chalmer with a Linux 4.1.16 kernel. Our modifications were introduced through a patch on OpenWRT.}

\subsection{Effectiveness}
\label{subsec:effectiveness}
To answer RQ1 and verify Hanguard's backward compatibility and its practicality, we performed attacks on real world smart-home devices, including a Belkin WeMo switch, a Belkin WeMo in.sight A1.C, a Belkin WeMo motion and a My N3rd device. The first two devices allow the user to connect them to any other electronic devices, which then the user can turn on/off through her WeMo app. The Belkin WeMo motion notifies the WeMo app when motion is detected. The My N3rd device can be connected to any other device, enabling remote control of it through the My N3rd app.

\vspace{5pt}\noindent\textbf{Hypothesis}. A malicious app on a HAN user phone might try to take advantage of the fact that the phone's \texttt{role} is allowed to access the \texttt{domain} of the target IoT device. However, the Monitor on that phone, will detect the offending flow and determine that the source application has a \texttt{category} different than the one of the target device. Thus it will not push a flow decision to the router, which means the packets on the data plane for that flow will be dropped by the router.

\vspace {5pt}\noindent\textbf{Experimental Setting}. We performed the following two experiments. (A) First we set up the target IoT devices over the ``Vanilla'' system (without Hanguard components), and further installed a repackaged version of their legitimate app on the phone to mimic the adversary. (B) Next, we updated the router with Hanguard-enhanced firmware, and also put our \textit{Monitor} app on the same phone.  Under this protected setting, we repeated the experiment (A), using the phone with the \textit{Monitor} app to set up the IoT devices.

\vspace {5pt}\noindent\textbf{Results}. During experiment (A), we found that both the official and the repackaged app on the first phone could see and interact with the target IoT devices. However, during experiment (B)---with our system in place---only the official app on the phone running the \textit{Monitor} app could communicate with the IoT device and any other attempt was successfully blocked. We conducted the above experiments on all devices listed in Table~\ref{tab:effectiveness}, which confirms the effectiveness of the access control enforced by Hanguard and its backward compatibility. Demos of Hanguard's success are posted on the project's website~\cite{hanguardgsites}.

\ignore{
\begin{table}[h]
\vspace{1em}
\scriptsize
\centering
\caption{Adversary's success on the unmodified (Vanilla) system and when Hanguard is in place.}
\label{tab:effectiveness}
\begin{tabular}{lcccc}
\textbf{Target Device} & \textbf{Description} & \textbf{\# App Installations} & \multicolumn{1}{l}{\textbf{Vanilla}} & \multicolumn{1}{l}{\textbf{Hanguard}} \\ \hline
WeMo Switch            &  Actuator & 100K - 500K &             \checkmark                 &           \ding{55}                 \\
WeMo Motion            &  Sensor   & 100K - 500K &              \checkmark                 &           \ding{55}                      \\
WeMo Insight Switch    &  Actuator & 100K - 500K &              \checkmark                 &           \ding{55}				\\
My N3rd			       &  Actuator & 100 - 500 &             \checkmark                 &           \ding{55}
\end{tabular}
\end{table}
}

\subsection{Phone-side Performance}
\label{subsec:phoneperform}

To answer RQ2, we focused on the overheads introduced by the \textit{Monitor} to the phone.

\begin{figure*}[t]
    \centering
    \null\hfill
    \subfloat[Android Monitor polling scheduled frequency vs Actual polling frequency.]{\includegraphics[height=3cm,width=0.23\textwidth]{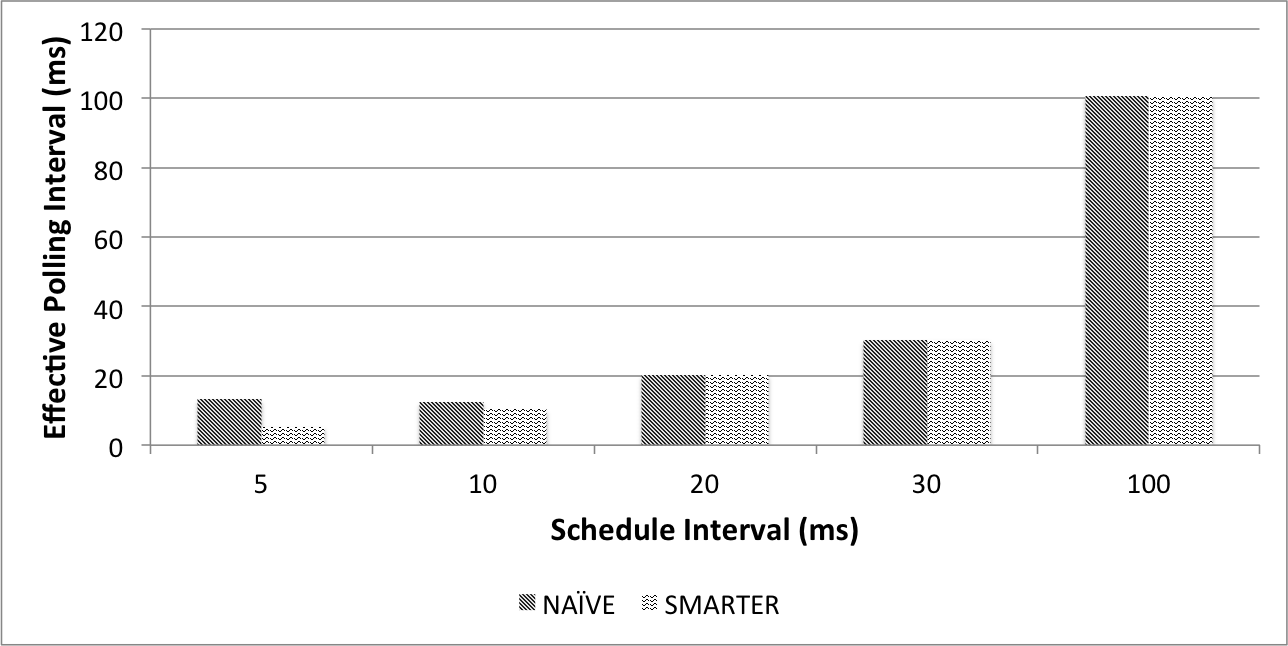}\label{fig:pollfreq}}
    \hfill
    \subfloat[File lines parsed for different Android Monitor configurations.]{\includegraphics[height=3cm,width=0.23\textwidth]{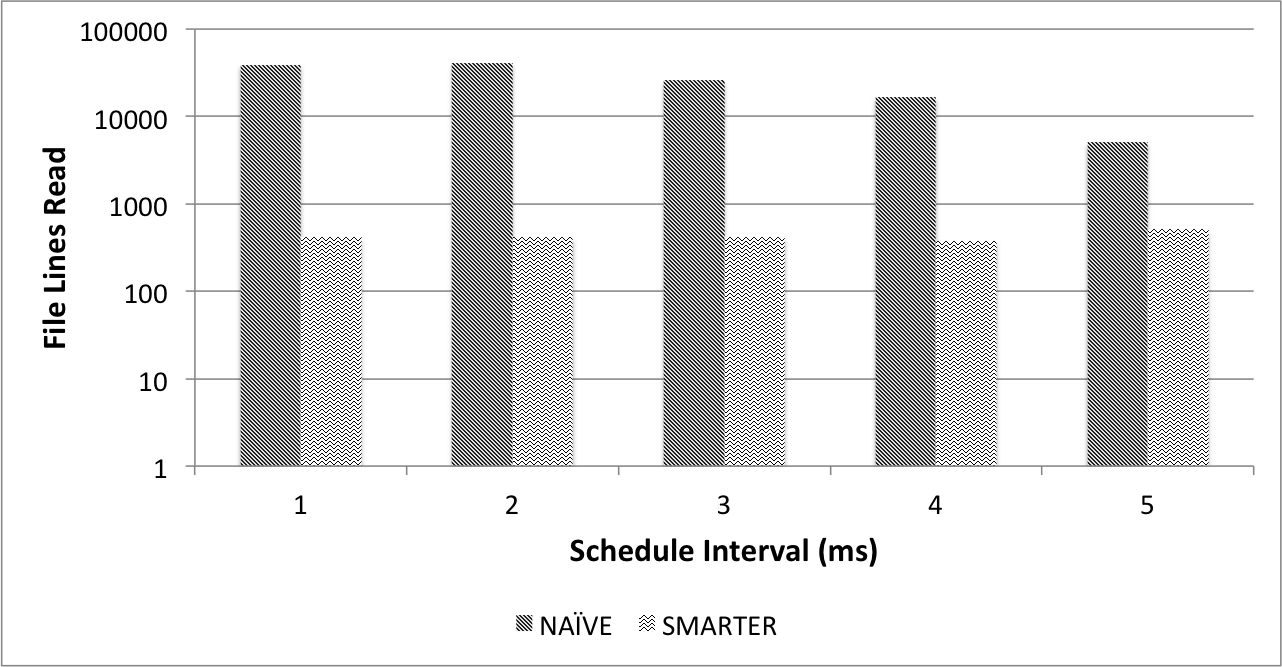}\label{fig:linesScanned}}
    \hfill
    \subfloat[Battery Power on Android.]{\includegraphics[height=3cm,width=0.23\textwidth]{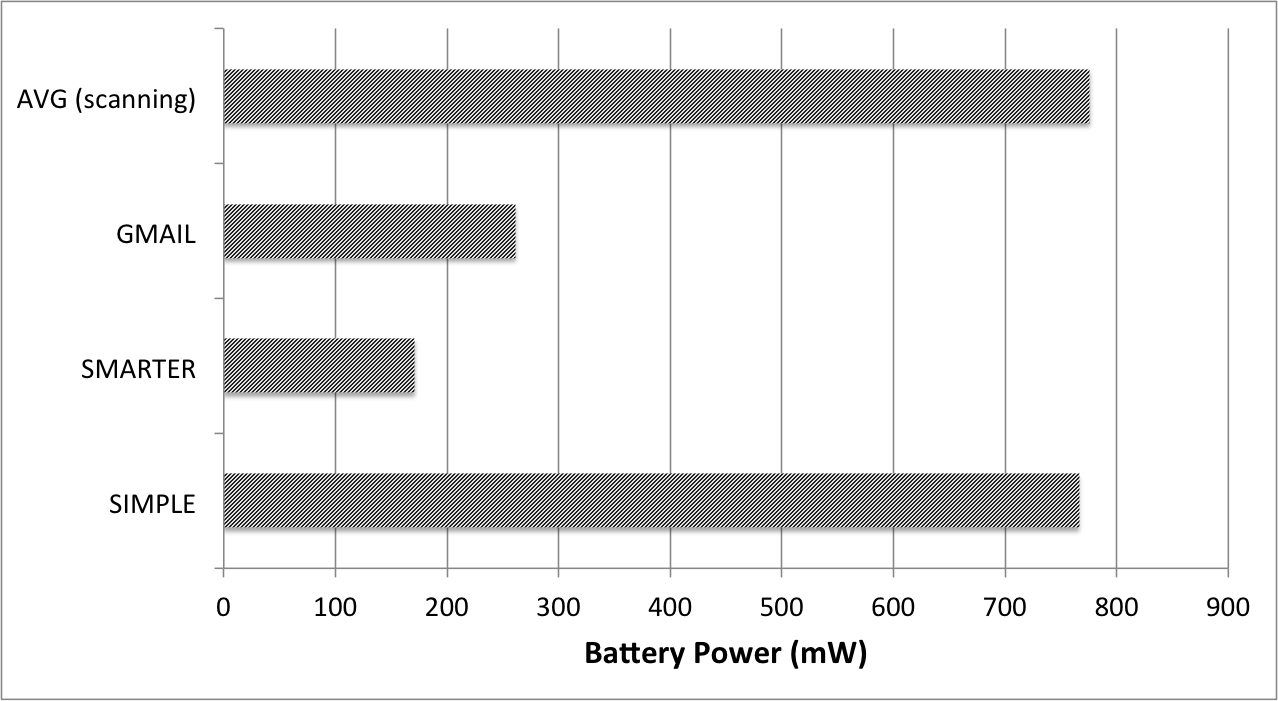}\label{fig:androidbattery}}
    \hfill
    \subfloat[CPU Load on Android and iOS(*).]{\includegraphics[height=3cm,width=0.23\textwidth]{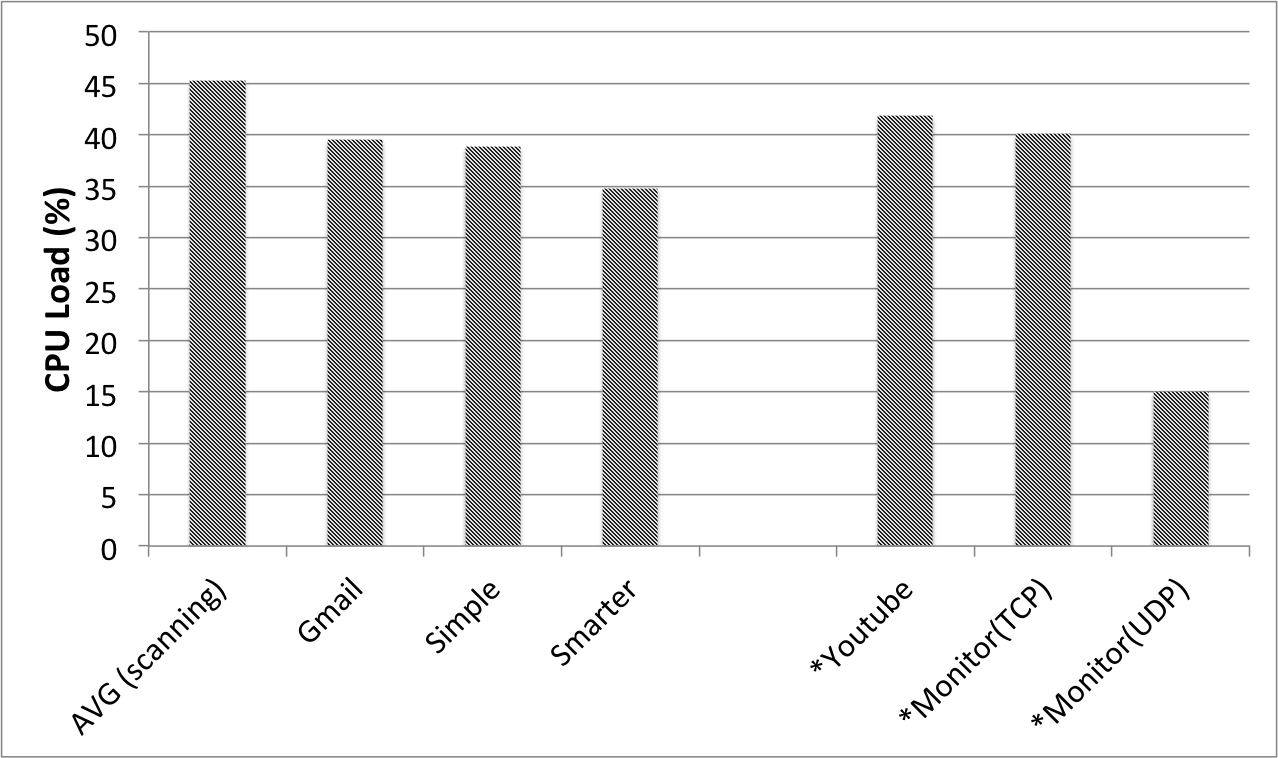}\label{fig:CPU}}
    \hfill\null
    \caption{\label{fig:monitoringCost}Monitoring Costs.}
    \minusminus
\end{figure*}

\ignore{
\begin{figure}[t]
\centering
\includegraphics[width=7cm]{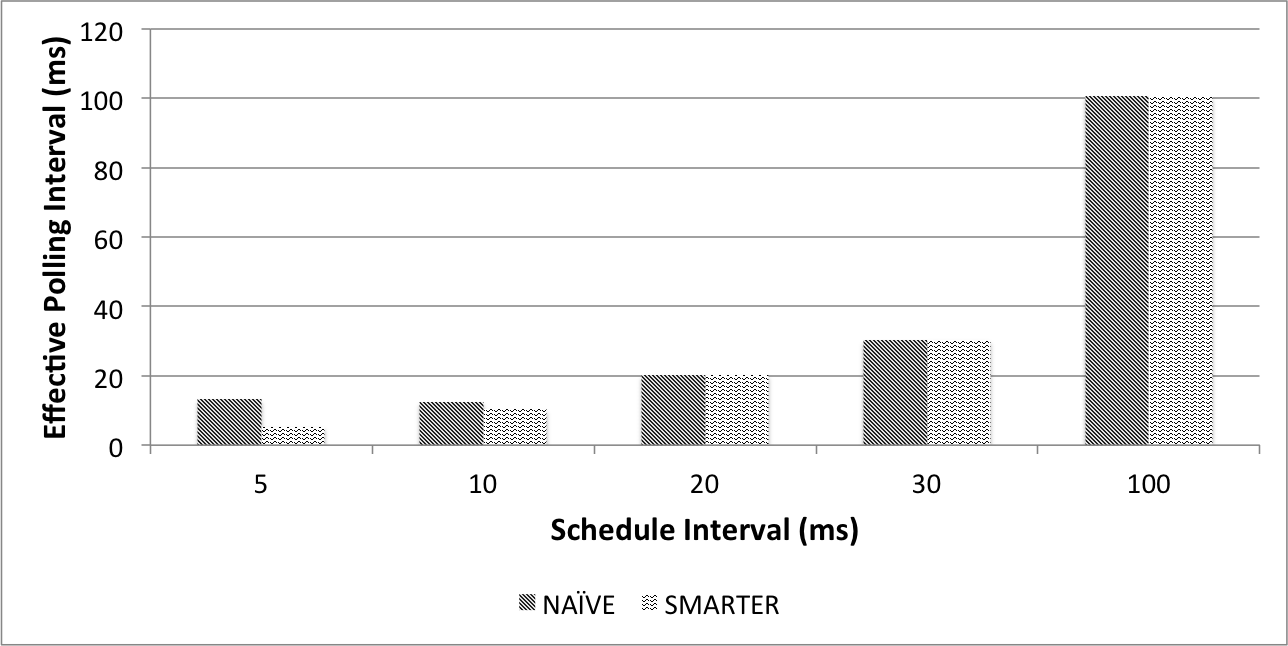}
\caption{Android Monitor polling scheduled frequency vs Actual polling frequency.}
\label{fig:pollfreq}
\end{figure}

\begin{figure}[t]
\centering
\includegraphics[width=7cm]{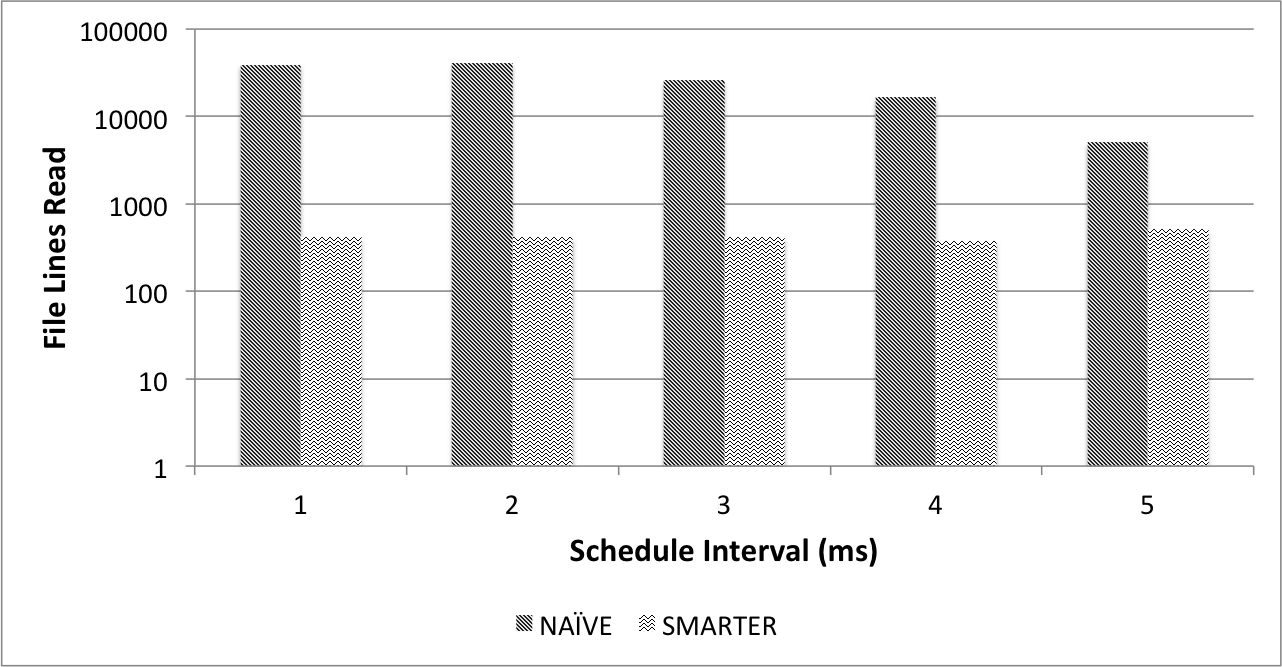}
\caption{Number of file lines parsed for different Android Monitor configurations.}
\label{fig:linesScanned}
\end{figure}

\begin{figure}[t]
\centering
\includegraphics[width=7cm]{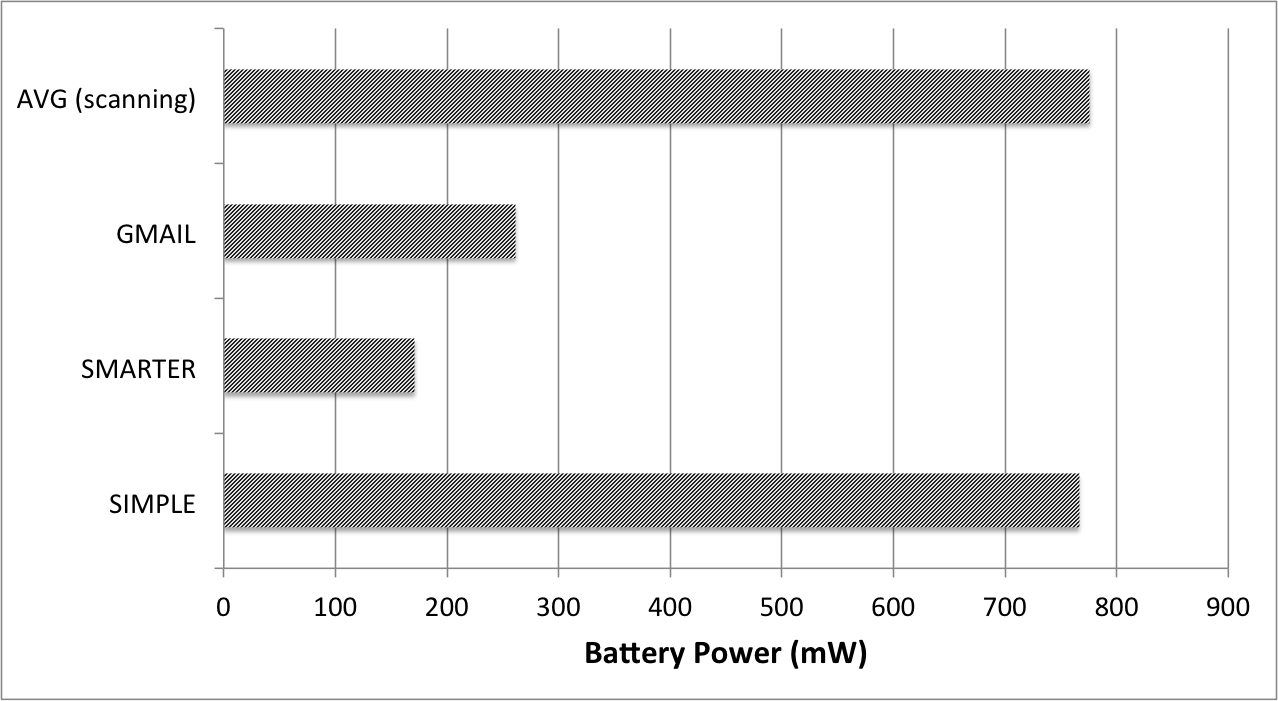}
\caption{Battery Power on Android.}
\label{fig:androidbattery}
\end{figure}

\begin{figure}[t]
\centering
\includegraphics[width=7cm]{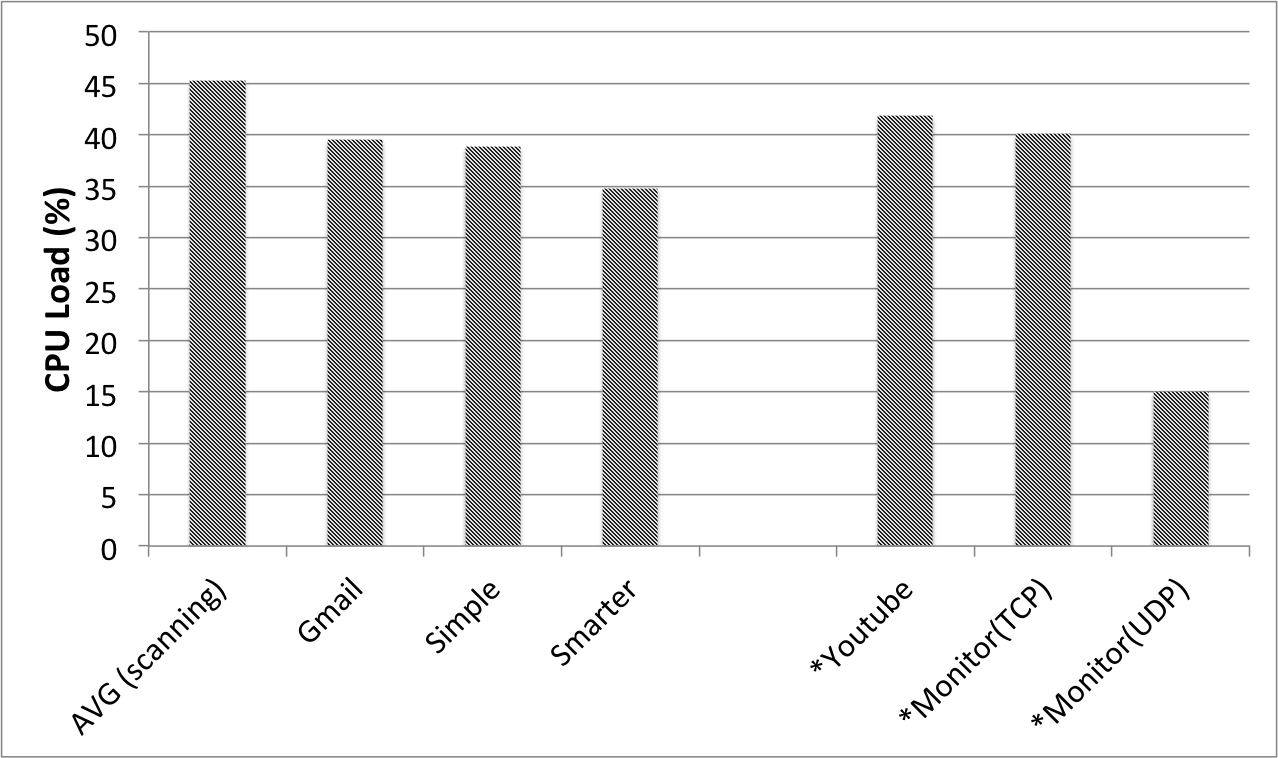}
\caption{CPU Load on Android and iOS(*).}
\label{fig:CPU}
\end{figure}
}

\vspace {5pt}\noindent\textbf{Monitoring cost on Android}. On Android, the \textit{Monitor} continuously polls the \texttt{procfs} file system to detect ongoing network connections. Here we report our study on two monitoring strategies and their performance impacts.  Specifically, we configured the Android \textit{Monitor} on a Nexus phone to inspect the \texttt{procfs} file system in different granularity (every 5ms, 10ms, 20ms, 30ms, 100ms). After running for 30 seconds, the \textit{Monitor} went through every single file line to check the presence of interesting network connections, a strategy called the \textit{Naive} mode. The approach was compared with another strategy, called the \textit{Smarter} mode, which first looked at the last modified time of a file before accessing its content. The outcomes of the study are illustrated in Figure~\ref{fig:pollfreq}. As we can see, the \textit{Smarter} strategy clearly can poll at a finer granularity (5 ms), given that it reads much fewer file lines compared with the \textit{Naive} approach (Figure~\ref{fig:linesScanned}), which is translated to less work per iteration in the common case.

We further looked into the resource consumption of the \textit{Monitor}. For this purpose, we configured the \textit{Monitor} to poll at 10 ms and recorded its CPU and battery consumption for both the Naive and Smarter mode. On the same Nexus 5 phone, we also ran Trepn~\cite{trepn} by Qualcomm to collect the baseline power profile of the phone for 30 seconds before running our \textit{Monitor} app for 2 minutes.  Figure~\ref{fig:androidbattery} illustrates the average battery consumption that can be attributed to the \textit{Monitor}, and Figure~\ref{fig:CPU} shows the average CPU usage (first 4 bars). To put things into perspective, we compared our \textit{Monitor} with a popular Antivirus app in scanning mode and the de facto mailing app on Android (Gmail).  As we can see from the figures, the power consumption of the naive approach is comparable to an antivirus app performing an expensive operation while the smarter mode's is comparable with Gmail which is optimized to always run in the background.

\vspace {5pt}\noindent\textbf{Monitoring cost on iOS}.  To evaluate the iOS \textit{Monitor}'s resource consumption, we used \textit{Instruments}~\cite{instruments}, a performance analysis and testing tool which is part of the official Apple IDE (Xcode~\cite{xcode}). Figure~\ref{fig:CPU} depicts the \% CPU utilization that can be attributed to a runtime process, where measurements on iOS are indicated with * (last 3 bars): the \textit{Monitor} when proxying a TCP app that sends 500 messages with payload size equal to one character; the \textit{Monitor} when proxying an equivalent UDP app; and YouTube while streaming a video configured to \textit{auto-select} its quality. The figure reflects the fact that the iOS \textit{Monitor} does a lot of work when proxying TCP traffic: this is expected as TCP is a connection oriented protocol and the \textit{Monitor} needs to guarantee reliable delivery of the packets. For UDP the \textit{Monitor} does very little work. In idle mode (not proxying), the \textit{Monitor} incurred no CPU overhead. \textit{Instruments} can also report the \textit{Energy Use Level} of an app at runtime as a value from 0 to 20. In all experiments the reported value was consistently 0/20. 

\vspace {5pt}\noindent\textbf{Decision Latency}. Our \textit{Monitor} is designed to detect an interesting outgoing connection (or its termination) and notify the router if the connection is allowed by the policy. By the time the data packet from the legitimate app reaches the router, the Controller module needs to have such information available. Otherwise it drops the packet, forcing the TCP layer to retry or the UDP application layer to handle the event.~\footnote{Note that during our manual analysis, we found that the studied devices consistently use TCP or other higher-level protocols build on top of it.} To quantify the impact of the possible delay (the control message arriving later than data packets), we measured the time difference between the first data packet arrival at the Controller Module's enforcing point and the moment the decision for that flow is stored in the Controller Modules's flow table (PFDC). We call our new metric the \textit{decision latency}. In the experiment, we ran a custom app (simulating an authorized IoT app) on the Nexus phone, sending packets of payload of 1 character to a TCP server on a PC connected to the router through Ethernet.

\ignore{
\begin{figure}[t]
\centering
\includegraphics[width=7cm]{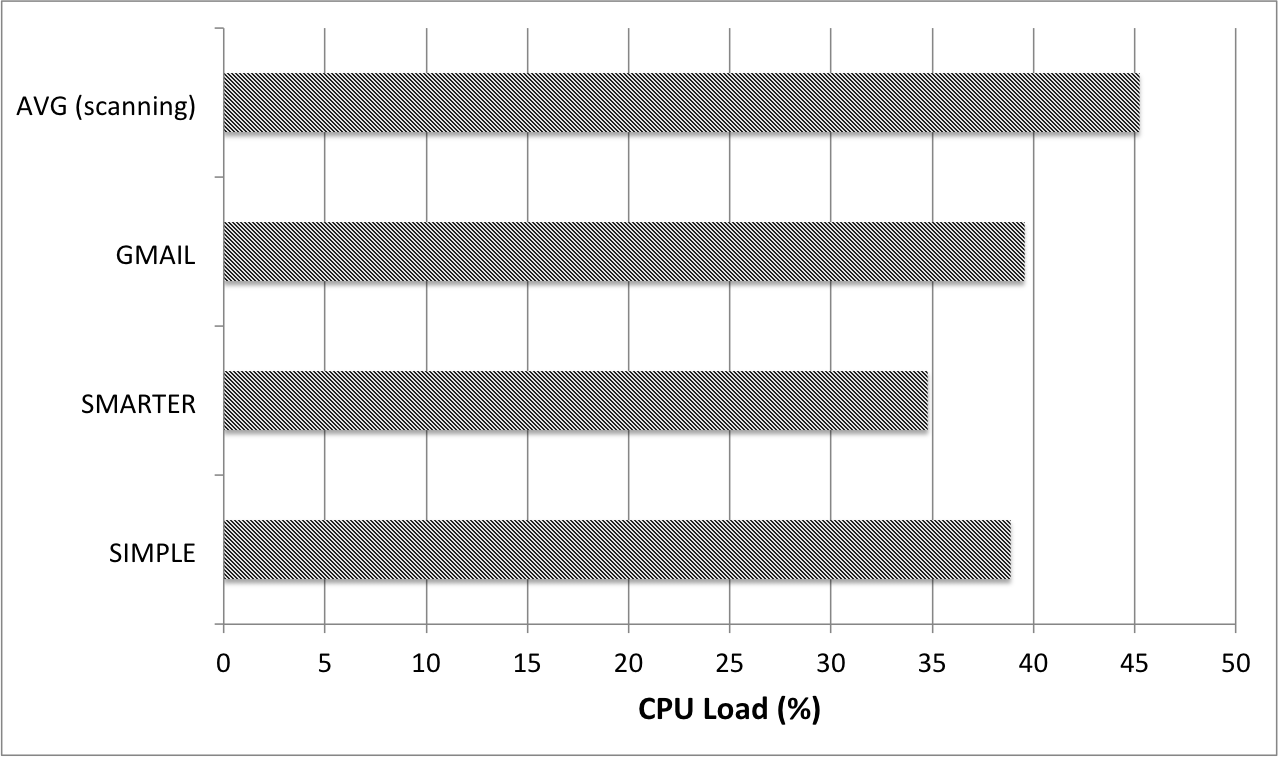}
\caption{CPU Load on Android.}
\label{fig:androidCPU}
\end{figure}

\begin{figure}[t]
\centering
\includegraphics[width=7cm]{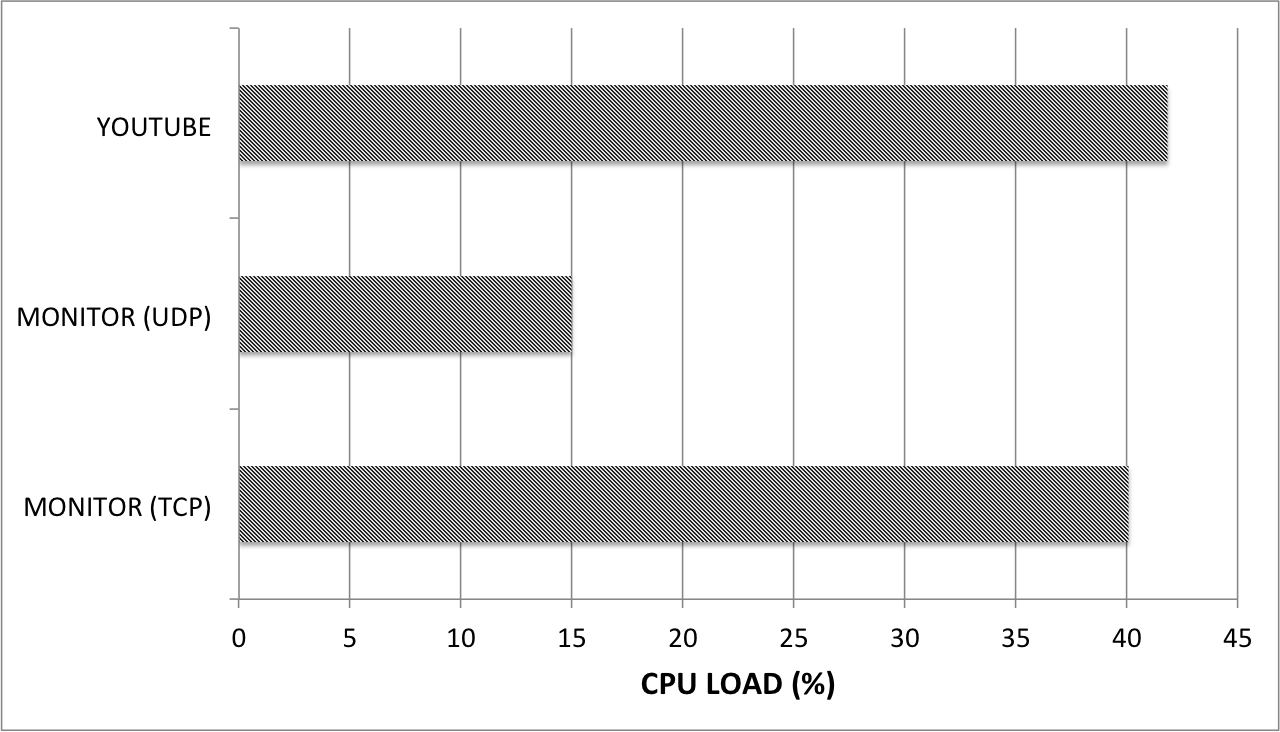}
\caption{CPU Load on iOS.}
\label{fig:iosCPU}
\end{figure}
}

On Android the experiment was performed with the \textit{Monitor} running in the \textit{Smarter} mode (with a polling interval of 10, 30 and 100ms respectively). We repeated the experiment 10 times for each interval and report our findings in Figure~\ref{fig:decisionLatencyAll}. Note that, since the control message and the flow packets leave the same device almost simultaneously, it is possible for the control message to arrive earlier at the Controller module than the flow packet (see the negative values in Figure~\ref{fig:decisionLatencyAll}).  As we can see from the figure, the decision latency does not depend on how fast we poll at the \textit{Monitor}. Other factors, such as link latency and congestion at the router dominate. Also note that this is a one-time cost: once the decision is stored, the router no longer has to wait for it again for other packets in the same flow. On iOS, the decision latency is more predictable\ignore{ and present our experimental results on Figure~\ref{fig:decisionLatencyIOS}}. This is because every packet of the target app is routed by the OS through the \textit{Monitor}. Thus, both the decision and the data packet are send in a more deterministic way. Nonetheless, similarly with Android, the decision is sent through a different channel, which in combination with the router's different treatment of packets to be forwarded compared with packets destined for its userspace, might cause the data and control packets to be stored and analyzed respectively by the router in a different order than their sending order. 

\ignore{
\begin{figure}[t]
\centering
\includegraphics[width=7cm]{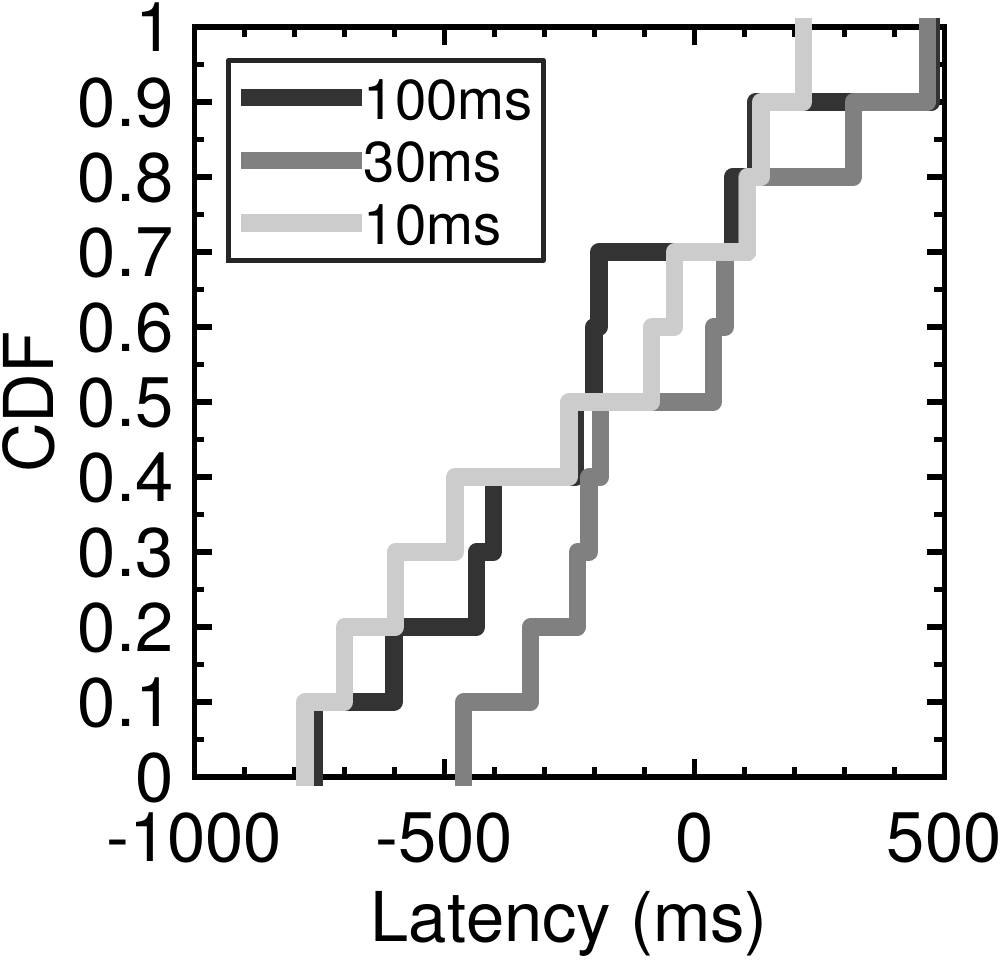}
\caption{Decision latency for the Android and iOS \textit{Monitor}.\ignore{: Time difference between a flow packet arriving at the router and the time a decision about that flow is available, over 10 trials, for different Android \textit{Monitor} polling intervals. Negative values denote a flow decision arriving earlier than the first packet of the flow.}}
\label{fig:decisionLatencyAll}
\end{figure}
}

\begin{figure}[t!]
	\minusminus
    \centering
    \null\hfill
    \subfloat[Android.]{\includegraphics[height=3cm]{figures/decision_latency.pdf}\label{fig:decisionLatencyAndroid}}
    \hfill
    \subfloat[iOS.]{\includegraphics[height=3cm]{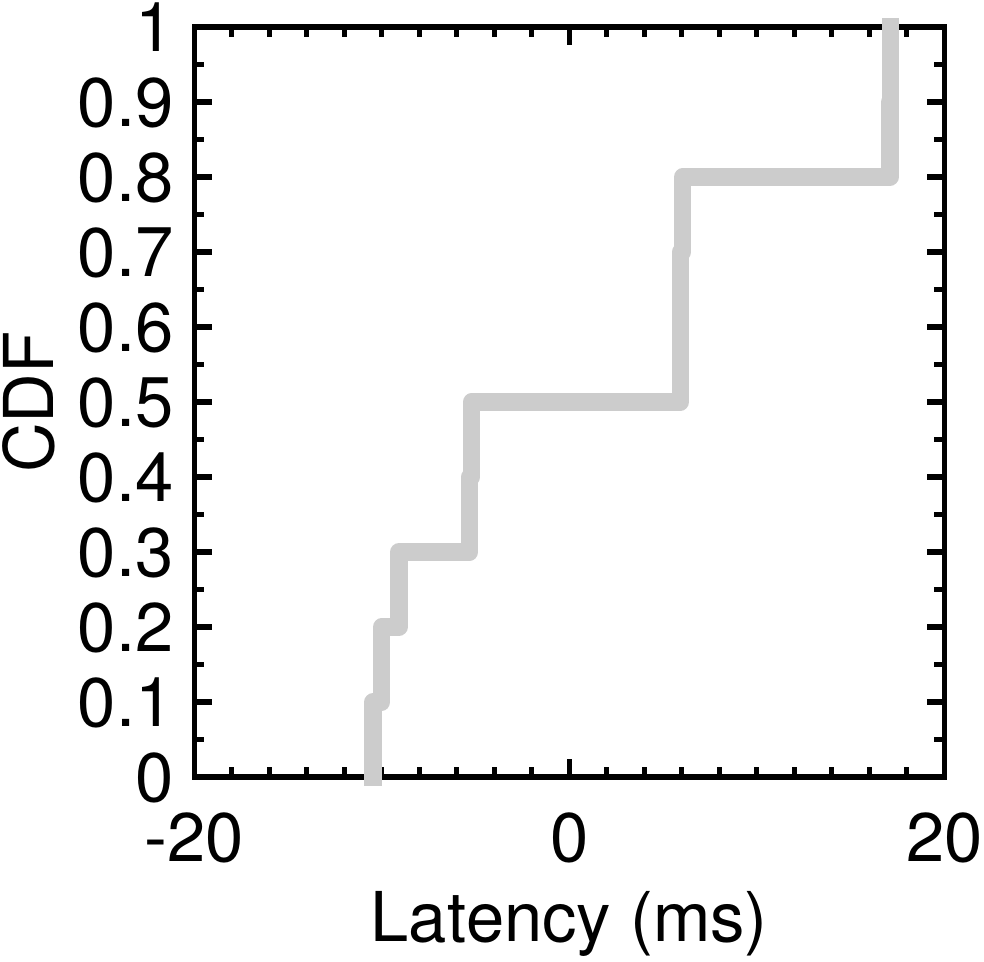}\label{fig:decisionLatencyios}}
    \hfill\null
    \caption{Decision latency for (a) Android and (b) iOS.}
    \label{fig:decisionLatencyAll}
    \minusminus
    \minusminus
\end{figure}

\ignore{
\begin{figure}[t]
\centering
\includegraphics[width=7cm]{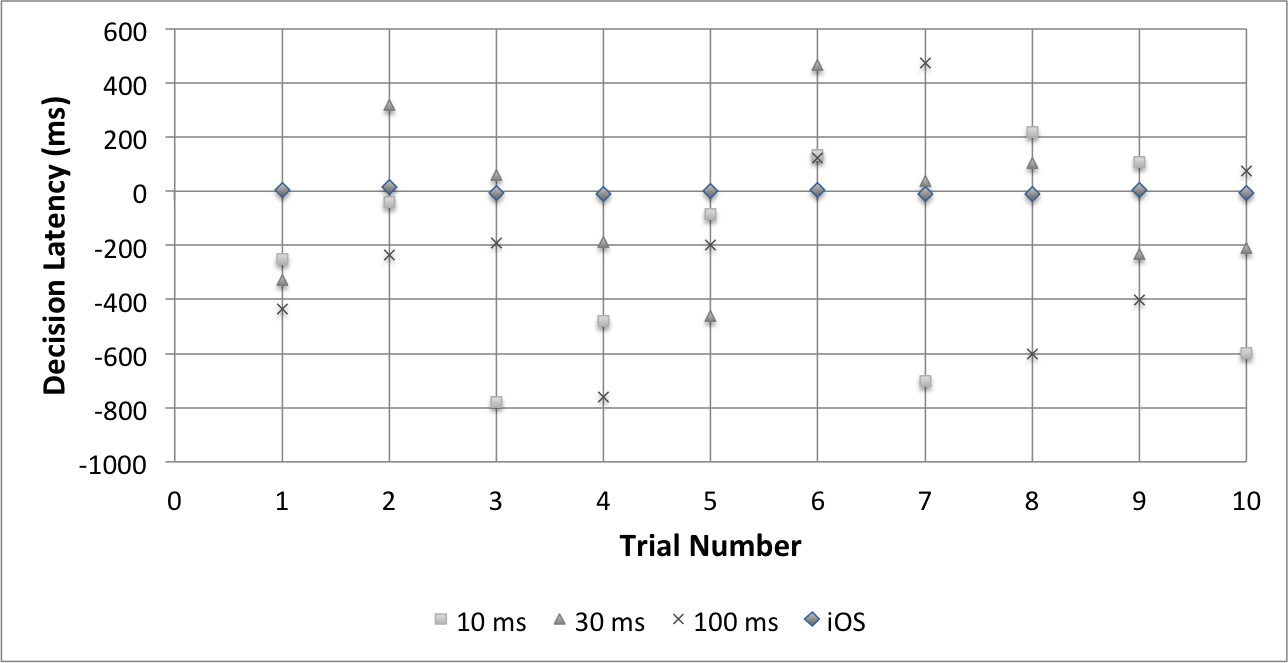}
\caption{Decision latency for the Android and iOS \textit{Monitor}.\ignore{: Time difference between a flow packet arriving at the router and the time a decision about that flow is available, over 10 trials, for different Android \textit{Monitor} polling intervals. Negative values denote a flow decision arriving earlier than the first packet of the flow.}}
\label{fig:decisionLatencyAll}
\end{figure}
}

\ignore{
\begin{figure}[t]
\centering
\includegraphics[width=7cm]{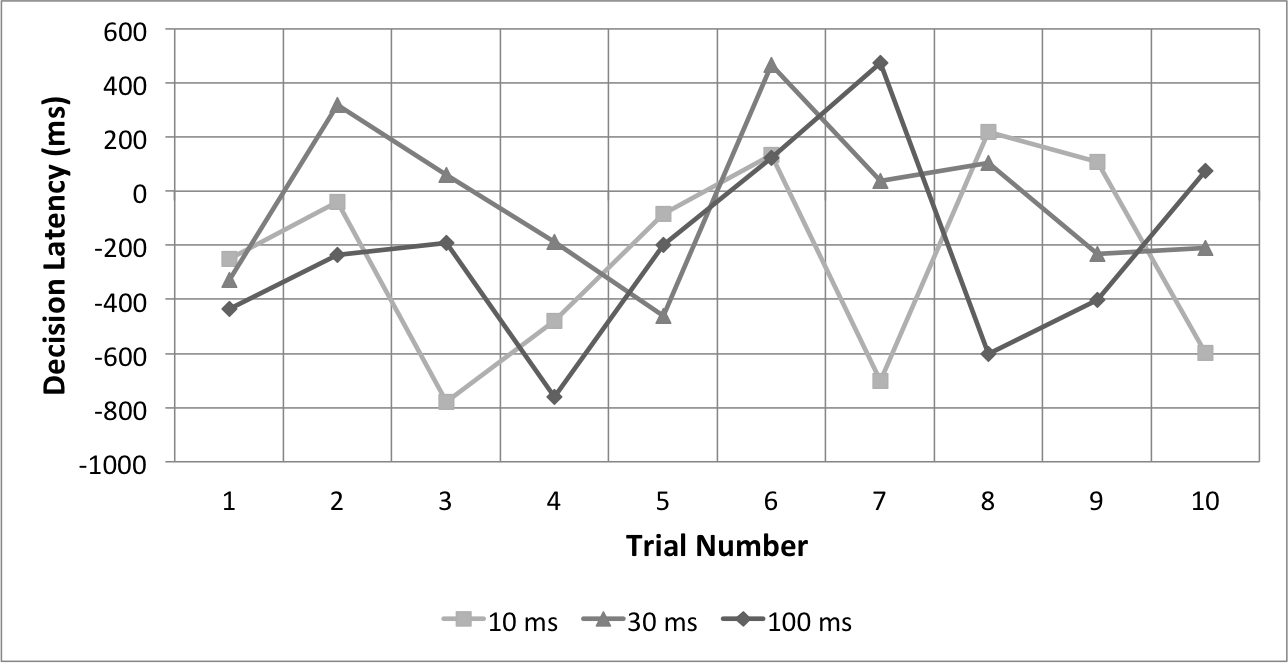}
\caption{Decision latency for the Android \textit{Monitor}.\ignore{: Time difference between a flow packet arriving at the router and the time a decision about that flow is available, over 10 trials, for different Android \textit{Monitor} polling intervals. Negative values denote a flow decision arriving earlier than the first packet of the flow.}}
\label{fig:decisionLatency}
\end{figure}

\begin{figure}[t]
\centering
\includegraphics[width=7cm]{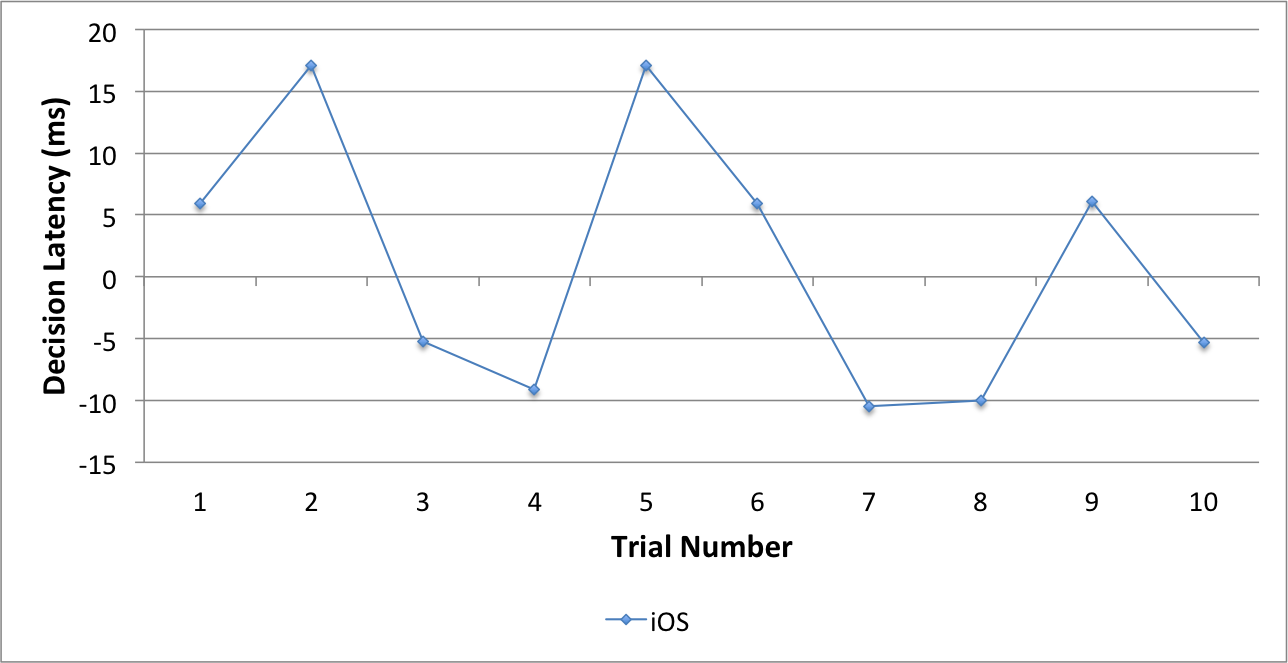}
\caption{Decision latency for the iOS \textit{Monitor}.}
\label{fig:decisionLatencyIOS}
\end{figure}
}

\begin{figure}[t]
\centering
\includegraphics[width=7cm]{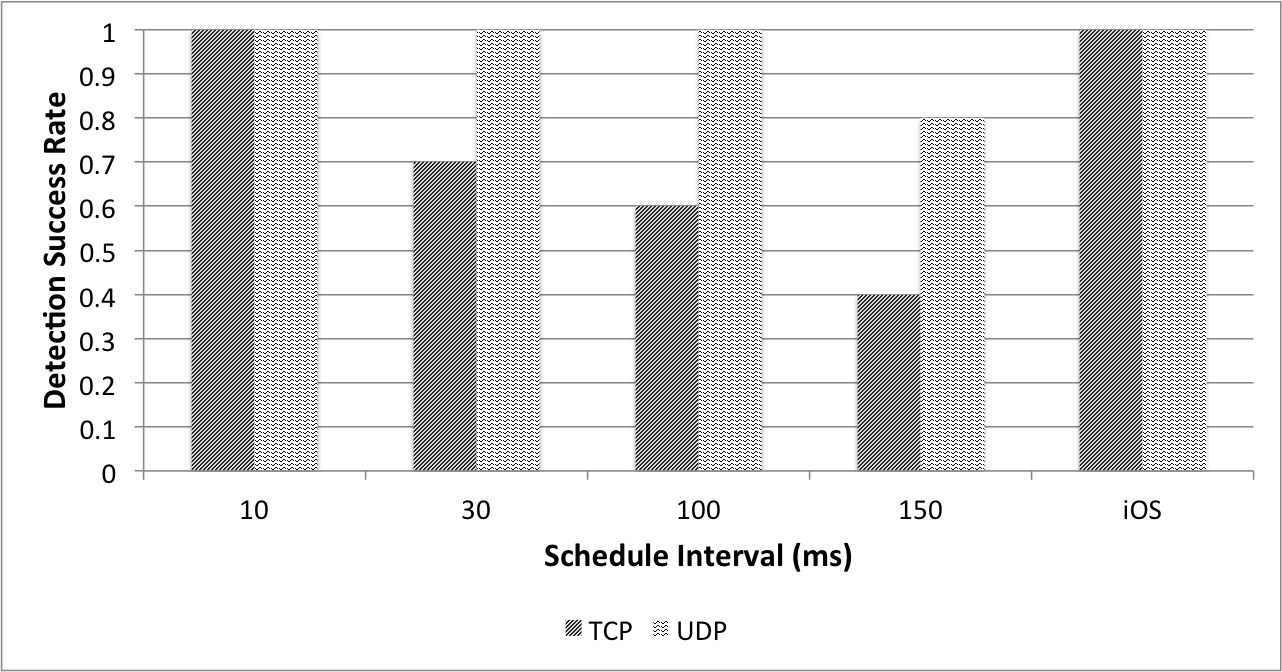}
\caption{TCP and UDP flow detection accuracy for different Android \textit{Monitor} configurations (10ms, 30ms, 100ms, 150ms) and for the iOS \textit{Monitor}.}
\label{fig:detection}
\end{figure}

\vspace {5pt}\noindent\textbf{Detection accuracy}. The \textit{Monitor}'s goal is to detect an interesting flow generated on the phone. For iOS the detection accuracy is 100\% since all packets from interesting apps are routed through the VPN the \textit{Monitor} runs. For the Android \textit{Monitor} though, the situation is more complicated. For example, an interesting app might quickly set up a socket, send a packet and then close the connection. The Android \textit{Monitor}'s detection accuracy depends on whether it can catch such events given its polling interval. To answer this question we created a micro-benchmark that includes a TCP and a UDP app connecting to a TCP and UDP echo server respectively. They both stop the communication once the server response is received. Again, we ran the \textit{Monitor} in the Smarter mode 10 times for each of the following polling configurations: 150ms, 100ms, 30ms and 10ms. We found (see Figure~\ref{fig:detection}) that the 10ms configuration could always detect outgoing TCP and UDP connections.

\subsection{Communication Overhead}
\label{subsec:communication}


To answer RQ3, we performed experiments to illustrate the overall performance overhead of Hanguard, as this can be observed from a target mobile app. We further evaluate the router's performance in handling non-interesting traffic given the Hanguard security enhancements. We created a baseline by performing our experiments below on the unmodified system (Vanilla). To evaluate Hanguard communication overheads we repeated the experiments with the respective benchmark app not being policy-protected (Unmanaged), and being protected (Managed).

\vspace {5pt}\noindent\textbf{Application latency}. We ran the TCP and UDP apps individually, configured to send 100 messages each. Figure~\ref{fig:appLatency} depicts the mean latency in milliseconds (ms) for a TCP message and a UDP message for Android. The latency is measured as round trip time (RTT) on the mobile app. In particular we measured the time interval between the API call to send the message and the time that the message is returned by the server and delivered to the application layer. As we can observe, Hanguard introduces negligible latency for Managed apps on Android.

\ignore{
In Figure~\ref{fig:iOSappLatency} we show that not only our approach can have no effect on Unmanaged apps but also the Managed apps' overhead is tolerable as it is not perceptible by the user.
}
In Figure~\ref{fig:appLatencyIOS} we can see that there is a big increase on TCP packet latency for the Managed apps on iOS. Nevertheless, in practice this is often tolerable\ignore{ and happens only for managed apps}, since most devices are actuators and sensors that create mice flows~\footnote{A mouse flow is a flow with a short number of total bytes sent on the network link.} delivering a small amount of information: for example, it is completely imperceptible when the delay for switching a light grows from a few milliseconds to tens of milliseconds. This Figure also reveals an important benefit of our design: the security controls have negligible impact on Unmanaged apps, on both Android and iOS devices, for both UDP and TCP.

\ignore{In both cases, Hanguard adds no statistically significant overhead for unmanaged apps which will be the most common use case scenario in a HAN. }

\begin{figure}[t!]
	\minusminus
    \centering
    \null\hfill
    \subfloat[Android.]{\includegraphics[height=3cm]{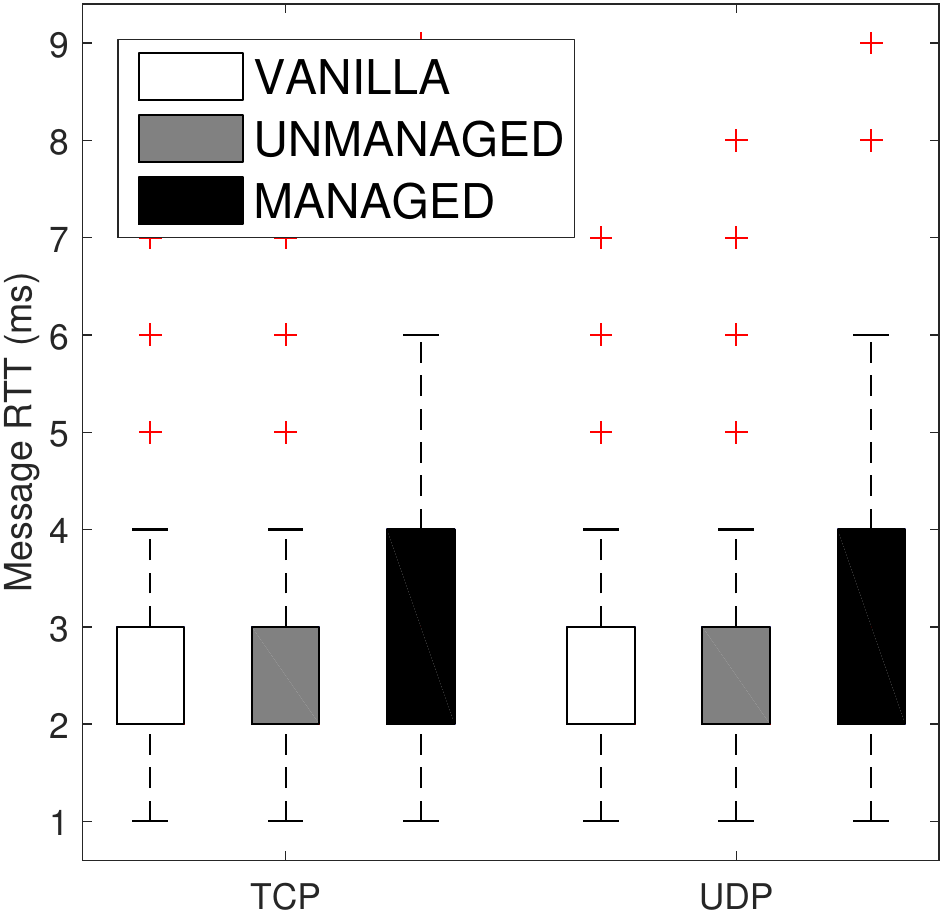}\label{fig:appLatency}}
    \hfill
    \subfloat[iOS.]{\includegraphics[height=3cm]{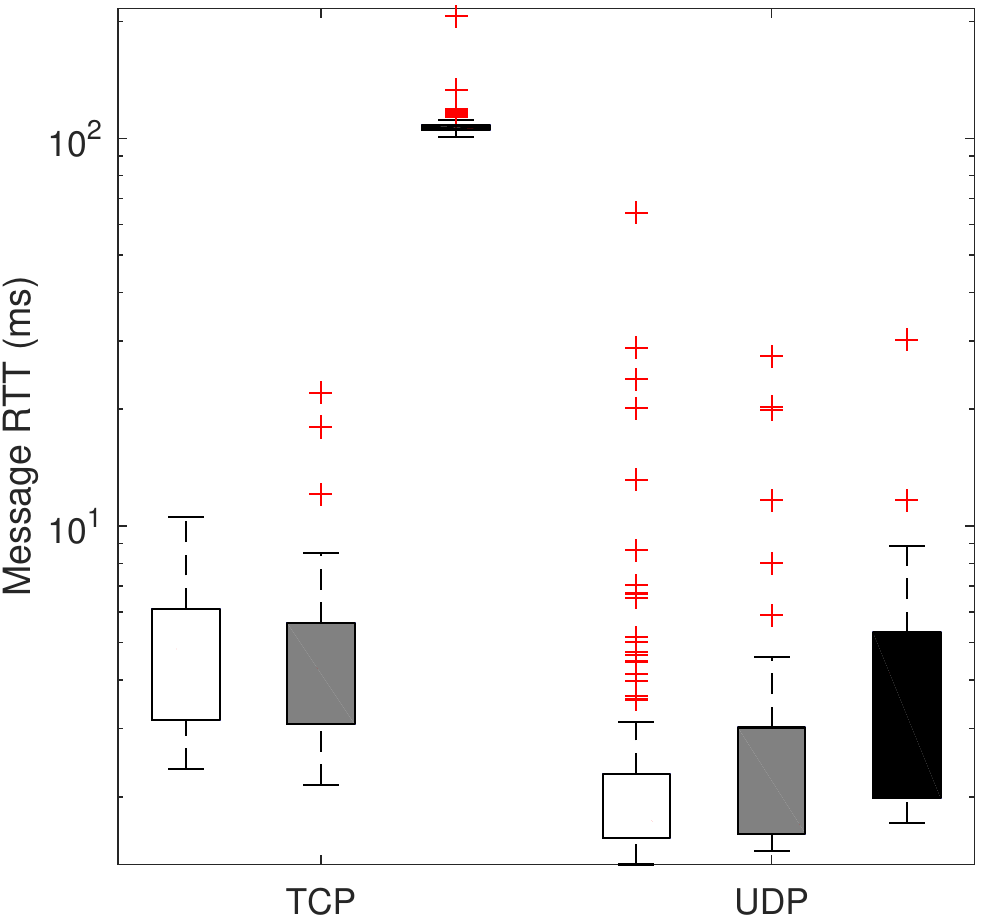}\label{fig:appLatencyIOS}}
    \hfill\null
    \label{fig:appLatencyAll}
    \caption{Application-level communication latency.}
    \minus
\end{figure}

\ignore{
\begin{figure}[t]
\centering
\includegraphics[width=7cm]{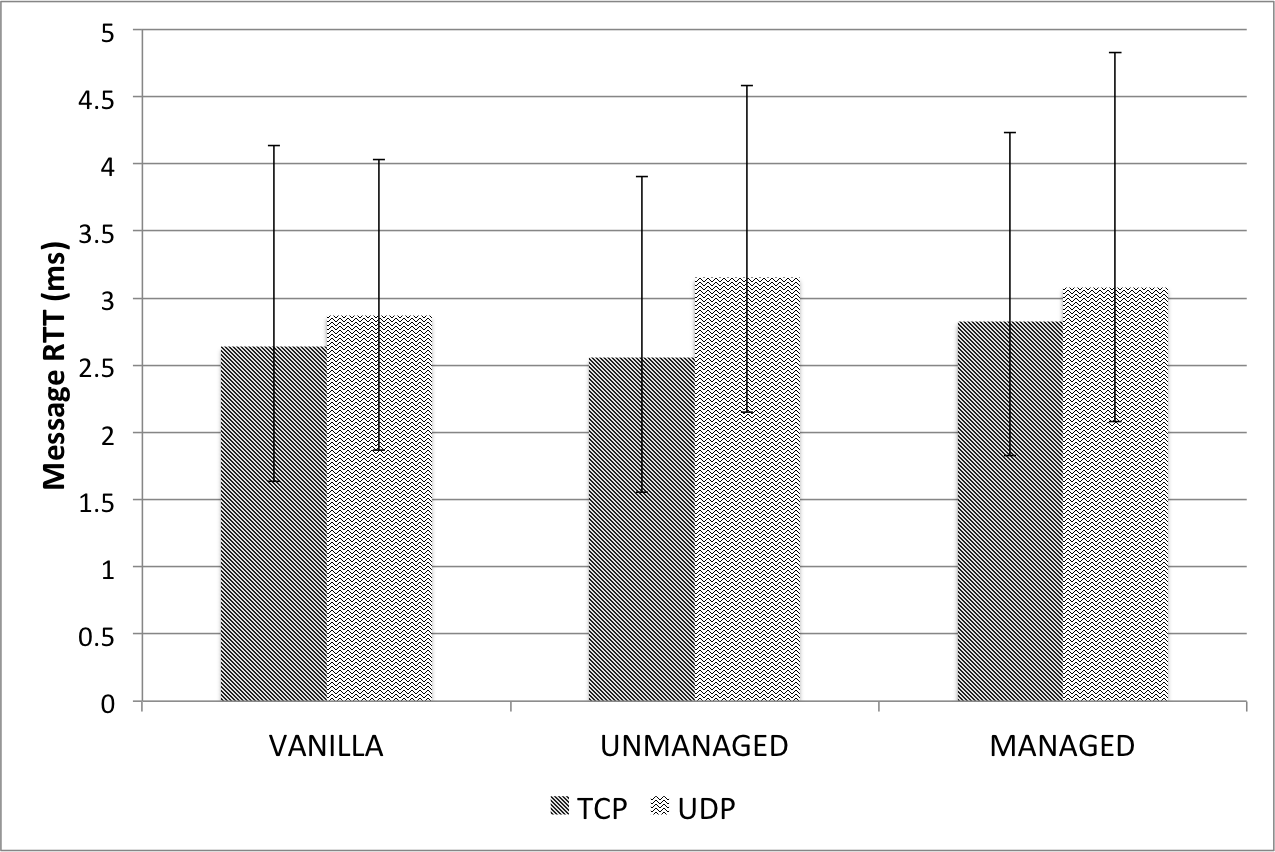}
\caption{Application-level communication latency for Android.}
\label{fig:appLatency}
\end{figure}

\begin{figure}[t]
\centering
\includegraphics[width=7cm]{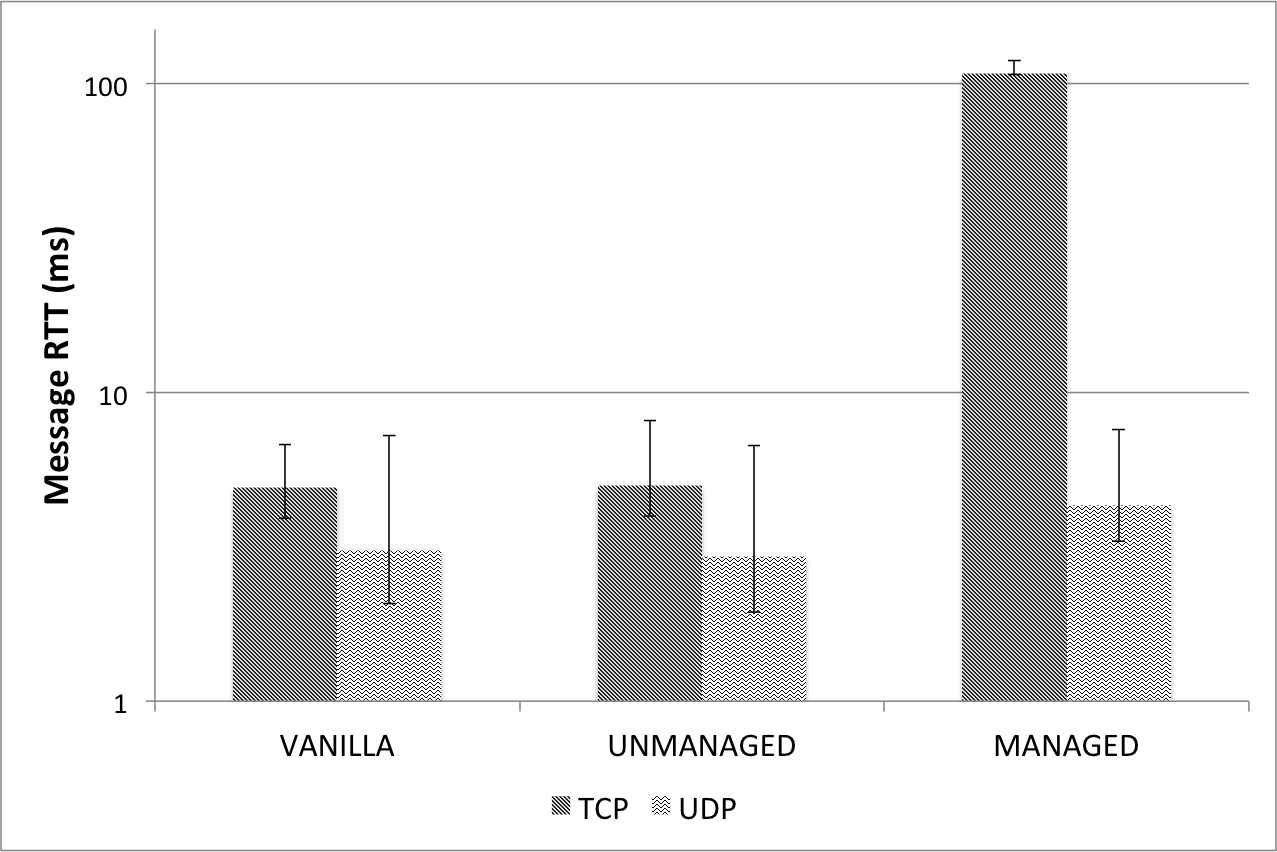}
\caption{Application-level communication latency for iOS. The $x-axis$ is in $log_{10}$ scale.}
\label{fig:appLatencyIOS}
\end{figure}
}

\vspace {5pt}\noindent\textbf{Application throughput}. To measure Hanguard's throughput overhead, we use our benchmark apps to transmit a file of 20MB to their server counterparts. We have repeated the experiment 10 times and in Figure~\ref{fig:throughputALL}\ignore{Figure~\ref{fig:throughput} and Figure~\ref{fig:throughputIOS}} we plot the CDF of the throughput for Android and iOS (*). Evidently, Hanguard has negligible impact on throughput for the Android apps and iOS unmanaged apps. Our evaluation also reveals an interesting case: throughput drops significantly for the iOS Managed apps. This happens because the iOS Monitor implementation uses the built in VPN utility of the OS. Thus, it has to inspect every packet for managed apps (see Figure~\ref{fig:iosMonitor}). This is a security, performance trade-off we had to address. We opted in for security.  Nonetheless, this will not affect performance of most managed  apps in practice: most devices are simple actuators and sensors; the performance penalty will only affect protected iOS-device communications involving real-time streaming services through the HAN. Since obviously this is an edge case, one could handle it differently. For example, when a HAN iOS communicates with such a service, the Monitor could opportunistically intercept traffic instead of checking every packet. Alternatively, in a less security stringent policy, real-time streaming services could get only phone-level protection which will eliminate all iOS overheads.

\ignore{In the more common situations (Unmanaged apps, control of actuators, reading sensor values), Hanguard offers a throughput comparable to the unmodified environment with enhanced security protection.}

\begin{figure}[t!]
	\minusminus
    \centering
    \null\hfill
    \subfloat[Android.]{\includegraphics[height=3cm]{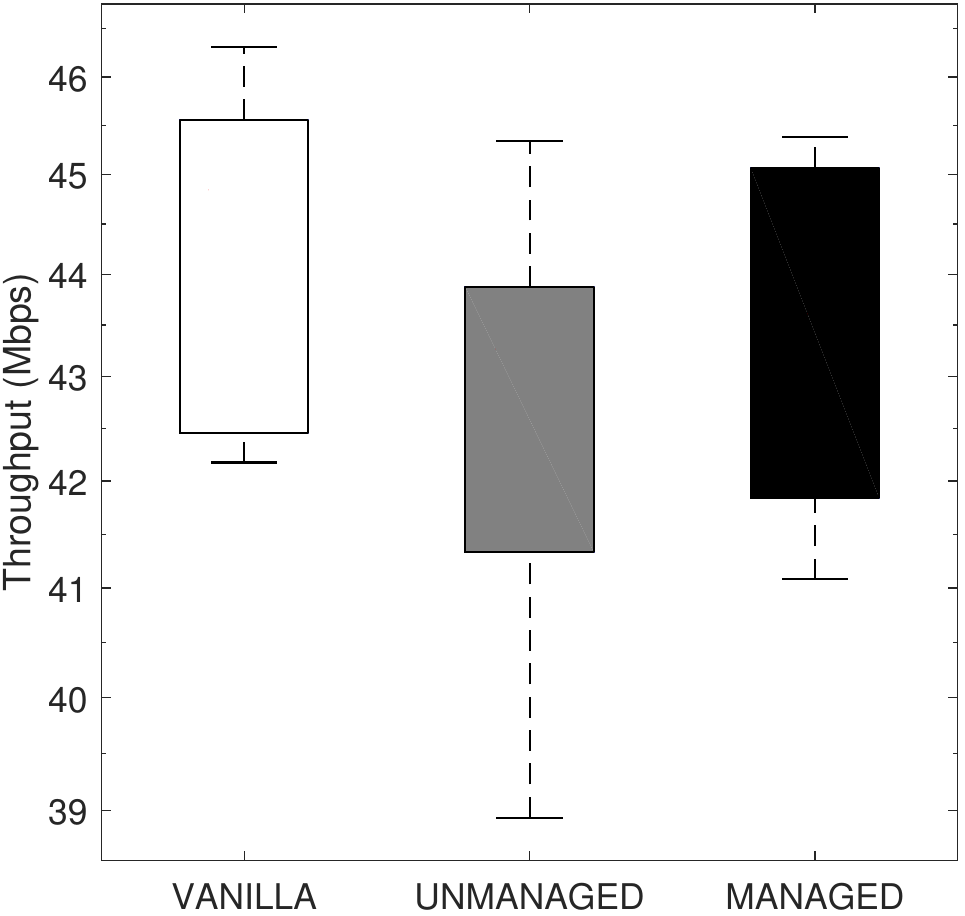}\label{fig:appThroughputAndroid}}
    \hfill
    \subfloat[iOS.]{\includegraphics[height=3cm]{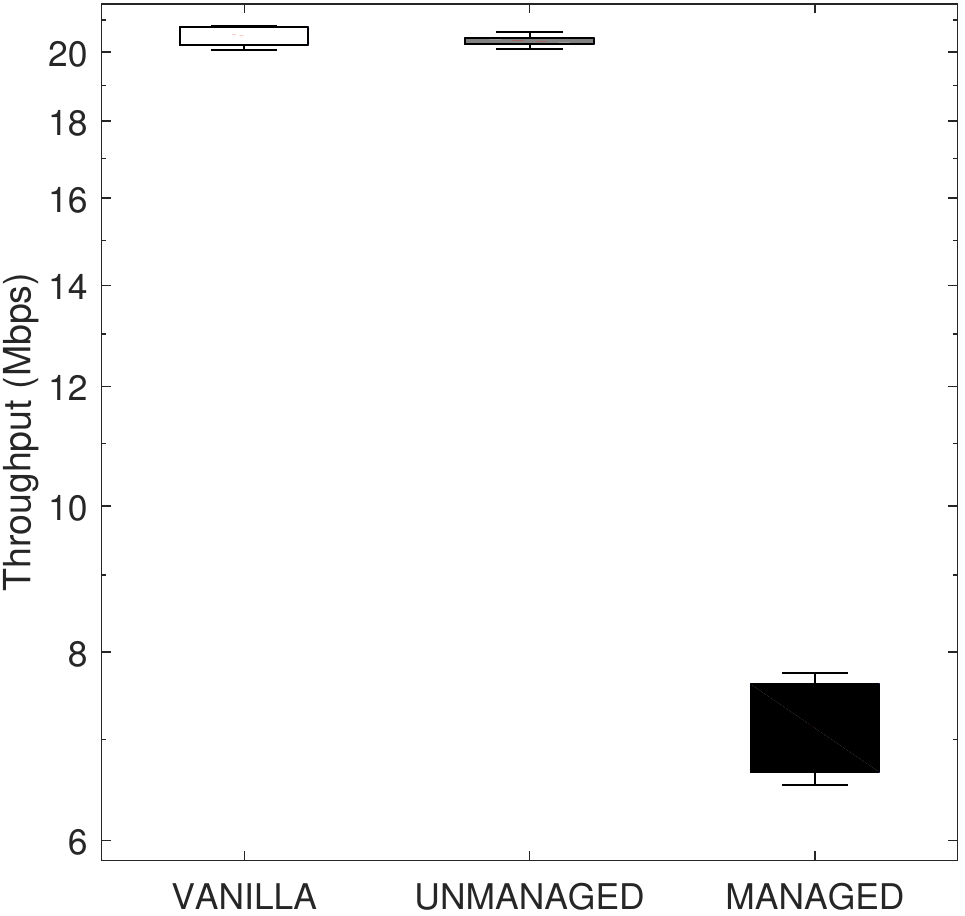}\label{fig:appThroughputAndroid}}
    \hfill\null
    \caption{Application-level throughput (TCP).}
      \label{fig:throughputALL}
    \minus
\end{figure}

\ignore{
\begin{figure}[t]
\centering
\includegraphics[width=7cm]{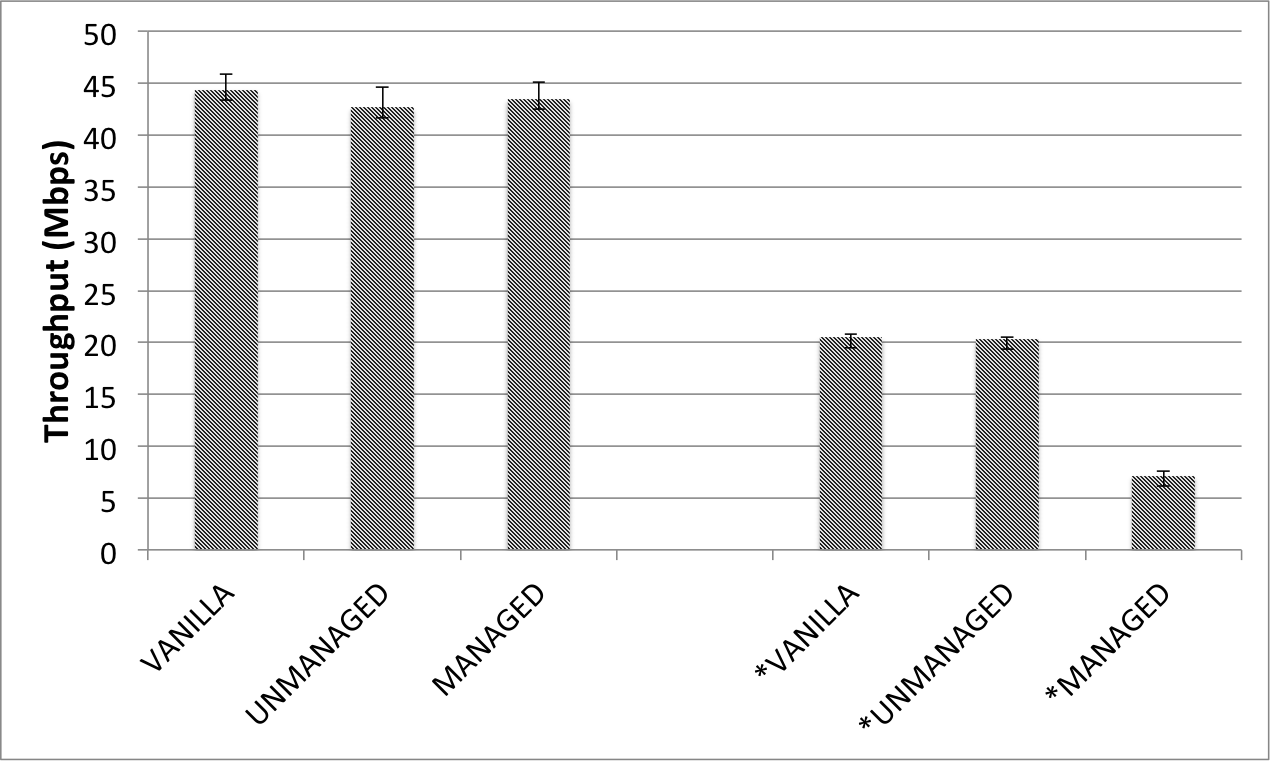}
\caption{Application-level throughput for Android and iOS (*).}
\label{fig:throughputALL}
\end{figure}
}

\ignore{
\begin{figure}[t]
\centering
\includegraphics[width=7cm]{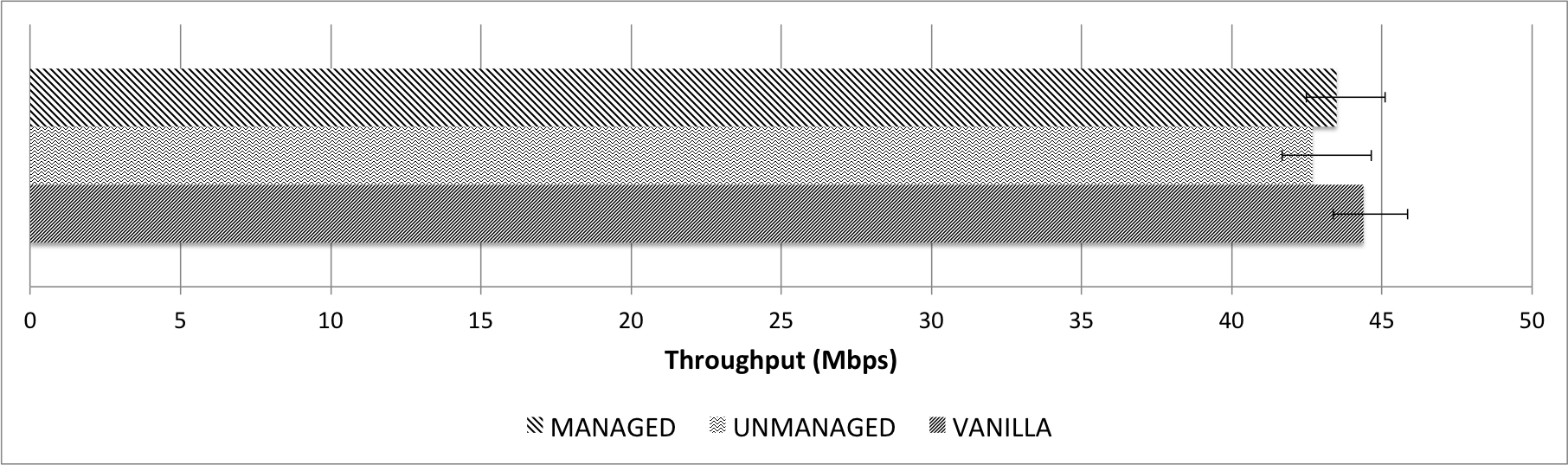}
\caption{Application-level throughput for Android.}
\label{fig:throughput}
\end{figure}

\begin{figure}[t]
\centering
\includegraphics[width=7cm]{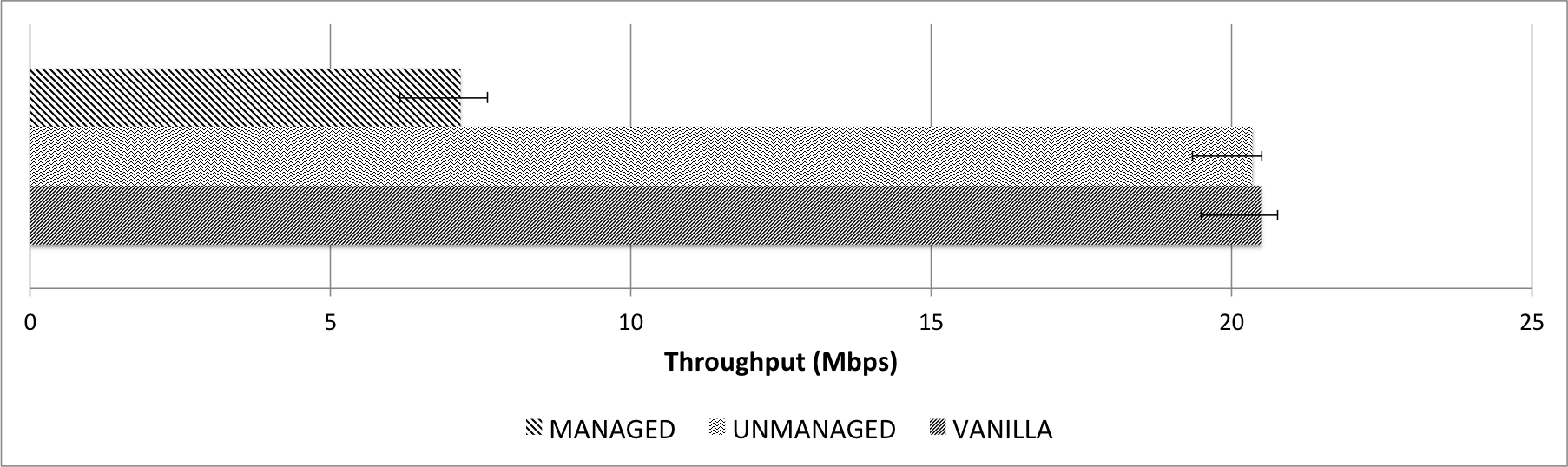}
\caption{Application-level throughput for iOS.}
\label{fig:throughputIOS}
\end{figure}
}

\section{Security Analysis}
\label{sec:analysis}


\subsection{App-level Attack Scenarios}
\label{subsec:scenarios}

\vspace {5pt}\noindent\textbf{Access from unauthorized HAN user phone.}  An app on an unauthorized HAN user phone might attempt to access an IoT device. Even if the app's category matches the IoT device's category, the Monitor on that phone will detect the outgoing flow and determine that the host phone's role cannot access the domain which encompasses the target IoT device. Thus it will not push a flow decision to the router for that app and packets on the data plane for that flow will be dropped by the router.

\vspace {5pt}\noindent\textbf{Access from unauthorized app on an authorized HAN user phone.} (a) A malicious app on a HAN user phone might try to take advantage of the phone rights and access an IoT device. However, the Monitor on that phone, will detect that the source application has a category different than the target device. Thus it will not push a flow decision to the router, and the malicious packets will be dropped by the router. (b) A repackaged app on Android could use the same package name as a Hanguard category-tagged app. However, the Android Monitor uses the package \textit{signature}) to verify an Android app. Therefore, the Monitor detects the discrepancy, notifies the user and does not authorize the data flow. \ignore{The malicious attempt is logged at the router and the admin user is notified.}

\subsection{Beyond the app-level adversary}
\label{subsec:beyond}
\vspace {5pt}\noindent\textbf{WPA2-PSK authentication and Hanguard network partition.} On a typical WLAN node, once a subnetwork is created it can be configured to use a Service Set Identifier (SSID) and the WPA2-PSK\ignore{ (WiFi Protected Access II - Pre Shared Key)} security protocol. WPA2-PSK derives a unique pairwise transient key to encrypt the communication traffic between individual nodes on a HAN and the router. However, all keys are derived from the same SSID and a secret passphrase shared across all the nodes. As a result, a compromised phone could potentially use the key to directly connect to an IoT device, bypassing the router-level protection. To address this threat, Hanguard partitions the HAN into two default subnetworks, each with their own SSID/passphrase pair, one for user phones, PCs and laptops, and the other for IoT devices. This ensures that even a fully compromised phone cannot acquire the secret key used by smart-home devices.

\vspace {5pt}\noindent\textbf{Access from unauthorized authenticated guest phone.}\ignore{Hanguard's threat model, assumes that guest phones could be compromised.}  (a) A guest phone's role is only allowed to access the \textit{unprotected} domain. If the target IoT device's type is in another domain, the router will reject the offending packets. (b) A guest phone might try to claim the identity of one of the HAN Monitor's to alter the policy rules. Since the guest phones are in a different local network and use different keys, the router will not accept rules from those phones. Further, a guest phone will not be able to authenticate to the router controller module, since it lacks both the client certificate and the user credentials. Lastly, Hanguard uses static IPs associated with phones/devices MAC addresses during setup by the admin role. Any attempt to claim a reserved IP (\textit{arp-spoofing}) or MAC address (\textit{MAC-spoofing}) is validated through the control channel and the admin user is notified out-of-band.

\vspace {5pt}\noindent\textbf{Compromised HAN user phones.} \ignore{Hanguard assumes that the HAN user phones are not compromised to guarantee application-level policy enforcement. However, even} In this case, Hanguard can still guarantee phone-level access control. (a) The phone might attempt to surreptisiously access an IoT device. Assuming it acquires the user credentials and the Monitor certificate, it can try to push a rule to the router to allow its flow. However, the router detects that the rule comes from a device whose \textit{role} is not allowed to access the \textit{type} of the target device. It rejects the rule and notifies the admin user. (b) The phone might attempt to update the policy in its favor. An attempted policy update by a non-admin device \ignore{(determined by the device MAC:IP address pair, Monitor certificate, and the user credentials)} will be rejected and the admin will be notified. Even if the admin device and its user credentials are compromised and a policy update is pushed, the admin user will be notified. Thus she can revoke the update and take action. (c) The phone, might try to flood the \textit{flow decision cache}. This would force the \textit{GCS} service to retire older flows, essentially invalidating benign flows and causing denial of service. To tackle this, we rate limit the flows a particular device can create. If that limit is surpassed, the device is penalized by having the router dropping all its packets for a few minutes. During that time, no flow entries will be added in the decision cache originating from that device. Furthermore, Hanguard triggers its \textit{notification mechanism}.


\vspace {5pt}\noindent\textbf{Remote adversary.} \ignore{Smart home devices are connected to the Internet which allows them to receive commands and other messages from their cloud services. To enable this the router needs to support port-forwarding or allow various NAT-traversal settings. }Commonly, an IoT device behind the NAT initiates a connection to its cloud, through which the cloud learns the device's external IP and port. However, if this information is exposed to an unauthorized party, that party can gain unfettered access to the device~\cite{shodan, shodanBaby, cctvVuln, shekyan2013}. \ignore{This happens because of the router's default full-cone NAT configuration that essentially allows all remote IPs to reach all local IPs on any port.}To shut down this exploitable channel, Hanguard by default configures the router to a \texttt{port-restricted cone NAT}, which ensures that only the flows from the remote IP/port pairs contacted before by a local device can reach that device. Note that this NAT mode is supported by most smart-home devices on the market~\cite{nabtoSpecs, nabtoUsedBy, weaved, kalay}.

\section{Discussion}
\label{sec:discuss}

\ignore{
\vspace {5pt}\noindent\textbf{Control on future iOT devices}. Our system assumes that the users' mobile phones are the system's subjects and the IoT devices are the system's objects. However, in a LAN home or enterprise environment, all WiFi devices can potentially communicate with each other. In particular, Google recently introduced Brillo~\cite{brillo}, a lightweight operating system for IoT devices, and Weave~\cite{weave}, a communication protocol enabling local interoperability among such devices. Actuators and sensors equipped with such capabilities can act not only as resources to be accessed but also as subjects consuming resources. In future work, we plan to extend our scheme to control interactions among all system's objects, in all possible directions. To do this we would simply need to allow IoT devices to be assigned not only domains but roles as well. Furthermore, all LAN devices can be assigned sensitivities to instantiate multi-level security distributed access control. An exciting direction would be to look into p2p access control schemes for IoT systems that do not have the capacity of a centralized enforcing point.
}

\vspace {5pt}\noindent\textbf{HAN users smartphones OS integrity}. Our application-level enforcement assumes that HAN user phones are not compromised. Preventing phone compromises is out of the scope of this work since other solutions already exist and even deployed on commodity smartphones~\cite{DBLP:conf/ndss/MarforioKSKC14,DBLP:conf/esorics/SunSWJJ14,DBLP:conf/ccs/Winter08,trustzone,Azab:2014:HAW:2660267.2660350,iOSsec}. For example, SELinux for Android~\cite{DBLP:conf/ndss/SmalleyC13} uses mandatory access control to ensure that even compromised system processes are restricted, and is deployed on all Android phones with version 4.4. and higher (more than 60\% in 2015~\cite{statista}). Most Android phones are equipped with ARM processors~\cite{armprocessors} with TrustZone~\cite{trustzone} which can be utilized for solutions stemming from the trusted computing domain. TZ-RKP~\cite{Azab:2014:HAW:2660267.2660350} is a real-time kernel protection technique deployed on Samsung Galaxy phones that ensures the kernel integrity using the ARM TrustZone secure world. iOS devices have the Secure Enclave, a secure co-processor that is used to guarantee secure boot~\cite{iOSsec}. However, even if a user device is compromised (and in the case of all guest phones), Hanguard can guarantee \textit{phone-level} protection (Section~\ref{subsec:beyond}). \ignore{Recall that Hanguard assigns a \texttt{domain} to each phone, domains to \texttt{roles}, and \texttt{types} to IoT devices (see Section~\ref{subsec:overview}). When a request is being made by a phone, all these are checked at the router which guarantees that every phone can only access IoT devices it is allowed by the policy.} Furthermore, Hanguard helps its users detect spurious access attempts by (1) keeping a log, (2) sending out-of-band notifications to the admin user when a violation of the policy is attempted.

\vspace {5pt}\noindent\textbf{Switching Between Information Gathering Approaches}. iOS follows a far more stringent approach than Android in isolating processes. In fact our Android solution for traffic monitoring does not work on iOS. Instead we utilize Apple's \texttt{NEPacketTunnel} \texttt{Provider} API with a per-app VPN configuration. The latter requires an MDM (Mobile Device Management) server: the \textit{router} vendor will need to enrol their users' iOS devices and push an over-the-air (OTA) configuration profile on the phone, just like the cell phone carriers (e.g. AT\&T, T-mobile e.t.c.) do. This process is already mature and streamlined for users who just need to accept the configuration. Apple does offer the non-enterprise \texttt{NEVPNManager} API but that would entail Hanguard iOS Monitors proxying traffic not only from a selected set of apps but from all apps, imposing the overheads we demonstrate in Section~\ref{sec:evaluation} for both unmanaged and managed apps. In our prototype we opted for security and runtime performance in the expense of an initial bootstrapping usability burden, that allows us to selectively proxy traffic only from a handful of apps when used in the HAN environment. Our work illustrates how such capabilities can facilitate novel solutions on the iOS platform. Also note that any of the two aforementioned techniques can be used in practice with Hanguard iOS Monitors. Hanguard's design, allows router vendors to readily switch between monitoring techniques with a mere application update. Similarly, if access to the Android \texttt{procfs} as a whole is forbidden in the future (not a straightforward decision since this would break a lot of legitimate apps), Hanguard can switch to a VPNService-based Android Monitor by merely pushing an app update.

\vspace {5pt}\noindent\textbf{Communication through cloud}.  Some apps still go through the cloud during the communication phase, even if they are on the same HAN as the device. Hanguard's RBAC can be applied to those apps as well by extending the NAT checks. In particular the Monitors can detect an app on a user phone contacting its respective cloud service and notify the router. Using the traditional port-restricted cone NAT, Hanguard allows the traffic from validated external IoT domains to reach their respective IoT devices as long as the connection was initiated from the IoT device itself. In this case, Hanguard can further utilize the information received from the Monitor, and allow the remote traffic only if the following also hold true: (a) the role of the phone generated the traffic to the device's IoT cloud is allowed by the policy to access that device, (b) the category of the app matches the category of the target IoT device and, (c) the traffic came within a time limit since the receipt of the control message (e.g. 1 sec). The last is needed to tackle the fact that someone could try to bypass the system while an accepted role also sent a remote command.

\ignore{
On Android, the ability of any app to readily get traffic information of other apps without any permission is indeed worrisome.~\footnote{Zhou et. al.~\cite{Zhou:2013:ILD:2508859.2516661} used other information from the \texttt{procfs} file system to demonstrate side-channel attacks on Android.} We believe that while this bootstraps creative applications on Android, access to the \texttt{procfs} file system should be mediated by the OS and allowed by the user. Also note that, Hanguard's Android Monitors can use the VPNService on Android instead of the \texttt{procfs} files if needed. Even if a Monitor is deployed using one of these techniques it can simply switch to another with an application update, a nice benefit stemming from our design which has the Monitors running as userspace apps.
}

\ignore{
\vspace {5pt}\noindent\textbf{Monitor platform dependence}. Our Monitor prototypes work well on the most popular mobile platforms which are the primary---and most times the only---interfaces to IoT devices. However, there might be cases where an IoT device is also controlled by a Windows phone/desktop/laptop or Mac desktop/laptop. In future work we plan to expand our implementations on those platforms as well. A promising direction would also be to build support in the OSes themselves to enable compatibility with IoT-aware HAN distributed access control schemes such our own.
}


\section{Related Work}
\label{sec:related}
\vspace {5pt}\noindent\textbf{IoT attacks}. Recent works have demonstrated attacks on IoT devices~\cite{ZhangY0ZW15, rapidReport, 2014notra, DBLP:conf/wimob/SivaramanGVBM15, fernandes2016security, Sivaraman:2016:SAS:2939918.2939925}. Fernandes et.al. found vulnerabilities on SmartThings' applications~\cite{fernandes2016security}. Their work focuses on a specific IoT hub that can integrate third-party IoT devices. Our work presents a solution applicable to an infrastructure that exists in almost all households with IoT devices. \cite{rapidReport,2014notra}, revealed vulnerabilities on the Philips Insight, iBaby baby cameras and Belkin devices.\ignore{ Notra et. al.~\cite{2014notra} also verified inadequate protections on the Belkin WeMo power switch and Motion Sensor,} However they consider an adversary on a separate device. \cite{Sivaraman:2016:SAS:2939918.2939925} considers an intricate mobile adversary which colludes with a cloud. We illustrate that the mobile adversary can succeed with minimal effort. \ignore{We further demonstrate and focus on more realistic adversaries that can place a malicious application on a user's phone.} All reported attacks further motivate the need for practical smart-home defenses.

\vspace {5pt}\noindent\textbf{Side-channels on Android and network monitors}. Several works focused on acquiring information for other processes using side-channels on Android~\cite{Zhou:2013:ILD:2508859.2516661, DBLP:conf/ndss/ZhouJ13, conf/sp/JanaS12a, ZhangY0ZW15}. \cite{ZhangY0ZW15} also utilized side channel information for defence purposes, avoiding system-level modifications. \cite{DBLP:conf/sigcomm/LeVLSGM15} used the VPN service on Android for passive monitoring of a selected set of mobile apps to collect user traffic information for analysis. However, it redirects all packets to a server that further routes the packets. This entails privacy concerns which we avoid by implementing the routing functionality locally.\ignore{In fact, after we check the packet headers, we forward the packets with their original structure to their original destination.}

\vspace {5pt}\noindent\textbf{Access control}. There have been various works on home access control\ignore{, even before the advent of smart devices that connect through the Internet. } which we classify in three major areas: surveys~\cite{gomez2010, ur2013, Denning2013CSM}; access control systems~\cite{Ahn2008, Das2011, kim2012, lioy2014, DBLP:conf/wimob/SivaramanGVBM15, fernandes2016flowfence}; and user studies for usable policy specifications~\cite{Kim:2010:CAR, Mazurek:2010:ACH}. More relevant to our work is the second. Nonetheless, most of these systems assume a clean-slate design where the OSes of participating nodes can be modified. Our solution is backward compatible: it requires just a software upgrade on the Home's router and downloading an app on the phone. Other work focused on access control enforced on the mobile phones~\cite{SmalleyNDSS13, bugiel2013, demetriouNDSS2015}. Demetriou et. al.~\cite{demetriouNDSS2015} enforced local policies to control access to personal devices while our target is to enforce a distributed policy on shared devices. \ignore{Jaeger et. al.\cite{jaegerIpsec2006, DBLP:conf/uss/JaegerMCCS06} studied such distributed enforcement, however his work assumes cooperation by the users' laptop/desktop OS. Our approach works on mobile phones and assumes no such cooperation. Finally our policy is consciously compatible with the SELinux specification. Nonetheless our application of \texttt{types} for IoT devices, \texttt{domains} for phones and \texttt{categories} for apps is unique.}\ignore{use of the category is unique: while Android uses it to separate users on the same device, we utilize it to enable app-level distributed control.}

\vspace {5pt}\noindent\textbf{IDS and Firewalls}. Work on intrusion detection systems (IDS), personal and application firewalls~\cite{Crotti:2007:TCT:1198255.1198257, judge2002, spinach, Cheswick:1994:FIS:200552}, focuses either solely at the host or at a network node, or only at the network layer. Our solution works in a distributed manner, consolidating application level semantics from hosts, and network level information from the network node. Furthermore, we do not require experts to set up policies.\ignore{In par with \ignore{Software Defined Networking}SDN concepts~\cite{sdnSurvey}, we employ a control channel to push flow decisions to a network node. However, our controller runs on the same device where the traffic is generated. This allows us to get application-level semantics and offer interaction capabilities with HAN users rather than datacenter admins.}

\section{Conclusion}
\label{sec:conclude}
In this work we presented Hanguard, a system that can enforce access control policies in a HAN among user phones and IoT devices. Hanguard uses a new software-defined networking (SDN) approach applied on home area networks using mobile phones as monitors: it employs situation awareness on users' phones through a userspace \textit{Monitor} app that detects whether an authorized app is establishing a network flow with a target IoT device; \textit{Monitors} push decisions to the HAN router bridging the gap between network and application-level semantics. This technique allows the router to enforce fine-grained access control based on a global policy protecting access to HAN IoT devices. Hanguard does not require mobile OSes modifications, any IoT device modifications, or new router hardware. It is backward compatible with the existing HAN infrastructure, and was implemented and evaluated in a realistic HAN setting, verifying both its practicality and effectiveness.






\bibliographystyle{./IEEEtran}
\bibliography{bibliography}
%




\section{APPENDIX A: IoT apps study}
\label{sec:appendix}

\subsection{IoT Product Functionality Categorization Summary}
On Section~\ref{subsec:analysis} we have elaborated our strategy for analyzing the existing state of IoT products. Here we provide a detailed categorization of our functionality categorization.

On Figure~\ref{fig:func_all_mirror} we depict the overall categorization of 353 IoT products listed on \url{iotlist.co}, on Figures \ref{fig:func_wearables}, \ref{fig:func_security}, \ref{fig:func_automation}, \ref{fig:func_entertainment} a further sub-categorization for the ``Wearables'', ``Home Security'', ``Home Automation / Appliances'' and ``Home Entertainment'' categories respectively.

\ignore{
\begin{figure*}[h!tb]
\centering
\begin{tabular}{cc}
  \includegraphics[height=3cm]{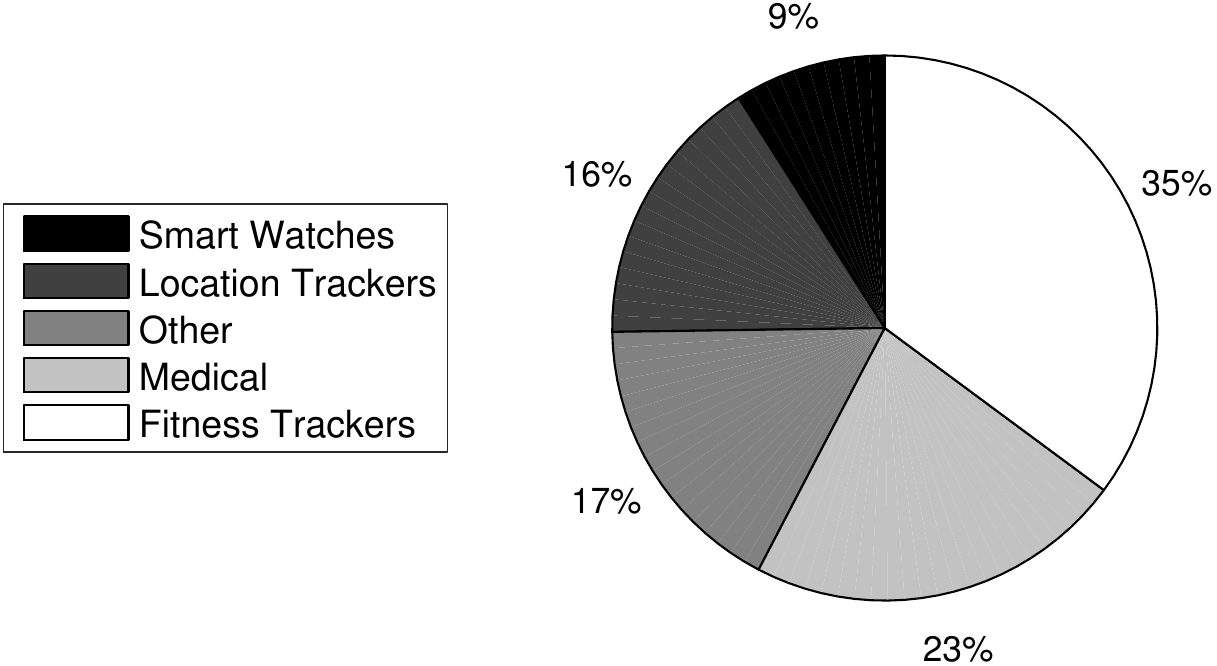}\label{fig:func_wearables} &   \includegraphics[height=3cm]{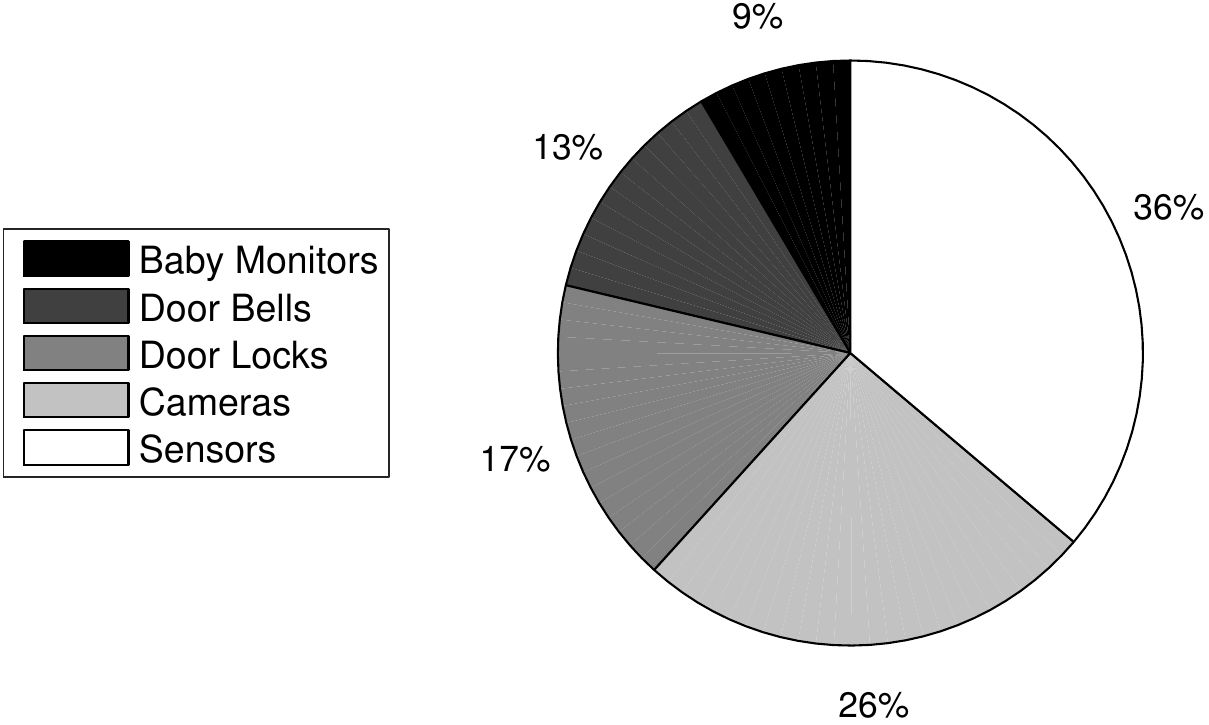}\label{fig:func_security}\\
(a) Wearables & (b) Home Security \\[6pt]
 \includegraphics[height=3cm]{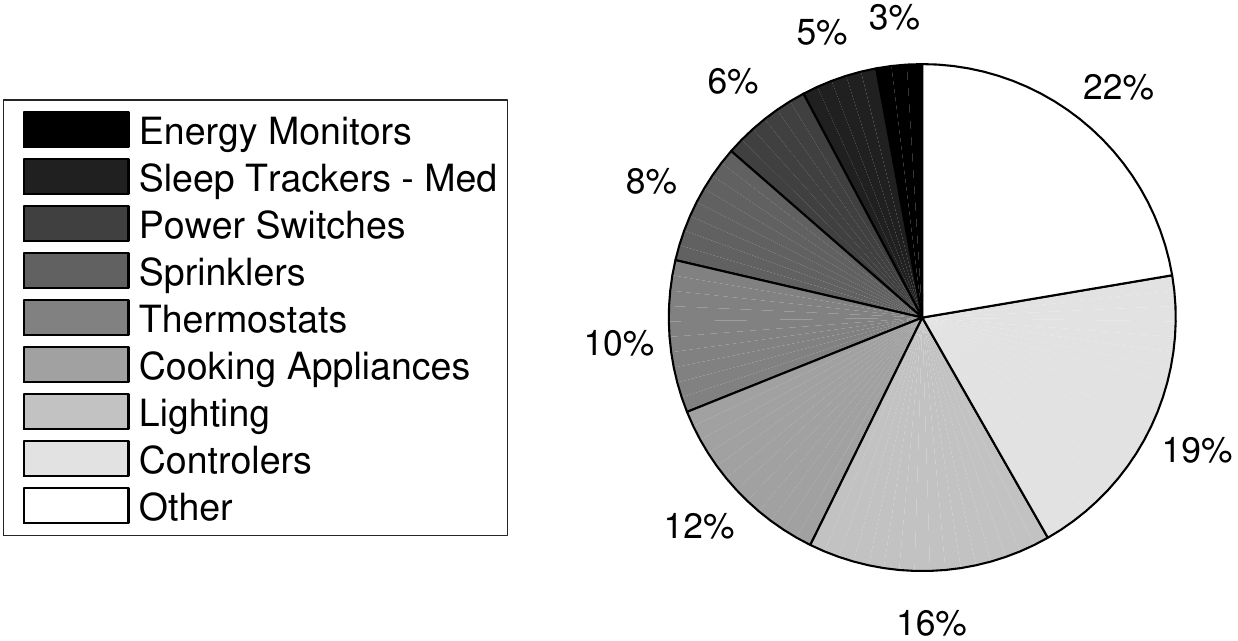}\label{fig:func_automation} &   \includegraphics[height=3cm]{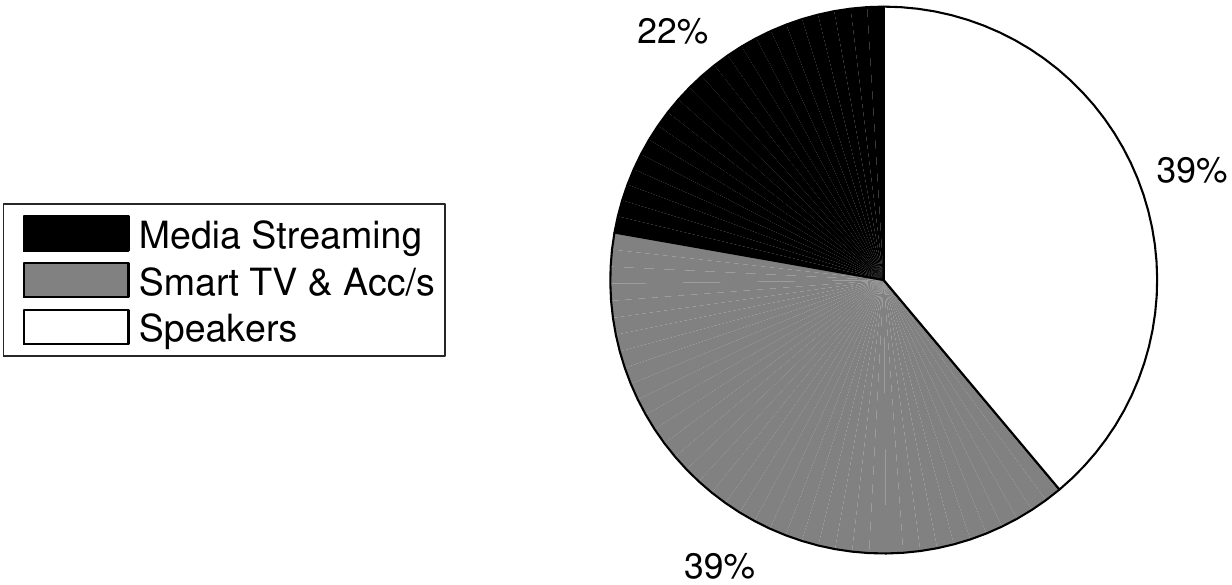}\label{fig:func_entertainment} \\
(c) Home Automation / Appliances & (d) Home Entertainment \\[6pt]
\end{tabular}
\caption{Functionality sub-categorization of 353 IoT products collected from \url{iotlist.co}.}
\label{fig:func_subcat}
\end{figure*}
}

\ignore{
\begin{figure}[h]
\centering
\includegraphics[width=6cm]{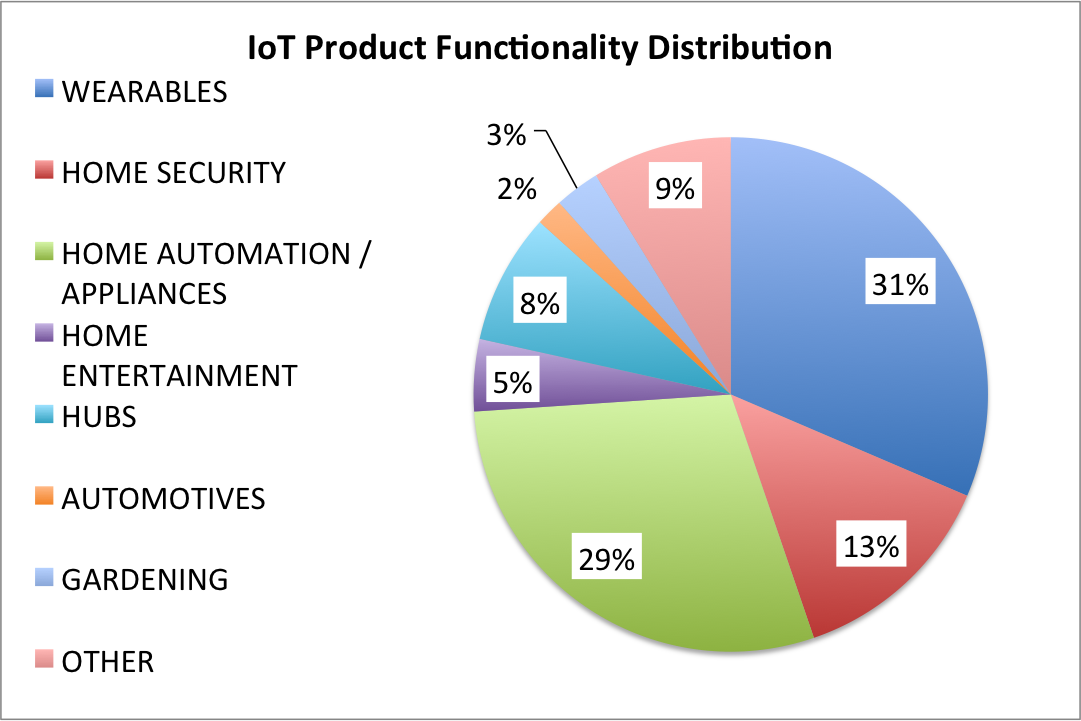}
\caption{IoT product functionality categorization.}
\label{fig:func_all}
\end{figure}
}

\begin{figure}[h]
\centering
\includegraphics[width=6cm]{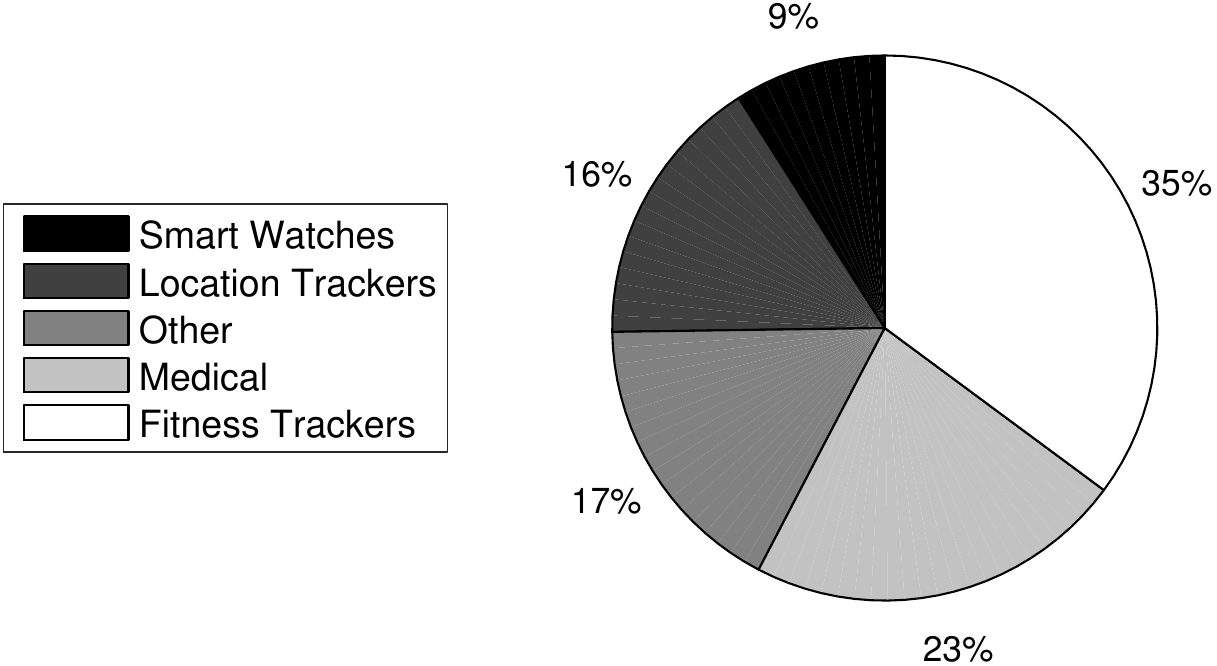}
\caption{Wearables functionality sub-categorization.}
\label{fig:func_wearables}
\end{figure}

\begin{figure}[h]
\centering
\includegraphics[width=6cm]{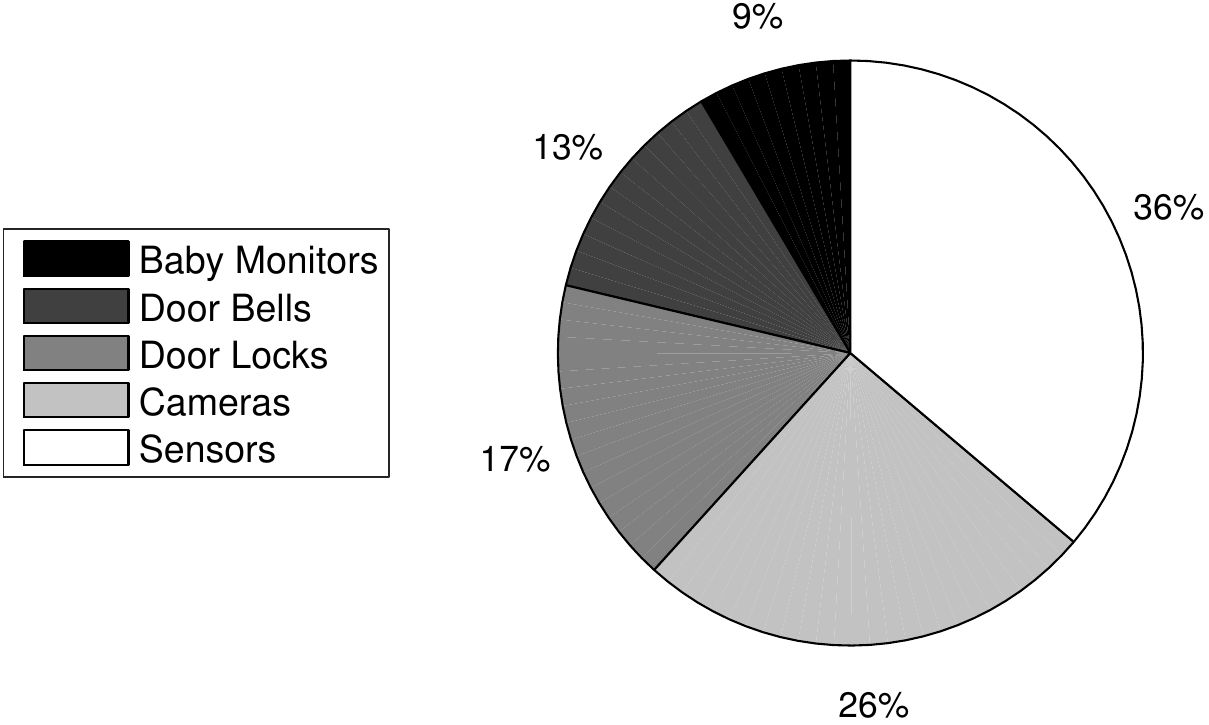}
\caption{Home Security product functionality sub-categorization.}
\label{fig:func_security}
\end{figure}

\begin{figure}[h]
\centering
\includegraphics[width=6cm]{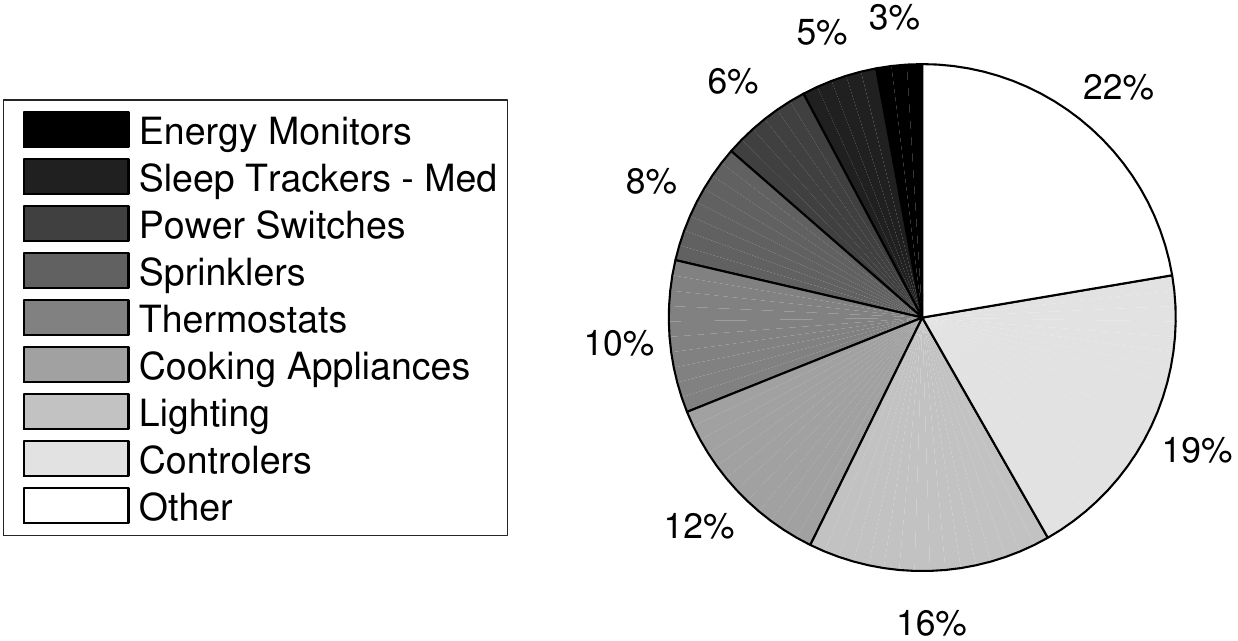}
\caption{Home Automation / Appliances product functionality sub-categorization.}
\label{fig:func_automation}
\end{figure}

\begin{figure}[h]
\centering
\includegraphics[width=6cm]{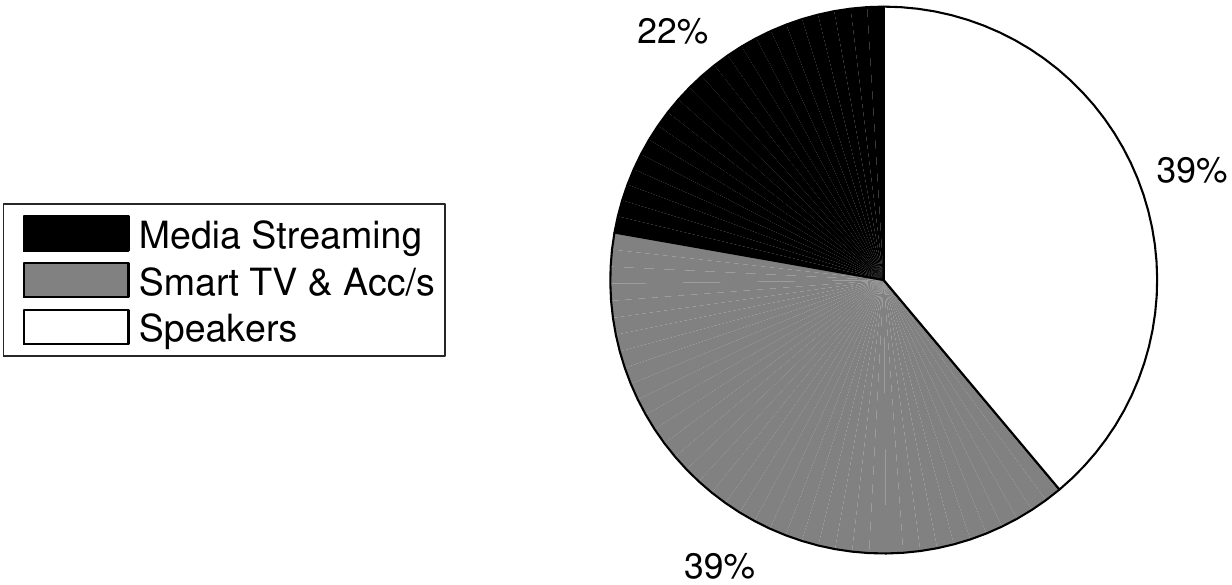}
\caption{Home Entertainment product functionality sub-categorization.}
\label{fig:func_entertainment}
\end{figure}

\subsection{IoT HAN Trust Model Study} As the IoT industry matures, more and more devices add local WiFi connections to avoid the cloud availability requirement and unnecessary latency. In our work we compared the authentication practices of IoT vendors when cloud connection is attempted vs when a local WiFi connection is attempted. 

\vspace{3pt}\noindent\textbf{App Selection.} Our IoT product collection (see Section~\ref{sec:threat}) resulted in 223 products having apps on Google Play. However, a lot of apps can control multiple products. For example, the WeMo app can control at least three Belkin WeMo products. In particular, we found 92 unique Android apps on Google Play. Note also that part of these 92 apps, are apps that control wearables such as smartwatches and fitness trackers. Thus we manually went through the 92 apps and selected only those that connect to smart-home devices. This resulted in 55 unique smart-home apps. Our collection is conservative: while it does not yield a complete set of smart-home Android apps, it ensures that the selected apps are related to WiFi smart-home devices. This is important since our null hypothesis is specific to devices that connect to the HAN. 

\vspace{3pt}\noindent\textbf{App Analysis.} See Section~\ref{subsec:security}.

\vspace{3pt}\noindent\textbf{Results.} Unfortunately, we found that smart-home apps tend to trust the local environment which exposes them to attacks from mobile adversaries---9/22 (41\%) apps with local WiFi connectivity perform no authentication when in the HAN. Table~\ref{tab:wifiApps} lists the 55 IoT systems we examined for our statistical significance test. This highlights the need for a solution which is independent of IoT vendors and application developers practices. Hanguard on top of its role-based management capabilities, also offers protection for IoT devices.



\begin{table*}[ht]
\centering
\scriptsize
\caption{IoT apps selected for trust model analysis. Y = YES; N = NO; ? = Undetermined.}
\label{tab:wifiApps}
\begin{tabular}{|c|l|l|l|l|l|l|l|}
\hline
\textbf{No} & \multicolumn{1}{c|}{\textbf{PACKAGE NAME}} & \multicolumn{1}{c|}{\textbf{CATEGORY}}                              & \multicolumn{1}{c|}{\textbf{\# Installations}} & \multicolumn{1}{c|}{\textbf{DECOMPILED}} & \multicolumn{1}{c|}{\textbf{OBFUSCATED}} & \multicolumn{1}{c|}{\textbf{LOCAL WIFI}} & \multicolumn{1}{c|}{\textbf{LOCAL AUTH}} \\ \hline
1           & is.yranac.canary                           & TOOLS                                                               & 100K - 500K                                    & N                                        & -                                        & -                                        & -                                        \\ \hline
2           & com.netgear.android                        & \begin{tabular}[c]{@{}l@{}}VIDEO PLAYERS \& \\ EDITORS\end{tabular} & 100K - 500K                                    & Y                                        & N                                        & N                                        & -                                        \\ \hline
3           & com.homeboy                                & LIFESTYLE                                                           & 1K - 5K                                        & Y                                        & N                                        & N                                        & -                                        \\ \hline
4           & com.belkin.android.androidbelkinnetcam     & \begin{tabular}[c]{@{}l@{}}VIDEO PLAYERS \& \\ EDITORS\end{tabular} & 100K - 500K                                    & Y                                        & N                                        & Y                                        & Y                                        \\ \hline
5           & com.petcube.android                        & LIFESTYLE                                                           & 10K - 50K                                      & Y                                        & N                                        & N                                        & -                                        \\ \hline
6           & com.blacksumac.piper                       & LIFESTYLE                                                           & 50K - 100K                                     & Y                                        & N                                        & Y                                        & Y                                        \\ \hline
7           & ibabymonitor.main                          & TOOLS                                                               & 10K - 50K                                      & Y                                        & N                                        & Y                                        & ?                                        \\ \hline
8           & com.philips.cl.insight                     & LIFESTYLE                                                           & 50K - 100K                                     & Y                                        & Y                                        & Y                                        & ?                                        \\ \hline
9           & com.chamberlain.myq.chamberlain            & LIFESTYLE                                                           & 100K - 500K                                    & Y                                        & N                                        & N                                        & -                                        \\ \hline
10          & com.garageio                               & TOOLS                                                               & 1K - 5K                                        & Y                                        & N                                        & N                                        & -                                        \\ \hline
11          & com.skybell                                & LIFESTYLE                                                           & 10K - 50K                                      & Y                                        & N                                        & N                                        & -                                        \\ \hline
12          & com.orvibo.kepler                          & LIFESTYLE                                                           & 1K - 5K                                        & Y                                        & N                                        & N                                        & -                                        \\ \hline
13          & com.kornersafe.secure.igg                  & LIFESTYLE                                                           & 1K - 5K                                        & Y                                        & N                                        & N                                        & -                                        \\ \hline
14          & com.lifx.lifx                              & LIFESTYLE                                                           & 100K - 500K                                    & Y                                        & Y                                        & Y                                        & N                                        \\ \hline
15          & com.philips.lighting.hue                   & LIFESTYLE                                                           & 500K - 1M                                      & Y                                        & Y                                        & Y                                        & N                                        \\ \hline
16          & com.allure\_energy.esmobile                & TOOLS                                                               & 1K - 5K                                        & Y                                        & N                                        & N                                        & -                                        \\ \hline
17          & com.honeywell.mobile.android.totalComfort  & LIFESTYLE                                                           & 500K - 1M                                      & Y                                        & N                                        & N                                        & -                                        \\ \hline
18          & com.ecobee.athenamobile                    & LIFESTYLE                                                           & 50K - 100K                                     & Y                                        & N                                        & N                                        & -                                        \\ \hline
19          & com.honeywell.android.lyric                & TOOLS                                                               & 10K - 50K                                      & Y                                        & N                                        & N                                        & -                                        \\ \hline
20          & com.wirelesstag.android3                   & TOOLS                                                               & 1K - 5K                                        & Y                                        & N                                        & N                                        & -                                        \\ \hline
21          & com.tado                                   & TOOLS                                                               & 10K - 50K                                      & Y                                        & N                                        & N                                        & -                                        \\ \hline
22          & com.albahra.sprinklers                     & TOOLS                                                               & 5K - 10K                                       & Y                                        & N                                        & Y                                        & Y                                        \\ \hline
23          & com.iconservo.blossom                      & LIFESTYLE                                                           & 1K - 5K                                        & Y                                        & N                                        & N                                        & -                                        \\ \hline
24          & com.skydrop.app                            & TOOLS                                                               & 5K - 10K                                       & Y                                        & N                                        & N                                        & -                                        \\ \hline
25          & com.hydrawise.android2\_2                  & PRODUCTIVITY                                                        & 5K - 10K                                       & Y                                        & N                                        & N                                        & -                                        \\ \hline
26          & com.rachio.iro                             & LIFESTYLE                                                           & 10K - 50K                                      & Y                                        & N                                        & N                                        & -                                        \\ \hline
27          & be.smappee.mobile.android                  & LIFESTYLE                                                           & 10K - 50K                                      & Y                                        & N                                        & N                                        & -                                        \\ \hline
28          & com.smart\_me                              & TOOLS                                                               & 1K - 5K                                        & Y                                        & N                                        & ?                                        & -                                        \\ \hline
29          & com.eyesight.singlecue                     & ENTERTAINMENT                                                       & 1K - 5K                                        & Y                                        & Y                                        & Y                                        & ?                                        \\ \hline
30          & com.amazon.dee.app                         & MUSIC \& AUDIO                                                      & 1M - 5M                                        & Y                                        & N                                        & N                                        & -                                        \\ \hline
31          & com.myn3rd.n3rdremote                      & TOOLS                                                               & 500 - 1K                                       & Y                                        & N                                        & Y                                        & N                                        \\ \hline
32          & com.sensibo.app                            & TOOLS                                                               & 5K - 10K                                       & Y                                        & N                                        & N                                        & -                                        \\ \hline
33          & net.wifisocket.advancedlumonicslabs        & LIFESTYLE                                                           & 100 - 500                                      & Y                                        & N                                        & ?                                        & -                                        \\ \hline
34          & us.hiku.android.app                        & SHOPPING                                                            & 1K - 5K                                        & Y                                        & Y                                        & ?                                        & -                                        \\ \hline
35          & com.dnm.heos.phone                         & MUSIC \& AUDIO                                                      & 100K - 500K                                    & Y                                        & Y                                        & ?                                        & -                                        \\ \hline
36          & com.musaic.musaiccontrol                   & MUSIC \& AUDIO                                                      & 1K - 5K                                        & Y                                        & N                                        & Y                                        & ?                                        \\ \hline
37          & com.beep.android                           & MUSIC \& AUDIO                                                      & 1K - 5K                                        & Y                                        & N                                        & Y                                        & N                                        \\ \hline
38          & com.wifiaudio                              & MUSIC \& AUDIO                                                      & 5K - 10K                                       & Y                                        & Y                                        & Y                                        & N                                        \\ \hline
39          & com.roku.remote                            & ENTERTAINMENT                                                       & 5M - 10M                                       & Y                                        & Y                                        & Y                                        & N                                        \\ \hline
40          & org.qtproject.example.EzeeSync             & \begin{tabular}[c]{@{}l@{}}VIDEO PLAYERS \& \\ EDITORS\end{tabular} & 100 - 500                                      & Y                                        & N                                        & ?                                        & -                                        \\ \hline
41          & com.insteon.insteon3                       & LIFESTYLE                                                           & 50K - 100K                                     & Y                                        & N                                        & Y                                        & ?                                        \\ \hline
42          & com.lutron.mmw                             & LIFESTYLE                                                           & 10K - 50K                                      & Y                                        & N                                        & Y                                        & Y                                        \\ \hline
43          & com.osram.lightify                         & LIFESTYLE                                                           & 10K - 50K                                      & Y                                        & N                                        & Y                                        & N                                        \\ \hline
44          & com.scoutalarm.android                     & TOOLS                                                               & 1K - 5K                                        & Y                                        & N                                        & N                                        & -                                        \\ \hline
45          & com.syabas.iot.iotmobile.android           & LIFESTYLE                                                           & 500 - 1K                                       & Y                                        & N                                        & N                                        & -                                        \\ \hline
46          & com.alyt.lytmobile                         & LIFESTYLE                                                           & 500 - 1K                                       & Y                                        & N                                        & Y                                        & ?                                        \\ \hline
47          & iSA.common                                 & TOOLS                                                               & 10K - 50K                                      & Y                                        & N                                        & Y                                        & N                                        \\ \hline
48          & com.wigwag.wigwag\_mobile                  & LIFESTYLE                                                           & 100 - 500                                      & N                                        & -                                        & -                                        & -                                        \\ \hline
49          & com.irisbylowes.iris.i2app                 & LIFESTYLE                                                           & 10K - 50K                                      & Y                                        & N                                        & N                                        & -                                        \\ \hline
50          & com.webee.mywebee                          & LIFESTYLE                                                           & 1K - 5K                                        & Y                                        & N                                        & N                                        & -                                        \\ \hline
51          & com.codeatelier.homee.smartphone           & PRODUCTIVITY                                                        & 5K - 10K                                       & Y                                        & N                                        & Y                                        & ?                                        \\ \hline
52          & com.zonoff.diplomat.staples                & LIFESTYLE                                                           & 5K - 10K                                       & Y                                        & Y                                        & Y                                        & ?                                        \\ \hline
53          & com.revolv.android.app                     & -                                                                   & -                                              & NOT FOUND                                & -                                        & -                                        & -                                        \\ \hline
54          & com.amazon.storm.lightning.client.aosp     & TOOLS                                                               & 1M - 5M                                        & Y                                        & N                                        & Y                                        & ?                                        \\ \hline
55          & com.belkin.wemoandroid                     & LIFESTYLE                                                           & 100K - 500K                                    & Y                                        & N                                        & Y                                        & N                                        \\ \hline
\end{tabular}
\end{table*}



\end{document}